\documentclass[journal]{IEEEtran}
\ifCLASSINFOpdf
  \usepackage[pdftex]{graphicx}
  \DeclareGraphicsExtensions{.pdf,.jpeg,.png}
\else
  \usepackage[dvips]{graphicx}
  \graphicspath{{Figures/eps/}}
  \DeclareGraphicsExtensions{.eps}
\fi
\usepackage{amsmath}
\usepackage{amssymb}
\usepackage{amsthm}
\newtheorem{definition}{Definition}
\newtheorem{lemma}{Lemma}
\newtheorem{theorem}{Theorem}

\usepackage{algorithmic}
\ifCLASSOPTIONcompsoc
\usepackage[caption=false,font=normalsize,labelfont=sf,textfont=sf]{subfig}
\else
\usepackage[caption=false,font=footnotesize]{subfig}
\fi
\usepackage{hyperref}

\newcommand{\rimulti}{\textit{MultiRI}}

\usepackage{color}
\newcommand{\high}{\textcolor{black}}

\renewcommand{\algorithmicrequire}{\textbf{Input:}}
\renewcommand{\algorithmicensure}{\textbf{Output:}}

\usepackage{paralist}
\begin{document}
%
\title{\rimulti: Fast Subgraph Matching in Labeled Multigraphs}
%
%
%

\author{Giovanni Micale,Vincenzo Bonnici, Alfredo Ferro, Dennis Shasha, Rosalba Giugno and Alfredo Pulvirenti
\thanks{\em GM, AF, and AP are with the Department of Clinical and Experimental Medicine of University of Catania, Italy, E-Mail {gmicale,ferro,apulvirenti@dmi.unict.it}. VB and RG are with the Department of Computer Science of University of Verona, Italy, E-mail: {vincenzo.bonnici, rosalba.giugno@univr.it}, DS is with the Department of Computer Science, New York University, NY, USA, E-Mail: {shasha@cs.nyu.edu}}
\thanks{Manuscript received ...; }}

\maketitle
\begin{abstract}
The Subgraph Matching (SM) problem consists of finding all the embeddings of a given small graph, called the query, into a large graph, called the target. The SM problem has been widely studied for simple graphs, i.e. graphs where there is exactly one edge between two nodes and nodes have single labels, but few approaches have been devised for labeled multigraphs, i.e. graphs having possibly multiple labels on nodes in which pair of nodes may have multiple labeled edges between them.

Here we present MultiRI, a novel algorithm for the Sub-Multigraph Matching (SMM) problem, i.e. subgraph matching in labeled multigraphs. MultiRI improves on the state-of-the-art by computing compatibility domains and symmetry breaking conditions on query nodes to filter the search space of possible solutions.
Empirically, we show that MultiRI outperforms the state-of-the-art method for the SMM problem in both synthetic and real graphs, with a multiplicative speedup between five and ten for large graphs, by using a limited amount of memory.

\end{abstract}


%
\IEEEpeerreviewmaketitle

%
%
%
%

\section{Introduction and Related Work}
Many graphs consist of labeled elements and many relations between elements. For example, two people in a social network can belong to different communities and interact in symmetric (e.g. as friends or colleagues) as well as asymmetric (e.g. one as fan of the other) ways.
Such social networks can be represented as graphs whose nodes are people (or types of people) and edges between two persons may represent multiple interactions.

Similarly, protein-protein interaction graphs consist of nodes representing genes with particular identifiers connected by undirected edges. Combining a protein-protein interaction graph with a directed transcriptional graph can give rise to a graph in which a transcription factor may both physically interact (at the protein level) with a gene and induce its transcription. Thus, such graphs have multiple labels on nodes (e.g., developmental gene and transcription factor) and multiple \high{edges between two nodes} (e.g., a directed transcriptional edge and an undirected protein-protein interaction edge). 

Analyzing such graphs can help to discover common features and underlying mechanisms that occur. Examples of graph analysis include: i) finding motifs, i.e. unexpectedly highly recurrent pattern of interactions, in a single graph \cite{Micale2018,Micale2019,Ribeiro2014,Rinnone2015}, ii) frequent subgraph mining with respect to a graph database or a single graph \cite{Ingalalli2018}, and iii) graph alignment, i.e. finding structurally conserved subgraphs in a collection of graphs \cite{Bayati2013,Klau2009,Micale2014Gasoline,Micale2014Cytoscape,Micale2014Protein,Micale2015}. 
In this paper, we focus on the Subgraph Matching  problem \high{on Multigraphs} called Sub-Multigraph Matching (SMM), i.e. finding all the occurrences of a small query sub-multigraph,  in a larger target multigraph,  a problem known to be NP-complete already on simple graphs \cite{Cook1971}.  
The matching problem can also be inexact, allowing missing nodes and/or edges or different node and/or edge labels (see \cite{Madi2016} for a comprehensive review), or deal with probabilistic graphs \cite{Lian2016}. However, the innovations in this paper pertain only to exact matching.  


While subgraph matching (SM) is NP-complete, the complexity of graph isomorphism is unknown, though there has been excellent work in the area, with applications to image analysis, document processing, biometric identification and natural language processing \cite{Conte2004,Foggia2014,Vento2015,Hu2018}.

The SM problem has attracted much research in the case of simple graphs, \high{i.e. graphs where there is exactly one edge between two nodes, which usually have single labels}. In the literature, several algorithms have been proposed and they can be classified into two major classes according to the paradigm used to solve the matching problem: Tree Search (TS) algorithms and Constraint Propagation (CP) algorithms.

Tree Search algorithms formulate the matching problem in terms of a State Space Representation (SSR), which consists of the exploration of a search space tree. Each state of the SSR corresponds to a partial solution. The search space is visited in a depth-first order and heuristics are devised to avoid exploring parts of the search space. Algorithms of this class include Ullmann's algorithm \cite{Ullmann1976}, VF2 \cite{Cordella2004}, VF3 \cite{Carletti2018}, RI and RI-DS \cite{Bonnici2013,Bonnici2017}.

In Constraint Propagation methods, the subgraph matching problem is represented as a Constraint Satisfaction Problem (CSP). Query nodes are represented as variables and target nodes represent values that can be assigned to such variables. Edges are translated into constraints that must be satisfied. CP algorithms first compute a compatibility domain for each node of the pattern graph and then iteratively propagate the constraints to reduce such domains and filter candidate nodes for matching. CP methods include nRF+ \cite{Larrosa2002}, Focus-Search \cite{Ullmann2010}, Zampelli et al. \cite{Zampelli2010} and LAD \cite{Solnon2010}.

While there are several works on subgraph matching in simple attributed or labeled graphs \cite{Hong2015,Roy2015,Saltz2014}, less attention has been paid to the same problem in \high{labeled multigraphs, i.e. graphs where nodes can have one or more labels and one or more labeled edges between two nodes may exist. The Subgraph Matching problem in labeled multigraph is the so-called} Sub-Multigraph Matching (SMM) problem. Solving an instance of the SMM problem means to find all occurrences of the query in the target, such that all the labels of query nodes \high{are present in the matched target nodes and all labeled edges in the query are present in the matched target edges}. To the best of our knowledge, SuMGra \cite{Ingalalli2016} is the only algorithm proposed to solve the SMM problem. 

SuMGra implements an efficient indexing strategy based on the multigraph properties of the target graph, such as node and edge multiplicities. In particular, two indexing structures are built before the searching phase: i) the vertex signature index, which captures information about the labels of edges incident on target nodes together with their multiplicities, ii) the vertex neighborhood index, which contains information about the neighbors of a given node $u$, the edges connecting $u$ to its neighbors and the labels of each edge. The two indexes are then used to filter candidate nodes for the initial query node and the subsequent nodes during the search phase.

In this paper we introduce MultiRI, a novel and fast subgraph matching algorithm, inspired by RI \cite{Bonnici2013,Bonnici2017}, for labeled \high{multigraphs}. MultiRI performs a series of pre-processing steps to speed up the matching process. These include the computation of: i) compatibility domains, i.e. a set of matchable target nodes, for each query node, ii) the ordering in which query nodes have to be processed during the matching phase, and iii) symmetry breaking conditions \cite{Codish2013} to avoid the outputting of redundant occurrences. Breaking conditions are an aspect that has been already taken into account in the context of motif search \cite{Grochow2007,Ribeiro2014} but that no previous subgraph matching  algorithm has used. Instead, current solutions for the SM and SMM problems ignore redundant occurrences or manage them at the end of the searching process with very naive and time consuming post-processing solutions.

The matching phase uses compatibility domains to scan target node candidates for the match with a query node and applies breaking conditions to avoid the generation of redundant occurrences. Our pre-processing and matching phase strategies result in significant improvements (up to one order of magnitude) across a variety of graphs and queries and enable scaling to multigraphs with millions of nodes and edges.

This paper compares MultiRI  with SuMGra on a comprehensive benchmark of graphs. Results show that MultiRI outperforms SuMGra by up to one order of magnitude. A further scalability analysis shows that MultiRI is capable of dealing with large graphs: it is able to retrieve sub-multigraph occurrences within multigraphs containing millions of nodes and edges in few seconds. 
The paper is organized as follows. In Section \ref{definitions} we give preliminary definitions. Section \ref{riMulti} introduces MultiRI providing details about its searching strategy. Section \ref{experiments} presents the experimental analysis. Section \ref{conclusion} concludes the paper by summing up the key innovations introduced by MultiRI.

\section{Preliminary definitions}
\label{definitions}
These preliminary definitions describe the data types and the problem.
\\
\begin{definition}[Labeled Multigraph]
\label{multigraph}
A labeled multigraph is a quintuple $G=(V,E,\Sigma,\Gamma,\sigma)$ where $\Sigma$ is an alphabet of node labels, $\Gamma$ is an alphabet of edge labels, $V$ is the set of nodes, $E \subseteq (V \times V \times \Gamma)$ is the set of edges, each of which links two nodes and has a label $l \in \Gamma$ and $\sigma:V \rightarrow \mathcal{P}(\Sigma) \setminus \emptyset$ ($\mathcal{P}(\Sigma)$ is the power set of $\Sigma$) is a function that assigns one or more distinct labels to each node in $V$.
\\
\end{definition}

\high{Each node in the graph has a unique \textit{identifier} or \textit{id} belonging to $\mathbb{N}$. With $id(v)$ we denote the id of node $v$. If $(u,v,l) \in E$ $\lor$ $(v,u,l) \in E$, we say that $v$ is a \textit{neighbor} of $u$. A labeled multigraph $G$ is \textit{undirected} iff $\forall (u,v,l) \in E \Rightarrow (v,u,l) \in E$, i.e. all neighbor relationships go both ways, \textit{directed} otherwise.  The \textit{neighborhood} of $u$ is defined as $N(u) = \{ v \in V : \exists\,l\,\in\,\Gamma\,\, s.t. \,\, (u,v,l) \in E\,\lor\,(v,u,l) \in E\}$.
The \textit{neighborhood} of a set of nodes $S$ is the set of nodes that are neighbors of at least one node in $S$, i.e. $N(S) = \{v \in V : \exists\,u\,\in\,$S$ : v \in N(u)\}$}.
For undirected multigraphs, the \textit{degree} of a node $v$, $deg(v)$, is the number of neighbors of $v$.
In the case of directed multigraphs, we distinguish between the out-degree, \high{$deg_{out}(v)=|\{u \in V: \exists\,l\,\in\,\Gamma\,\, s.t. \,\, (v,u,l) \in E\}|$, and the in-degree, $deg_{in}(v)=|\{u \in V: \exists\,l\,\in\,\Gamma\,\, s.t. \,\, (u,v,l) \in E\}|$.}  The \textit{multiplicity} of a node $u$, denoted as $nm(u)$, is the number of labels of $u$. \high{If $e=(u,v,l) \in E$, we define $\gamma(e)=l$. The \textit{multiplicity} of a pair of nodes $u$ and $v$, is the number of edges between $u$ and $v$, i.e. $|\{l \in \Gamma : (u,v,l) \in E\}|$.}

The Sub-Multigraph Matching (SMM) is defined as follows:
\\
\begin{definition}[Sub-Multigraph Matching (SMM)]
\label{MSM}
\high{Let $Q = (V_Q,E_Q,\Sigma,\Gamma,\sigma_Q)$ and $T = (V_T,E_T,\Sigma,\Gamma,\sigma_T)$ two labeled multigraphs}, named \textit{query} and \textit{target}, respectively. The Sub-Multigraph Matching (SMM) problem aims to find an injective function $f: V_Q \rightarrow V_T$, called a \textit{mapping}, which maps each node in $Q$ to a node in $T$, such that the following conditions hold:
\begin{enumerate}
\item \high{$\forall\, q \in V_Q: \sigma_Q(q) \subseteq \sigma_T(f(q))$;}
\item \high{$\forall\, e_Q=(q', q'',l) \in E_Q$: $e_T=(f(q'),f(q''),l) \in E_T$;}
\end{enumerate}
\end{definition}

In other words, in the SMM problem all the labels of a  query node $q$ must be present in the matched target node, but not necessarily conversely (condition 1). Further, each \high{query edge $e_Q$ must have a match with a target edge having the same label of $e_Q$ (condition 2), again not necessarily conversely. If $Q$ and $T$ have the same number of nodes and edges, the mapping is also called an \textit{isomorphism} from $Q$ to $T$ and $Q$ and $T$ are said to be isomorphic.} 

\high{The SMM problem can have more than one solution, i.e. there may exist one or more mappings.} Given a mapping $f$, a \textit{match} of $Q$ in $T$ is defined as the set of pairs of query and target matched nodes \high{$\mathcal{M} = \{(q_1,f(q_1)),(q_2,f(q_2)), ..., (q_k,f(q_k)\}$}, where $k=|V_Q|$.  

\begin{figure}[!ht]
\centering
\includegraphics[width=0.4\textwidth]{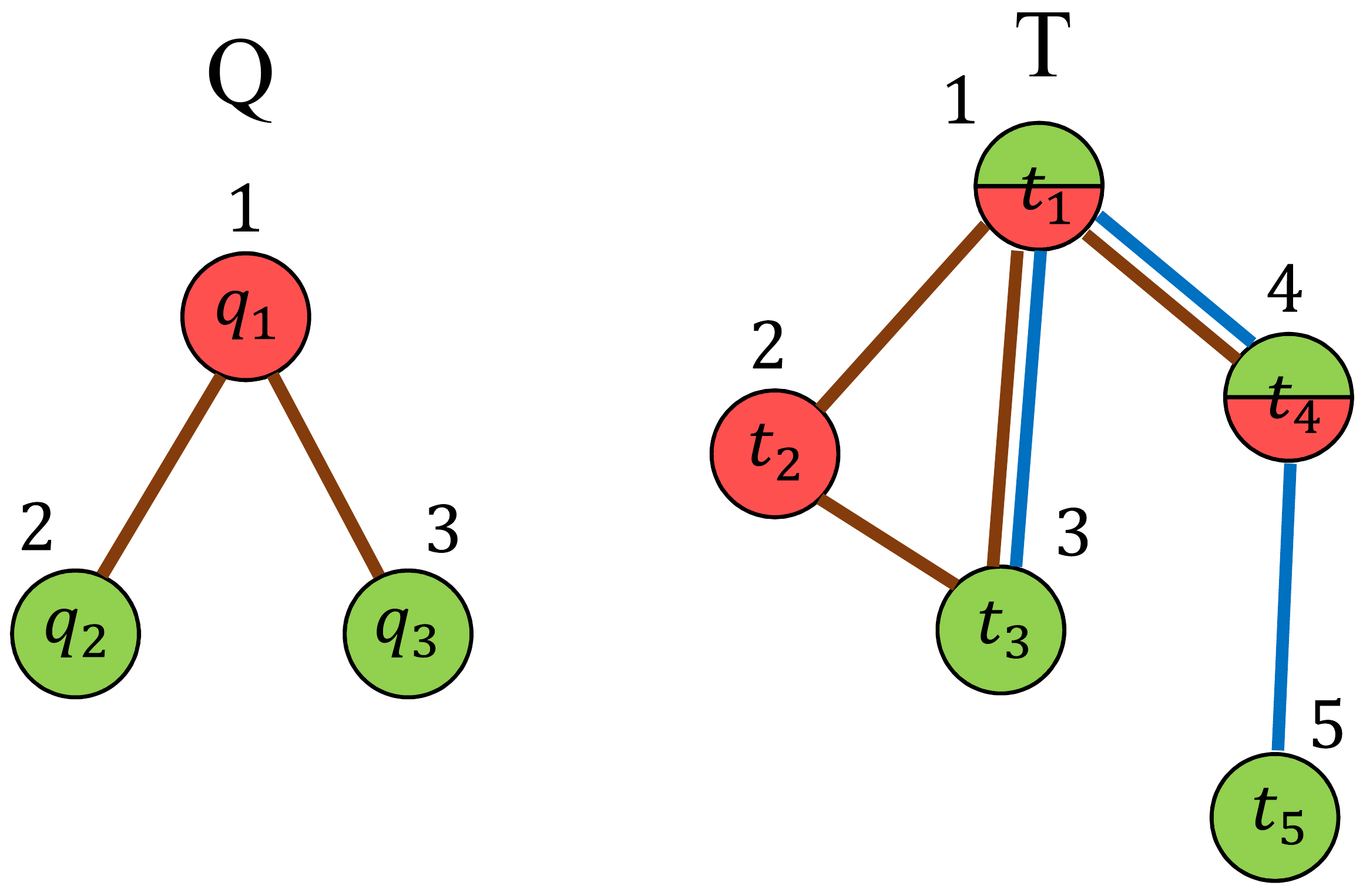}%
\caption{Example of SMM with a query $Q$ and a target $T$. Node ids are drawn outside the circles. \high{Query nodes $q_2$ and $q_3$ each has a green label and a brown labeled outgoing edge pointing to a red labeled node. Therefore, the target nodes that can be matched to query nodes $q_2$ and $q_3$ must have at least a green label and at least a brown labeled outgoing edge pointing to a red labeled node. The target nodes satisfying this condition are $t_1$, $t_3$ and $t_4$. Likewise, query node $q_1$ can be matched only to target nodes $t_1$ and $t_2$. So, there are 4 matches of $Q$ in $T$, namely $\mathcal{M}_1=\{(q_1,t_1),(q_2,t_3),(q_3,t_4)\}$, $\mathcal{M}_2=\{(q_1,t_1),(q_2,t_4),(q_3,t_3)\}$, $\mathcal{M}_3=\{(q_1,t_2),(q_2,t_3),(q_3,t_1)\}$ and $\mathcal{M}_4=\{(q_1,t_2),(q_2,t_1),(q_3,t_3)\}$.}}
\label{MSM-Example}
\end{figure}

\high{Fig \ref{MSM-Example} shows a toy example of SMM with a query $Q$ and a target $T$. In this case there are four matches of $Q$ in $T$, namely $\mathcal{M}_1=\{(q_1,t_1),(q_2,t_3),(q_3,t_4)\}$, $\mathcal{M}_2=\{(q_1,t_1),(q_2,t_4),(q_3,t_3)\}$, $\mathcal{M}_3=\{(q_1,t_2),(q_2,t_3),(q_3,t_1)\}$ and $\mathcal{M}_4=\{(q_1,t_2),(q_2,t_1),(q_3,t_3)\}$}.

\high{An \textit{occurrence} of $Q$ in $T$ is the subgraph $O$ of $T$ formed by nodes $f(q_1), f(q_2), ..., f(q_k)$ and all edges $e_T=(f(q_i),f(q_j),l)$ such that $e_Q=(q_i,q_j,l) \in E_Q$ for all $1 \leq i < j \leq k$. In other words, $O$ is isomorphic to $Q$. \\}

\begin{definition}[Automorphism]
\label{autoDef}
Given a graph $G=(V,E,\Sigma,\Gamma,\sigma)$, an \textit{automorphism} of $G$ is a permutation $\rho$ of the set of nodes $V$ such that:
\begin{enumerate}
\item $\forall v \in V : \sigma(v)=\sigma(\rho(v))$;
\item $\forall u, v \in V : (u,v,l) \in E \Leftrightarrow (\rho(u),\rho(v),l) \in E$ \\
\end{enumerate}
\end{definition}

In other words, an automorphism is a rearrangement of the set of nodes that preserves the structure of a graph and the labels of its nodes and edges. The result of the application of an automorphism is a new graph $G'=(V,E,\Sigma,\Gamma,\sigma)$, obtained from $G$ by permuting the ids of its nodes based on $\rho$. $G'$ is said to be automorphic to $G$.\\

\begin{definition}[Automorphism Matrix]
\label{autoMatrix}
Given a graph $G=(V,E,\Sigma,\Gamma,\sigma)$ with $k$ nodes $v_1, v_2, ..., v_k$ and $h$ automorphisms $\rho_1, \rho_2, \ldots, \rho_h$. The $A$ automorphims matrix, is a matrix containing $h$ rows and $k$ columns, where $A[i,j]=\rho_i(v_j)$. 
\end{definition}

\begin{definition}[Orbit]
\label{orbit}
Given an automorphims matrix $A$, for a given a column of the matrix, $A[,j]$, its set of nodes is an orbit of $G$.\\  
\end{definition}

We denote with $Orb(u)$ the orbit to which node $u$ belongs. 

\begin{figure}[!ht]
\centering
\includegraphics[width=0.4\textwidth]{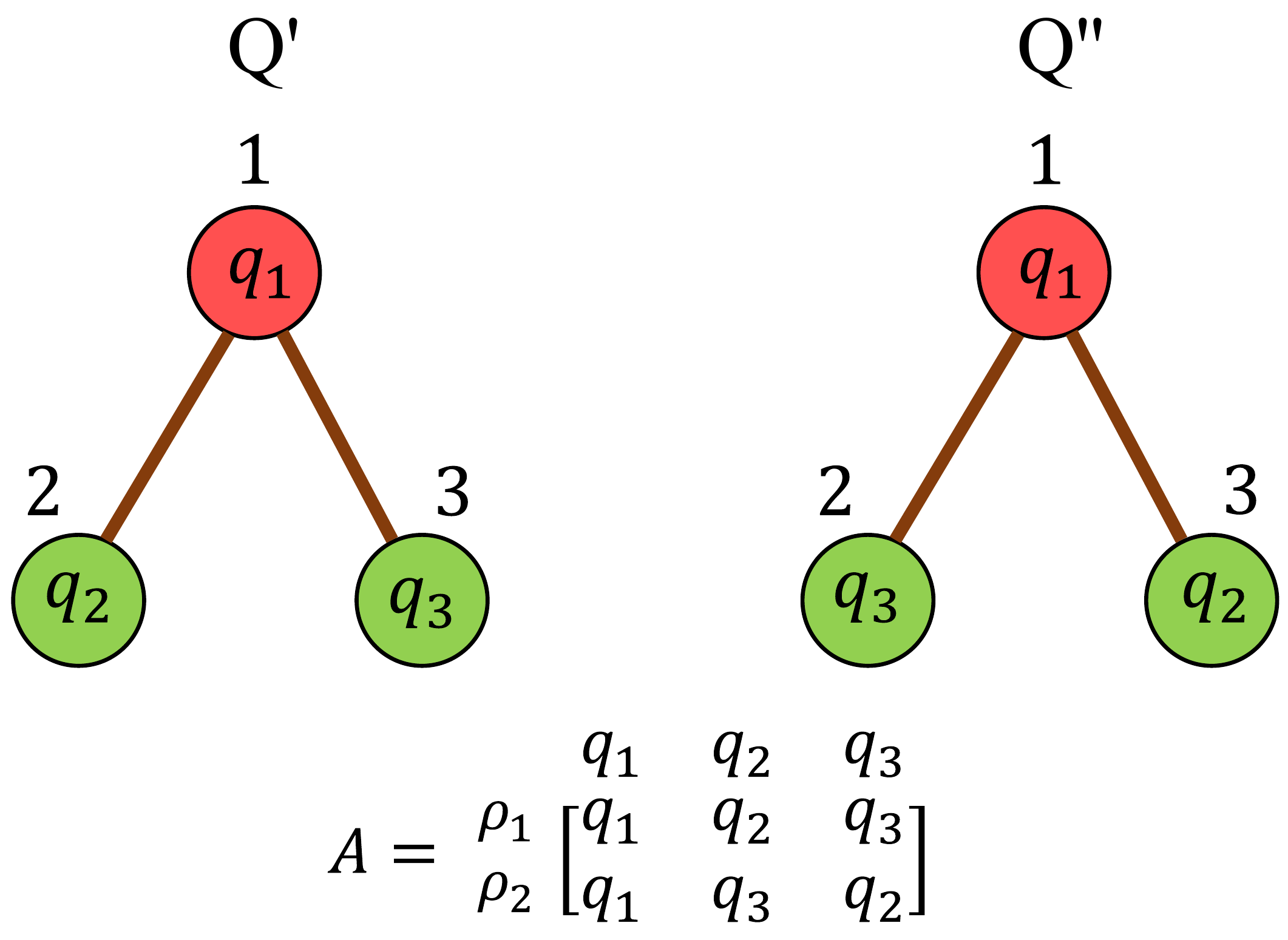}
\caption{\high{The query graph $Q$ of Fig. \ref{MSM-Example} has two automorphisms: a) the identity function $\rho_{1}$ that maps each node to itself, b) the permutation $\rho_{2}$ that swaps node $q_2$ with node $q_3$. The application of the two automorphisms results in graphs $Q'$ and $Q''$, where ids of nodes are permuted based on $\rho_{1}$ and $\rho_{2}$, respectively. The corresponding automorphism matrix $A$ has 2 rows and 3 columns. The first column of $A$ yields one orbit formed by only node $q_1$. The second and third columns yield a second orbit formed by nodes $q_2$ and $q_3$.}}
\label{automorphism}
\end{figure}

Fig. \ref{automorphism} illustrates a toy example of automorphisms and orbits. As the example shows, the number of distinct orbits may be less than the number of columns since two columns may have the same set of nodes. Furthermore orbits may have different cardinalities.

\section{MultiRI algorithm description}
\label{riMulti}
In this section we describe MultiRI, a novel algorithm for solving the SMM problem.

\high{Fig. \ref{multirialgo} describes the main steps of the MultiRI algorithm}: (i) computation of compatibility domains, (ii) ordering of query nodes, (iii) computation of symmetry breaking conditions and (iv) matching process. Below, we describe each step in detail. MultiRI has been implemented in the Java and C++ languages. Both versions are freely available at \url{https://alpha.dmi.unict.it/~gmicale/MultiRI}.

\begin{figure}[!ht]
\begin{algorithmic}[1]
\renewcommand{\algorithmicrequire}{\textbf{Algorithm:}}
\REQUIRE \textsc{MultiRI}
\renewcommand{\algorithmicrequire}{\textbf{Input:}}
\renewcommand{\algorithmicensure}{\textbf{Output:}}
\REQUIRE $Q$: query, $T$: target
\ENSURE $k$: number of occurrences of $Q$ in $T$
\STATE $Dom :=$ \textsc{ComputeDomains} ($Q, T$)
\STATE $\mu :=$ \textsc{OrderQueryNodes} ($Q, Dom$)
\STATE $\mathcal{C} :=$ \textsc{ComputeSymmBreakCond} ($Q$)
\STATE $Matches := \textsc{SubgraphMatching} (Q, T, Dom, \mu, \mathcal{C})$
\RETURN $Matches$
\end{algorithmic}
\caption{Pseudocode of MultiRI algorithm.}
\label{multirialgo}
\end{figure}

\subsection{Computation of compatibility domains}

The first step of MultiRI computes for each query node $q$ the compatibility domain $Dom(q)$ which is the set of nodes in the target graph that \high{could match} $q$ based on node labels and degree. This step speeds up the matching process, because only target graph nodes in $Dom(q)$ are considered to be possible candidates for a match to $q$ during the search. 

Formally, let \high{$Q = (V_Q,E_Q,\Sigma,\Gamma,\sigma_Q)$ be a query labeled multigraph and $T = (V_T,E_T,\Sigma,\Gamma,\sigma_T)$ be a target labeled multigraph}. A node $t \in V_T$ is compatible to a node $q \in V_Q$ iff the following conditions hold:

\begin{enumerate}
\item $\sigma_Q(q) \subseteq \sigma_T(t)$;
\item $\deg(q) \leq \deg(t)$.
\end{enumerate}

Therefore, $t$ is compatible to $q$ iff all the labels of node $q$ are also labels of $t$ (condition 1) and the degree of $q$ is less than or equal to the degree of $t$ (condition 2). In the case of directed multigraphs, condition 2 must hold for both the out-degree and the in-degree of $q$ and $t$.
\high{Computation of domains is detailed in the pseudocode of Fig. \ref{domainsalgo} in the Appendix.}

Referring to the toy example of Fig. \ref{MSM-Example}, target node $t_1$ is compatible to query node $q_1$, because the red node label of $q_1$ is among the labels of $t_1$ and $deg(q_1) = 2 \leq deg(t_1) = 3$. Target nodes $t_2$ and $t_4$ are compatible to $q_1$ too, so the compatibility domain of $q_1$ is $Dom(q_1)=\{t_1,t_2,t_4\}$.

After computing compatibility domains, an Arc Consistency (AC) technique is applied \cite{Boussemart2004,Mackworth1977}.  The AC procedure states that if there exists an edge between two pattern nodes, \high{$q'$ and $q''$}, then a target node in \high{$Dom(q')$} must have at least one neighbor in \high{$Dom(q'')$}, and vice versa. This implies that if a target node $t$ belongs to the domain of a query node $q$ but does not satisfy the AC condition, it can be removed from $Dom(q)$. \high{The AC procedure is detailed in lines 11-19 of Fig. \ref{domainsalgo}}.

\subsection{Ordering of query nodes}

In the next step, MultiRI computes the order in which query nodes have to be processed for the search during the matching. This step is done in the same way as in the RI algorithm \cite{Bonnici2013,Bonnici2017}. The idea is to define a preferred order of processing query nodes, without regard to the target graph. We have found that this works well in practice.

In MultiRI, query nodes that both have high degree and are highly connected to nodes already present in the partial ordering come earlier in the final ordering. \high{Fig. \ref{orderalgo} in the Appendix illustrates the main steps of the algorithm to compute the ordering of query nodes.}

Formally, let $n$ be the number of nodes in the query and \high{$\mu^{i-1} = (q_1, q_2, ..., q_{i-1})$} be the partial ordering up to the $(i-1)$-th node, with $i<n$. \high{We also define the set $\mathcal{U}^{i-1}$ of unordered query nodes, i.e. nodes that are not in the partial ordering $\mu^{i-1}$.} To choose the next node of the ordering, we define for each candidate query node \high{$q$} three sets \high{(Fig. \ref{orderalgo}, lines 10-23)}:

\begin{enumerate}
\item \high{$V_{q,vis}$}: the set of nodes in \high{$\mu^{i-1}$} and neighbors of \high{$q$}; 
\item \high{$V_{q,neig}$}: the set of nodes in \high{$\mu^{i-1}$} that are neighbors of at least one node \high{in $\mathcal{U}^{i-1}$} and connected to \high{$q$};
\item \high{$V_{q,unv}$}: the set of neighbors of \high{$q$} that are not in \high{$\mu^{i-1}$} and are not even neighbors of nodes in \high{$\mu^{i-1}$}. 
\end{enumerate}

The next node in the ordering \high{(Fig. \ref{orderalgo}, lines 24-28)} is the one with: (i) the highest value of \high{$|V_{q,vis}|$}, (ii) in the case of a tie in (i), the highest value of \high{$|V_{q,neig}|$}, (iii) in the case of a tie in (ii), the highest value of \high{$|V_{q,unv}|$}. In case of a tie according to all criteria, the next node is arbitrarily chosen.

\subsection{Computation of symmetry breaking conditions}

One issue arising in subgraph matching is that the matching process may yield redundant occurrences.

\begin{figure}[!ht]
\centering
\subfloat[]
{\includegraphics[width=3in]{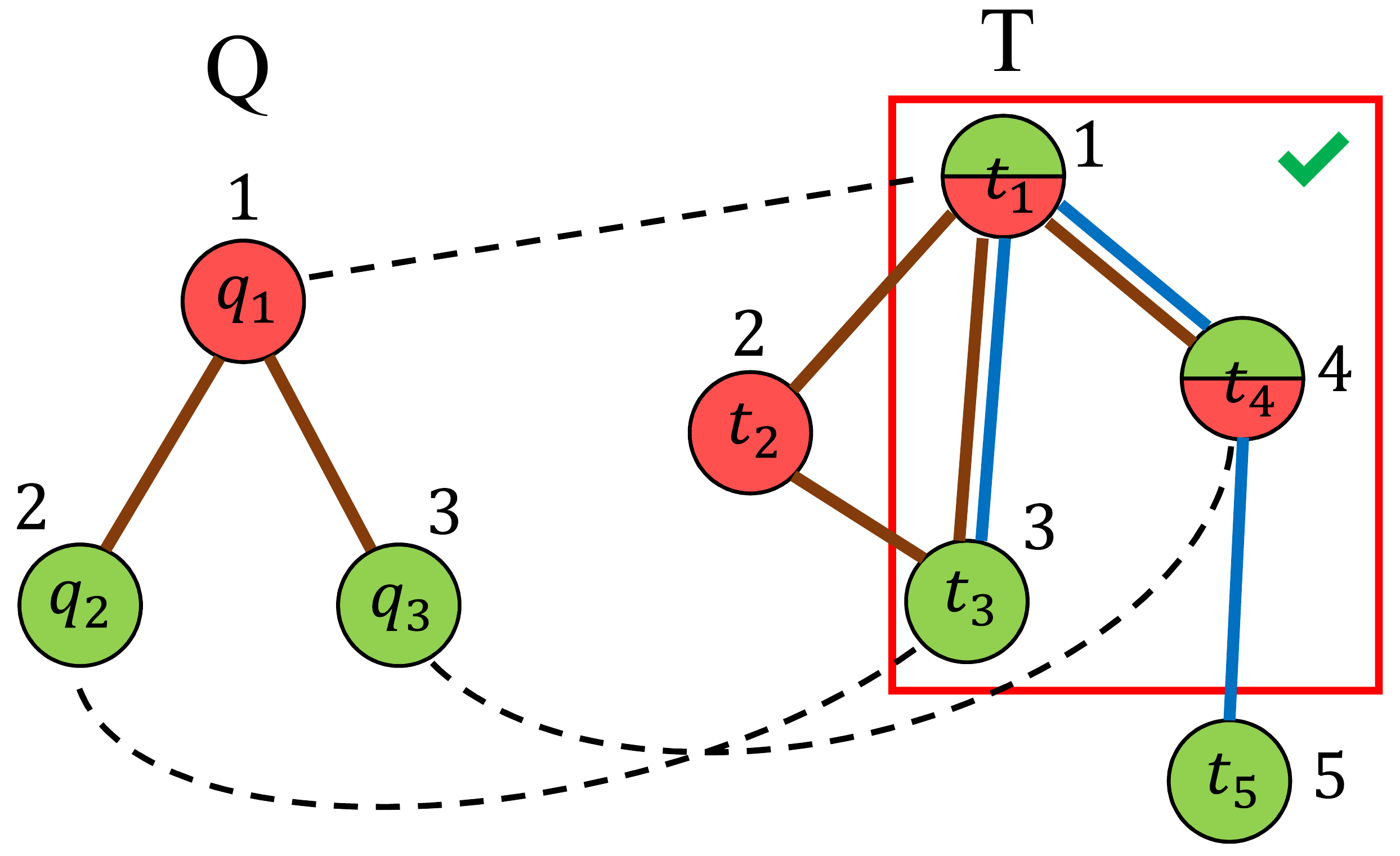}%
\label{BreakCond1}}
\hfill
\subfloat[]
{\includegraphics[width=3in]{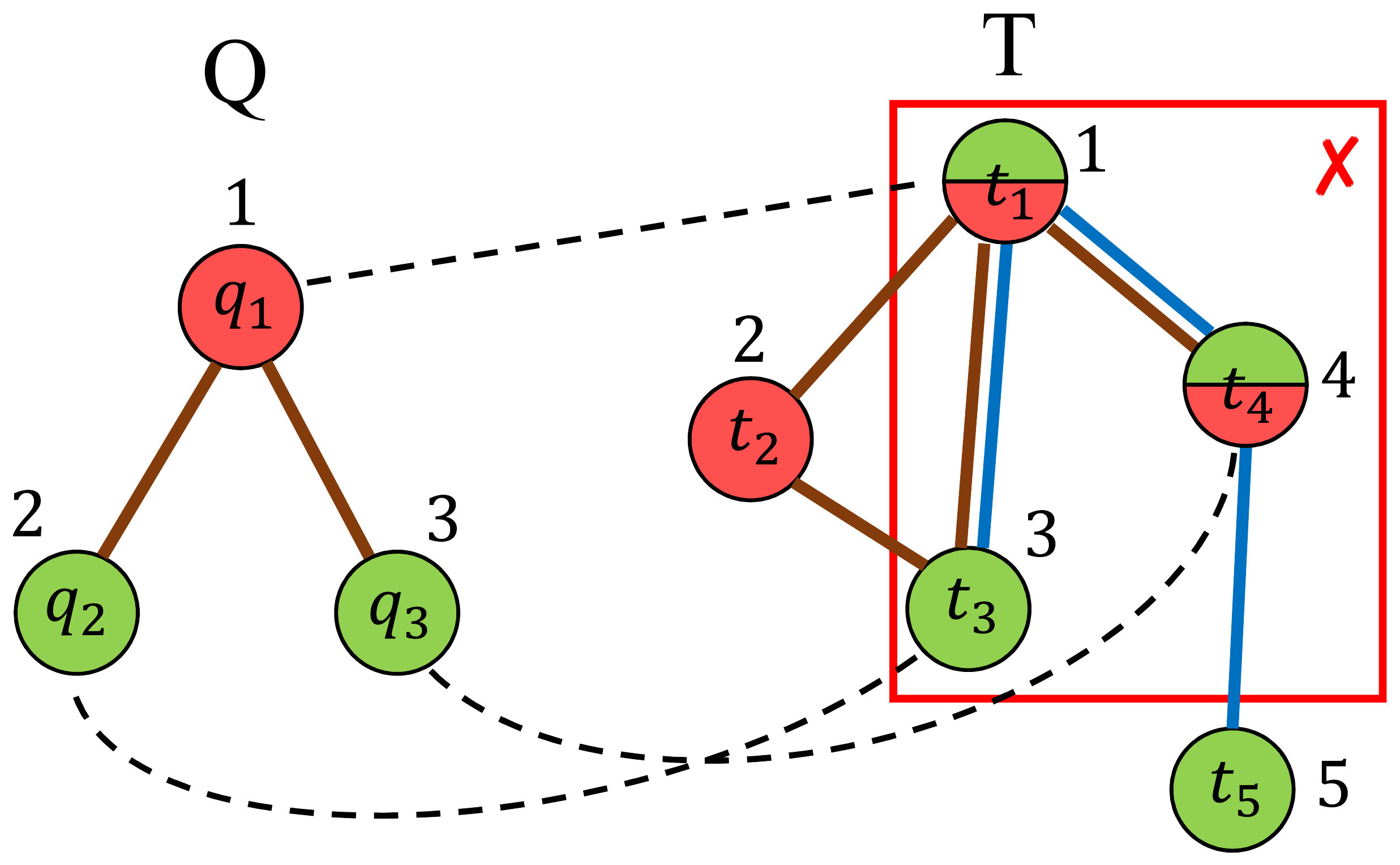}%
\label{BreakCond2}}
\caption{Example of usage of breaking conditions \high{for the query $Q$ and target $T$ of Fig. \ref{MSM-Example}}. Matched query and target nodes are linked by dashed edges. \high{As a result of the automorphism calculation for $Q$, query nodes $q_2$ and $q_3$ are in the same orbit. Since $id(q_2)=2<id(q_3)=3$ we can define a symmetry breaking condition $q_2 \prec q_3$}. By applying the SBC rule for $q_2 \prec q_3$, a) the match $\mathcal{M}_1=\{(q_1,t_1),(q_2,t_3),(q_3,t_4)\}$ is included as an occurrence (since $id(t_3)=3 < id(t_4)=4$), b) the match $\mathcal{M}_2=\{(q_1,t_1),(q_2,t_4),(q_3,t_3)\}$ is discarded (since $id(t_4)=4 > id(t_3)=3$).}
\label{breakCond}
\end{figure}

Fig. \ref{breakCond} depicts a toy example with \high{the query $Q$ and the target $T$ of Fig. \ref{MSM-Example}}. Query nodes $q_2$ and $q_3$ have the same sets of labels, so they are indistinguishable. Since we can map the two nodes indifferently to target nodes with ids $t_3$ and $t_4$, \high{the two matches of $Q$ in $T$, $\mathcal{M}_1=\{(q_1,t_1),(q_2,t_3),(q_3,t_4)\}$ and $\mathcal{M}_2=\{(q_1,t_1),(q_2,t_4),(q_3,t_3)\}$} actually correspond to the same occurrence in $T$. Therefore, one of them should be excluded in the final occurrence count. \high{The same exclusion applies to matches $\mathcal{M}_3=\{(q_1,t_2),(q_2,t_3),(q_3,t_1)\}$ and $\mathcal{M}_4=\{(q_1,t_2),(q_2,t_1),(q_3,t_3)\}$}.

In order to exclude redundant occurrences during the matching process and prune the search space, MultiRI defines symmetry breaking conditions based on the node identifiers within the query graph.

Symmetry breaking conditions are related to the concepts of automorphisms and orbits of the query graph.

Given two query nodes $q$ and $q'$ belonging to the same orbit, with $id(q)<id(q')$, a symmetry breaking condition is an inequality of the form $q \prec q'$, indicating that node $q$ must precede node $q'$. In other words, a symmetry breaking condition is a condition that imposes a relative order between two query nodes belonging to the same orbit.

In the query graph $Q$ of \high{Fig. \ref{automorphism}} nodes $q_2$ and $q_3$ belong to the same orbit and their ids are 2 and 3, respectively. So, we can define a symmetry breaking condition $q_2 \prec q_3$.

The algorithm for computing the set of all symmetry breaking conditions in a query is provided in Fig. \ref{breakingalgo}. It is based on an iterative computation of query automorphisms and orbits starting from the current set of symmetry breaking conditions discovered by the algorithm. Given a set of breaking conditions $\mathcal{C}$, we denote with $Aut_{|\mathcal{C}}$ and $Orb_{|\mathcal{C}}$ the automorphism matrix and the set of orbits respecting the breaking conditions in $\mathcal{C}$, respectively. When $\mathcal{C}$ is empty, $Aut_{|\mathcal{C}}$ corresponds to the automorphism matrix of $Q$ and $Orb_{|\mathcal{C}}$ is the corresponding set of orbits. A breaking condition $q' \prec q$ in $\mathcal{C}$, prevents $q'$ from being mapped to $q$ in any automorphism. Then, $Aut_{|\mathcal{C}}$ becomes the automorphism matrix obtained from the set of all query automorphisms where $q'$ is not mapped to $q$, for all nodes $q'$ and $q$ such that $q \prec q' \in \mathcal{C}$. $Orb_{|\mathcal{C}}$ is the set of orbits relative to $Aut_{|\mathcal{C}}$. 

The first step of the algorithm is the computation of the automorphism matrix of $Q$ (Fig. \ref{breakingalgo}, line 2). This can be accomplished by using any graph matching algorithm (e.g. the NAUTY algorithm \cite{Mckay2014}). Then, the orbits of $Q$ are calculated (Fig. \ref{breakingalgo}, line 3) and the query node $q'$ with minimum id across all orbits with at least two equivalent nodes is computed (Fig. \ref{breakingalgo}, line 5). For each node $q \neq q'$ belonging to the same orbit of $q'$, a new symmetry breaking condition $q' \prec q$ is defined and added to the final set of conditions (Fig. \ref{breakingalgo}, line 7). This step is equivalent to preventing node $q'$ from being mapped to any other node and putting $q'$ in a separate orbit. To be consistent with that, we need to retain from the current automorphism matrix only the rows corresponding to the automorphisms mapping $q'$ to itself (Fig. \ref{breakingalgo}, line 9) and update the relative set of orbits $Orb_{|\mathcal{C}}$ (Fig. \ref{breakingalgo}, line 10). 

Note that line 10 of Fig. \ref{breakingalgo} does not always consist in just creating a new orbit containing $q'$, because putting $q'$ in a separate orbit may also change the orbits of the remaining nodes. Consider, for example, an unlabeled square query with nodes $q_1, q_2, q_3$ and $q_4$ such that $q_1$ is linked to $q_2$, $q_2$ is linked to $q_3$, and so on. At the beginning we have (i) two automorphisms: the identity mapping $\rho_1$ and the permutation $\rho_2$ which swaps $q_1$ with $q_2$ and $q_3$ with $q_4$, and  (ii) two orbits: the first one containing $q_1$ and $q_2$ and the second one containing $q_3$ and $q_4$. If we set the breaking condition $q_1 \prec q_2$, then we discard $\rho_2$. As a result, both orbits change and each node is put in a separate orbit. Steps in lines 5-10 of Fig. \ref{breakingalgo} are iterated until $Aut_{|\mathcal{C}}$ has only one row (corresponding to the identity mapping) (Fig. \ref{breakingalgo}, line 4). This is equivalent to having $k$ orbits with a single node, where $k$ is the number of query nodes.

\begin{figure}[!ht]
\begin{algorithmic}[1]
\renewcommand{\algorithmicrequire}{\textbf{Procedure:}}
\REQUIRE \textsc{ComputeSymmBreakCond}
\renewcommand{\algorithmicrequire}{\textbf{Input:}}
\renewcommand{\algorithmicensure}{\textbf{Output:}}
\renewcommand{\algorithmicprint}{\textbf{break}}
\REQUIRE $Q$: query
\ENSURE $\mathcal{C}$: set of symmetry breaking conditions of $Q$
\STATE $\mathcal{C} := \emptyset$
\STATE $Aut_{|\mathcal{C}} := \textsc{ComputeAutomorphismMatrix}(Q)$
\STATE $Orb_{|\mathcal{C}} := \textsc{ComputeOrbits}(Aut_{|\mathcal{C}})$
\WHILE{$|Aut_{|\mathcal{C}}|>1$}
\STATE $q' := \underset{q \in V_Q : |Orb_{|\mathcal{C}}(q)|>1}{arg\,min} \{id(q)\}$
\FORALL{$q \in V_Q: q \neq q' \land Orb_{|\mathcal{C}}(q)=Orb_{|\mathcal{C}}(q')$}
\STATE $\mathcal{C} := \mathcal{C} \cup \{q' \prec q\}$
\ENDFOR
\STATE $Aut_{|\mathcal{C}} := \{Aut_{|\mathcal{C}}[i] : \rho_{i}(q')=q'\}$
\STATE $Orb_{|\mathcal{C}} := \textsc{ComputeOrbits}(Aut_{|\mathcal{C}})$
\ENDWHILE
\RETURN $\mathcal{C}$
\end{algorithmic}
\caption{Pseudocode for the computation of symmetry breaking conditions.}
\label{breakingalgo}
\end{figure}

\begin{figure}[!ht]
\centering
\includegraphics[width=0.5\textwidth]{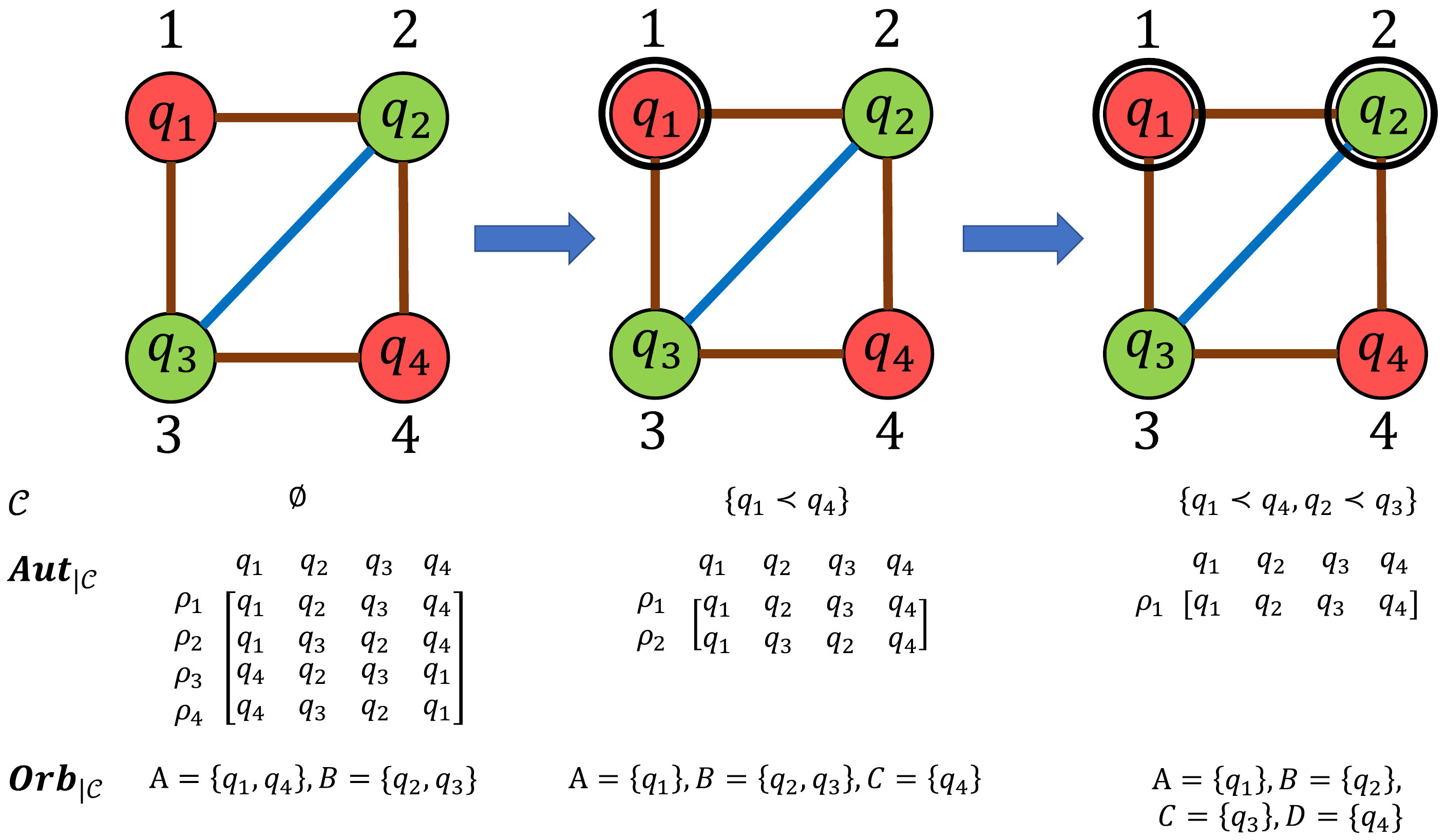}
\caption{Example of application of the Algorithm of Fig. \ref{breakingalgo}, showing the content of $\mathcal{C}$, $Aut_{|\mathcal{C}}$ and $Orb_{|\mathcal{C}}$ sets at each iteration of the while loop (lines 4-11). $\mathcal{C}$ is the set of symmetry breaking conditions discovered by the algorithm, $Aut_{|\mathcal{C}}$ is the automorphism matrix respecting $\mathcal{C}$ and $Orb_{|\mathcal{C}}$ is the set of orbits with respect to $Aut_{|\mathcal{C}}$. Nodes enclosed by black circles are nodes that cannot be permuted by any automorphism.}
\label{CondAlgoExample}
\end{figure}

Fig. \ref{CondAlgoExample} illustrates the execution of the algorithm for computing symmetry breaking conditions in a toy example.

In the Appendix we formally prove that the algorithm of Fig. \ref{breakingalgo} terminates and returns the set of all query's breaking conditions (Lemma \ref{lemmaBreaking}).

\subsection{Matching process}
Following the previously defined ordering of query nodes, MultiRI starts the matching process to find occurrences of the query within the target, using breaking conditions for pruning as it proceeds. \high{Figs. \ref{submatchingalgo} and \ref{matchalgo} in the Appendix detail the steps of the matching process.}

\high{Matching is done by building a mapping function $f: V_Q \rightarrow V_T$ (initially undefined for all query nodes) and the corresponding match $\mathcal{M}$, which is initially empty (lines 3-6 of Fig. \ref{submatchingalgo}).} Whenever a new match between a query node $q$ and a target node $t$ is found, the pair $(q,t)$ is added to $\mathcal{M}$ \high{(Fig. \ref{matchalgo}, line 4)}. When all query nodes have been matched, $\mathcal{M}$ constitutes a new match of $Q$ in $T$, \high{so it can be added to the list of matches found (Fig. \ref{matchalgo}, line 7)}.

\high{The core of the matching process is the recursive \textsc{Match} procedure illustrated in Fig. \ref{matchalgo}}.
Let $\mathcal{M}$ be the partial match found and $q$ a query node that has not been matched yet. \high{If $q'$ is the node that preceeds $q$ in the ordering $\mu$,} the set \high{$Cand(q)$} of candidate target nodes to be matched to $q$ is given by $N(f(q')) \cap Dom(q)$, i.e. the set of nodes which are neighbors of the target node that has been already matched to \high{$q'$ and are in the compatibility domain of $q$ (Fig. \ref{matchalgo}, line 12)}. \high{If $q$ is the first node in $\mu$,} then the set of candidate nodes is just the compatibility domain of $q$ \high{(line 8 of Fig. \ref{submatchingalgo})}.

If some feasibility rules are satisfied, MultiRI adds the pair $(q,t)$ to the partial match $\mathcal{M}$ \high{and updates the mapping function $f$ (Fig. \ref{matchalgo}, lines 3-5)}.

When all query nodes have been matched, a new occurrence of $Q$ in $T$ is found and the corresponding \high{match is added to the list of matches found (Fig. \ref{matchalgo}, line 7)}. 

Whenever all query nodes have been matched or no match has been found for a query node, the algorithm backtracks and continues the search from the last matched node (Fig. \ref{matchalgo}, lines 8-9 and 18-19). \high{Backtracking implies removing both the last pair of matched query and target nodes from $\mathcal{M}$ and the mapping between such nodes using $f$.} When no other matches can be built for any query node, MultiRI stops \high{(Fig. \ref{matchalgo}, line 21)}.

At the end of the matching process, the algorithm returns \high{the list of all matches found (Fig. \ref{submatchingalgo}, line 10)}. 
\\

{\bf Checking for Feasibility.} Feasibility rules are defined in order to i) take into account the links of $t$ to already matched nodes and ii) ensure that the partial match does not result in a redundant occurrence. The latter condition can be satisfied by using symmetry breaking conditions and the following rule:\\

\high{\textit{Symmetry Breaking Condition (SBC) rule}: Let $Q$ be a query with $k$ nodes $q_1, q_2,..., q_k$, $T$ a target, $f$ the mapping between nodes of $Q$ and $T$ and $\mathcal{M}$ the corresponding match. Let $q_i$ and $q_j$ be any two query nodes. If $q_i \prec q_j$ but $id(f(q_i))>id(f(q_j))$, then discard $\mathcal{M}$.}
\\

In other words, the SBC rule prevents certain mappings. Specifically, the SBC rule discards any mapping whose identifier order among the target nodes of a particular query orbit is inconsistent with the identifier  of the query nodes of that orbit.

\high{In the query of Fig. \ref{breakCond}, $q_2 \prec q_3$ is the only symmetry breaking condition. By applying the SBC rule, the match $\mathcal{M}_1=\{(q_1,t_1),(q_2,t_3),(q_3,t_4)\}$ is included as a solution, because $id(f(q_2))=id(t_3)=3 < id(f(q_3))=id(t_4)=4$, whereas the match $\mathcal{M}_2=\{(q_1,t_1),(q_2,t_4),(q_3,t_3)\}$ is discarded, because $id(f(q_2))=id(t_4)=4 > id(f(q_3))=id(t_3)=3$.}

A candidate target node $t$ is matched to a query node $q$ iff the following \textit{feasibility rules} hold:

\begin{enumerate}
\item \high{$\forall \, q' \in V_Q$:  $q \prec q' \land (q',f(q')) \in \mathcal{M} \Rightarrow id(t) < id(f(q'))$;}
\item \high{$\forall \, e_Q=(q,q',l) \in E_Q \,\,s.t.\,\,(q,f(q)) \in \mathcal{M} : e_T=(t,f(q'),l) \in E_T$;}
\end{enumerate}
 
Condition 1 applies the SBC rule and eventually discards a partial match that does not satisfy the order between already matched target nodes imposed by the rule. Condition 2 guarantees that every outgoing edge \high{$e_Q$} from $q$ to already matched query nodes has a corresponding matched edge \high{$e_T$ in the target graph with the same label}.

Thanks to the SBC rule, at the end of the matching process no redundant occurrences will be returned.  Appendix (Lemma \ref{lemmaMatching}) proves this.

\subsection{MultiRI complexity analysis}
\label{complexityAnalysis}

In this section we analyze the time and space complexity of MultiRI. Let $n_Q$ the number of query nodes, $n_T$ the number of target nodes.

The first step of MultiRI is the computation of compatibility domains. The building of domains (Fig. \ref{domainsalgo}, lines 4-10) requires $O(n_Q n_T)$, while the AC procedure (Fig. \ref{domainsalgo}, lines 11-19) takes $O(n_Q^2 n_T^2)$ because, in the worst case, for each query outgoing edge from $q'$ we need to check all outgoing edges from target nodes in $Dom(q'')$. So, computation of compatibility domains requires $O(n_Q^2 n_T^2)$.

Then, MultiRI computes the ordering of query nodes for the matching process. The core of this step is the building of sets $V_{q,vis}$, $V_{q,neig}$ and $V_{q,unv}$ (Fig. \ref{orderalgo}, lines 13-21), which requires $O(n_Q^2)$.

The third step of MultiRI is the computation of symmetry breaking conditions. The calculation of the automorphism matrix for the query (Fig. \ref{breakingalgo}, line 2) is at least as difficult as solving graph isomorphism \cite{Mathon1979}, and its complexity is bound by $O(2^{n_Q})$. Orbits are computed (Fig. \ref{breakingalgo}, line 3) by scanning the automorphism matrix column by column. Since the number of automorphisms of a graph with $n$ nodes is at most $n!$ (in the case the graph is complete), calculating orbits requires $O(n_Q!n_Q)$. In the worst case at each step of the while loop (Fig. \ref{breakingalgo}, lines 4-11) the number of automorphisms decreases only by one and the loop is executed $n_Q!$ times. So, the computation of breaking conditions takes $O(n_Q!^2 n_Q)$.

Finally, we evaluate the computational complexity of the matching process (Fig. \ref{matchalgo}). For ease of simplicity, we start with the case in which we do not have labels on nodes and edges and breaking conditions are not applied. In Section \ref{symmetryExp}, we will empirically show how breaking conditions reduce the computation time of MultiRI. The computational complexity of the matching process depends on: i) the number of examined candidate pairs for the matching and ii) the time needed to check the feasibility rules for each candidate pair. In the worst case, the compatibility domain of each query node has $n_T$ nodes and the number of neighbors of each query node is at most $n_T-1$. Therefore, the set of candidate nodes for the initial query node in the ordering has at most $n_T$ nodes ( Fig. \ref{submatchingalgo}, line 8). Once a new pair has been added to the partial match, the set of candidate nodes for the next query node in the ordering (Fig. \ref{matchalgo}, line 12) is at most $n_T-i$, where $i$ is the number of already matched target nodes. By summing up, the total number of examined candidate pairs is $n_T+n_T(n_T-1)+n_T(n_T-1)(n_T-2)+ ... +n_T(n_T-1)...(n_T-n_Q)$. Since the last term of the summation is dominant, this sum is $O(n_T!/(n_T - (n_Q+1)!))$, i.e. $O(n_T!/(n_T-n_Q)!)$. For each candidate pair, checking the feasibility rules (Fig. \ref{matchalgo}, line 2) requires $O(n_Q)$, because we are ignoring breaking conditions. Therefore, the complexity of the matching process is $O(n_Q \times n_T!/(n_T-n_Q)!)$, which is also the time complexity of MultiRI.

When we also consider breaking conditions, checking the feasibility rules in the matching process now requires $O(2n_Q)$, which is again $O(n_Q)$. The number of explored candidate pairs scales by a factor $f$, which depends on the topology of the query. Therefore, the complexity of MultiRI becomes $O(n_Q \times n_T!/(n_T-n_Q)!)/f$.

For some special classes of topologies, we can provide theoretically an upper bound for $f$. For instance, in the case of paths with any number of nodes, $f$ is at most 2. Therefore, a path can be traversed twice in two opposite directions, following two different orders of nodes. In this case, however, breaking conditions require the traversal of the path in only one direction. For stars, $f$ is at most $(n_Q-2)!$, because we need to match the central node and one of the external nodes of the star in all possible ways, while the remaining $n_Q-2$ nodes will be matched iff the ids of the candidate target nodes for the match will follow the order imposed by breaking conditions. Likewise, for a clique, $f$ is bounded by $(n_Q-1)!$, because the first query node can be matched in all possible ways, but matches for the remaining $n_Q-1$ nodes are subjected to breaking conditions. 

Compared to MultiRI, the computational complexity of SuMGra is the same except for the factor $f$ related to breaking conditions, because in the worst case the number of examined candidate pairs for matching and the time needed to check if a pair can be added to the partial match are the same.

For general labeled multigraphs, a theoretical evaluation of the computational complexity of MultiRI becomes very unwieldy, because there are many ways in which node and edge labels can be combined and this has to be related to the multiplicity of nodes and edges and query's topology too. For such general graphs, our experimental results show the impact of each of these features on the time performance of our algorithm. Please see Section \ref{symmetryExp}.

Regarding the spatial complexity of MultiRI, given the edge multiplicity in the query $m_Q$ and the target graph $m_T$ respectively, the query and the target graph require $O(n_Q^2 \times m_Q)$ and $O(n_T^2 \times m_T)$ space in the worst case, respectively. Additional data structures used by the algorithm during the search include the set of compatibility domains for each query node ($O(n_Q n_T)$ space in the worst case), the set of symmetry breaking conditions for the query ($O(n_Q^2)$ space), the mapping and the partial match (both require $O(n_Q)$ space) and the set of candidate target nodes for the matching for each query node (which costs $O(n_Q n_T)$). Therefore, the total spatial complexity of MultiRI is $O(n_Q^2 \times m_Q)+O(n_Q n_T)+O(n_T^2  \times m_T)$. Assuming $n_Q << n_T$ and $m_Q \simeq m_T$, this is dominated by $O(n_T^2*m_T)$, i.e. the space needed to store the target graph.

\section{Experimental Analysis}
\label{experiments}
In this section we first analyze the performance of MultiRI on a benchmark of synthesized \high{labeled multigraphs} having varying number of nodes, varying number of node and edge labels, and varying \high{node and edge multiplicities}. We next test our algorithm on a dataset of real graphs of medium size (tens of thousands nodes and edges). For both real and synthetic graphs, we compare MultiRI with SuMGra \cite{Ingalalli2016}, the state-of-the art algorithm for the SMM problem. We use a customized version of SuMGra provided by the authors that can handle multiple labels on nodes. \high{In all experiments we evaluate the running times and the memory usage of both algorithms}. Finally, we test the scalability of our method on a large actor association graph with millions of nodes and edges. 

\high{For the experimental analysis} presented here we used the Java implementation of MultiRI. (Java is both more portable and is usually faster than our C++ version.)
All experiments have been performed on an Intel Core i3-3240 CPU 3.40 Ghz with 32 GB RAM.

\subsection{Synthetic dataset}
\label{synthData}

To test the performance of MultiRI and SuMGra on graphs of different size and \high{variable number of node labels and edge multiplicities,} we generated a benchmark of artificial undirected \high{labeled multigraphs}.

We used Barabasi-Albert \cite{Barabasi1999} as the generative model. This model adopts the preferential attachment principle: a new node $u$ enters the graph and creates a link to an existing node $v$ with a probability proportional to the out-degree of $v$ (which is the same as the degree for the undirected case). The Barabasi-Albert model describes the structure of graphs having few nodes with high degree, called hubs, and many nodes with low degree.

Artificial graphs were generated based  on three values for each of the following features in the target graph:

\begin{itemize}
\item Number of  nodes $N$: 10,000, 20,000 and 50,000;
\item Density  $d$, i.e. the ratio between the number of edges and the number of nodes: 2, 5 and 10 \high{(multiple edges between two nodes are counted only once)};
\item Number of \high{distinct} node labels $\sigma$: 2, 10 and 20;
\item Maximum node multiplicity $NM$ (i.e. maximum number of labels that any single node has): 2, 4 and 6;
\item Number of \high{distinct} edge labels $\gamma$: 2, 10 and 20;
\item Maximum edge multiplicity (i.e. maximum number of edges there can be between two nodes) $EM$: 2, 4 and 6.
\end{itemize}

We generated 10 graphs for each combination of values of these features. When $\sigma=2$, $NM$ is set to 2. Likewise, when $\gamma=2$, $EM$ is set to 2. Labels for a node $u$ are numeric values between 1 and $\sigma$ and are chosen as follows: first, the node multiplicity is set to a random value $m$ between 1 and $NM$, then $m$ distinct random labels are assigned to $u$. Edge labels are numeric values between 1 and $\gamma$ and are chosen similarly to node labels.

Queries for the experiments were randomly extracted from synthetic graphs as follows. We considered queries with $k=4,10,16$ nodes and, for each value of $k$, we extracted 10 queries with $k$ nodes from graph $G$ as follows:

\begin{enumerate}
\item Compute the set $S$ of all connected components of $G$ with at least $k$ nodes in the graph;
\item Select uniformly and randomly a connected component $C$ in $S$;
\item Select  a node $u$ randomly with uniform distribution from $C$ and do a random walk with restart which ends when exactly $k$ nodes in $C$ have been visited;
\item Consider the graph $Q$ formed by all nodes \high{and edges traversed during random walk};
\item Call $R$, the set of edges between nodes of $Q$ that are not present in $E_Q$, the \textit{remaining set}. Pick a random number $r$ uniformly between 0 and $|R|$, select randomly and without replacement $r$ edges in $R$ and add them to $E_Q$.
\end{enumerate}

The final dataset includes 5,880 synthetic graphs and 132,300 queries.

\subsection{Real dataset}

The performance of MultiRI and SuMGra have also been evaluated on a dataset of five real \high{labeled multigraphs}. Tab. \ref{realNets} describes the main features of these graphs.

\begin{table}[!ht]
\scriptsize
\centering
\begin{tabular}{|c|c|c|c|c|c|c|c|}
\hline
Graph & Type & \textbf{$|N|$} & \textbf{$|E|$} & \textbf{$|\Sigma|$} & \textbf{$|\Gamma|$} & \textbf{$\max$} & \textbf{$\max$} \\
& & & & & & \textbf{$NM$} & \textbf{$EM$} \\
\hline
\textsc{openflights} & U & 2,712 & 15,549 & 3 & 5 & 3 & 5 \\ \hline
\textsc{foldoc} & D & 6,667 & 17,853 & 20 & 6 & 5 & 1 \\ \hline
\textsc{ppiyeast} & U & 2,327 & 64,342 & 18 & 2 & 14 & 2 \\ \hline
\textsc{swissleaks} & D & 278,669 & 504,672 & 4 & 53 & 2 & 6 \\ \hline
\textsc{imdb} & U & 692,052 & 920,406 & 28 & 3 & 3 & 3 \\ \hline
\end{tabular}
\vspace{6pt}
\label{realNets}
\caption{Database of real multigraphs. For each multigraph, the number of nodes, number of edges, number of node and edge labels and maximum multiplicity of nodes and edges are reported, respectively. U = Undirected, D = Directed.}
\end{table} 

\textsc{openflights} is an undirected graph extracted from OpenFlights.org\footnote{\url{https://openflights.org}}, a repository of data about airports, airlines, airplanes and routes. Nodes of \textsc{openflights} are world cities and node labels are the types of airports that are present in that city: air terminals, train stations and/or ferry terminals. Edges connect two cities if there is at least one air route linking them. Edges are labeled with the types of airplanes that fly between the two cities: 'short range', 'short to medium range', 'medium range', 'medium to long range' and 'long range' planes.

\textsc{foldoc} is a semantic directed graph taken from the on-line computing dictionary FOLDOC\footnote{\url{http://foldoc.org}}, where nodes are computer science terms and edges connect two terms $X$ and $Y$ iff $Y$ is used to explain the meaning of $X$ \cite{Batagelj2002}. Computing terms in \textsc{foldoc} have been labeled according to their domains (e.g. 'computer science', 'hardware', 'programming', 'operating systems'). Edge labels represent the number of times a term is used to explain the meaning of another term.

\textsc{ppiyeast} is a protein-protein interaction network downloaded from Saccharomyces Genome Database (SGD)\footnote{\url{http://www.yeastgenome.org}}. Proteins in \textsc{ppiyeast} have been annotated using the MIPS functional catalogue FunCat \cite{Ruepp2004}. So, node labels represent one or more functions or processes in which a protein is involved. Edges are labeled according to the type of interaction (physical and/or genetic), which is included in the SGD interaction data.

\textsc{swissleaks} is a relationship directed graph of people involved in the Swiss Leaks, a journalistic investigation in 2015 about giant tax evasion involving the British bank HSBC and its Swiss subsidiary, the HSBC Private Bank. Data are provided by the International Consortium of Investigative Journalists (ICIJ)\footnote{\url{https://github.com/swissleaks/swiss\_leaks\_data}}. Nodes are people, companies and addresses and they are labeled according to their type: 'officer' and/or 'master client' for people, 'address' for addresses and 'entities' for companies. Edges denote the relationships between two nodes (e.g. 'director of','shareholder of' or 'registered address').

\textsc{imdb} is an association graph of movies extracted from the Internet Movie DataBase (IMDB)\footnote{\url{http://www.imdb.com}}. Nodes are movies and an edge connects two movies if they share one or more actors, directors, producers and/or writers. Movies are labeled according to their genres (e.g. 'horror', 'comedy', 'drama', 'thriller'). Edges are labeled according to the role of people shared by two movies (i.e. 'actors', 'directors', 'writers' and/or 'producers').

\subsection{Comparison between MultiRI and SuMGra \high{on synthetic graphs}}

We first tested MultiRI and SuMGra on the graphs of the synthetic dataset. For each graph and each query of the dataset, we ran both algorithms and measured the corresponding running time \high{and memory usage}. In each experiment we fixed a timeout of 5 minutes for the execution of an algorithm.

To analyze the impact of each graph feature used to build the synthetic dataset on the performance, we varied one feature at a time and fixed all other features to their median value. Therefore, for example, if the feature of interest is the number of nodes $N$, all experiments with $d=5$, $\sigma=10$, $NM=4$, $\gamma=10$ and $EM=4$ are considered and partitioned into three groups according to the value of $N$. Running times and \high{memory consumption} for MultiRI and SuMGra were then plotted as boxplots. \high{Running times were considered only for} experiments in which both algorithms ended before the timeout, \high{while for memory usage we took into account all tests}. Considering all experiments on synthetic graphs, both algorithms ended before the 5 minute timeout in 95.1\% of the cases. In 0.3\% of the experiments only MultiRI finished before the timeout, while in 0.3\% of the cases only SuMGra ended before the timeout. In 4.3\% of the experiments neither algorithm completed before the timeout. 

\begin{figure*}[!ht]
\centering
\begin{tabular}{c}
\subfloat[]
{\includegraphics[width=0.24\linewidth]{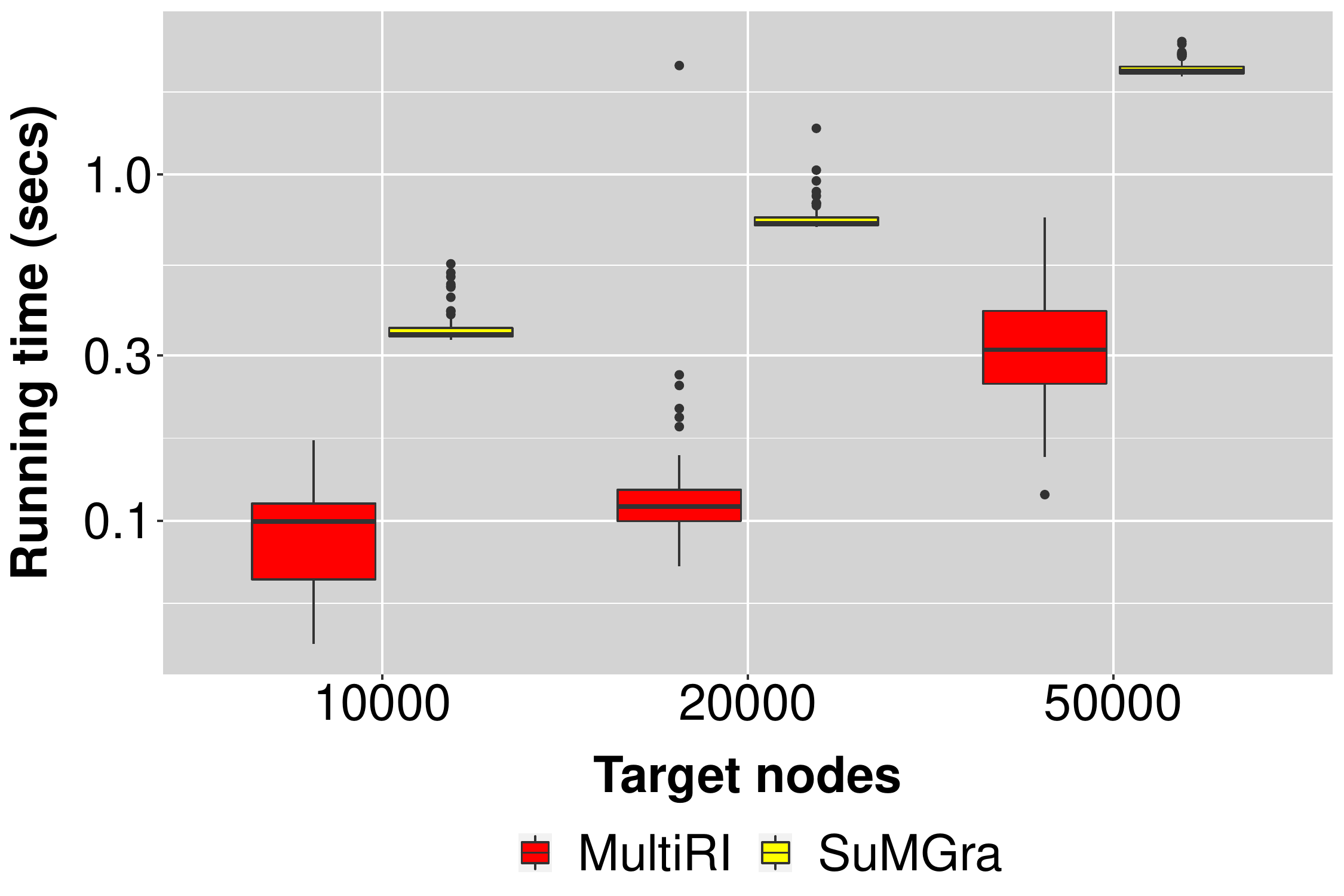}
\label{barabasi_num_nodes}}
\subfloat[]
{\includegraphics[width=0.24\linewidth]{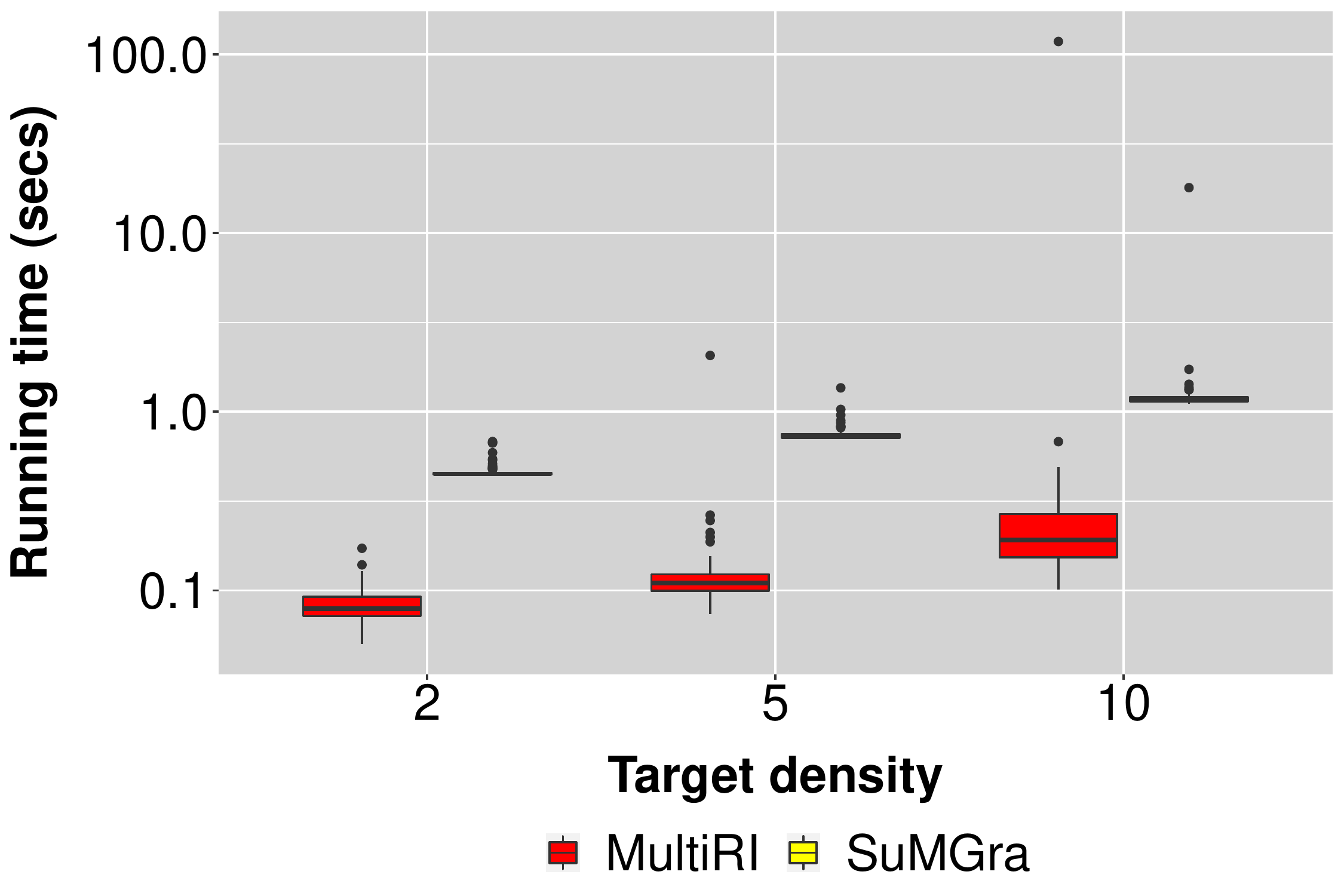}
\label{barabasi_density}}
\subfloat[]
{\includegraphics[width=0.24\linewidth]{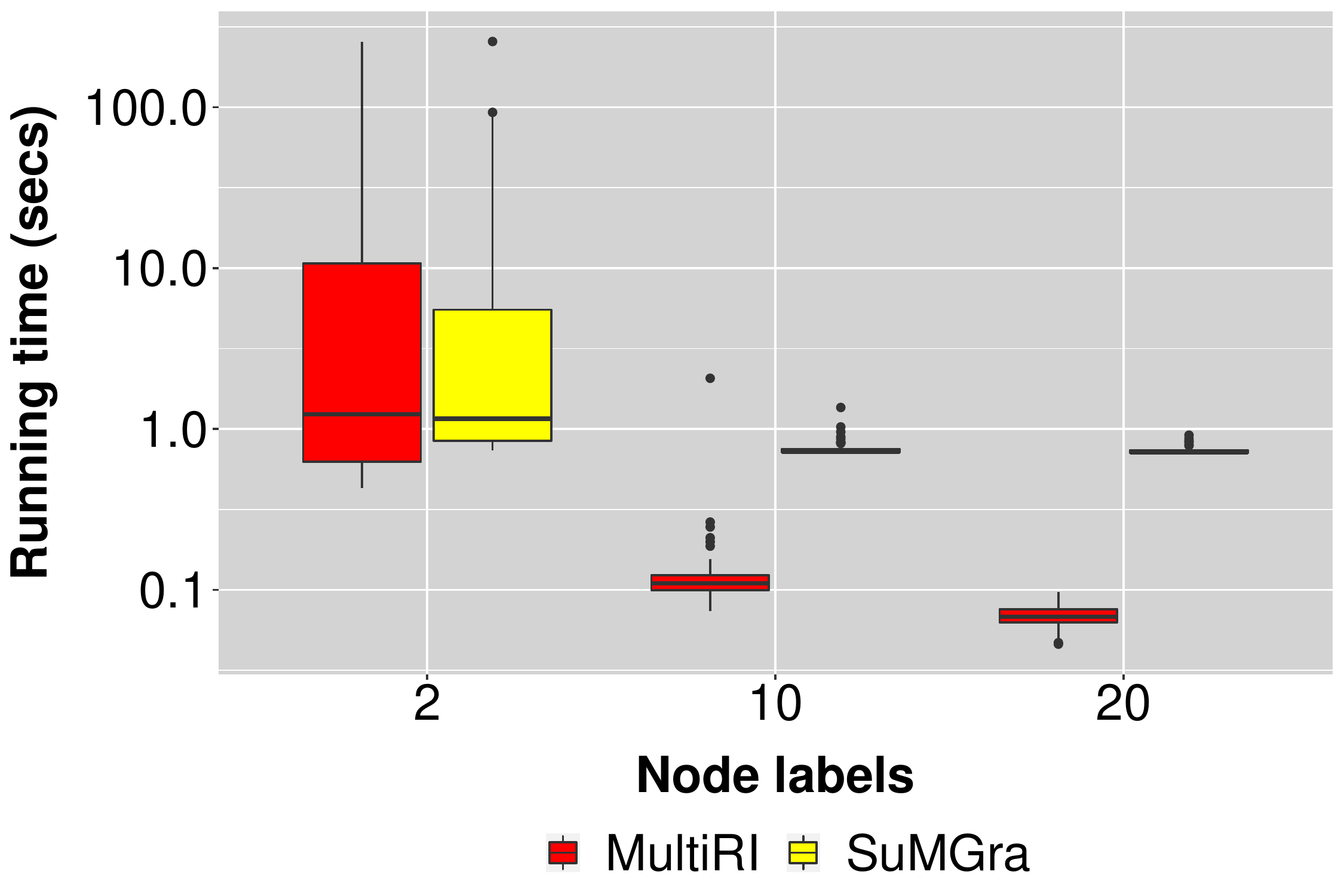}
\label{barabasi_num_node_labs}}
\subfloat[]
{\includegraphics[width=0.24\linewidth]{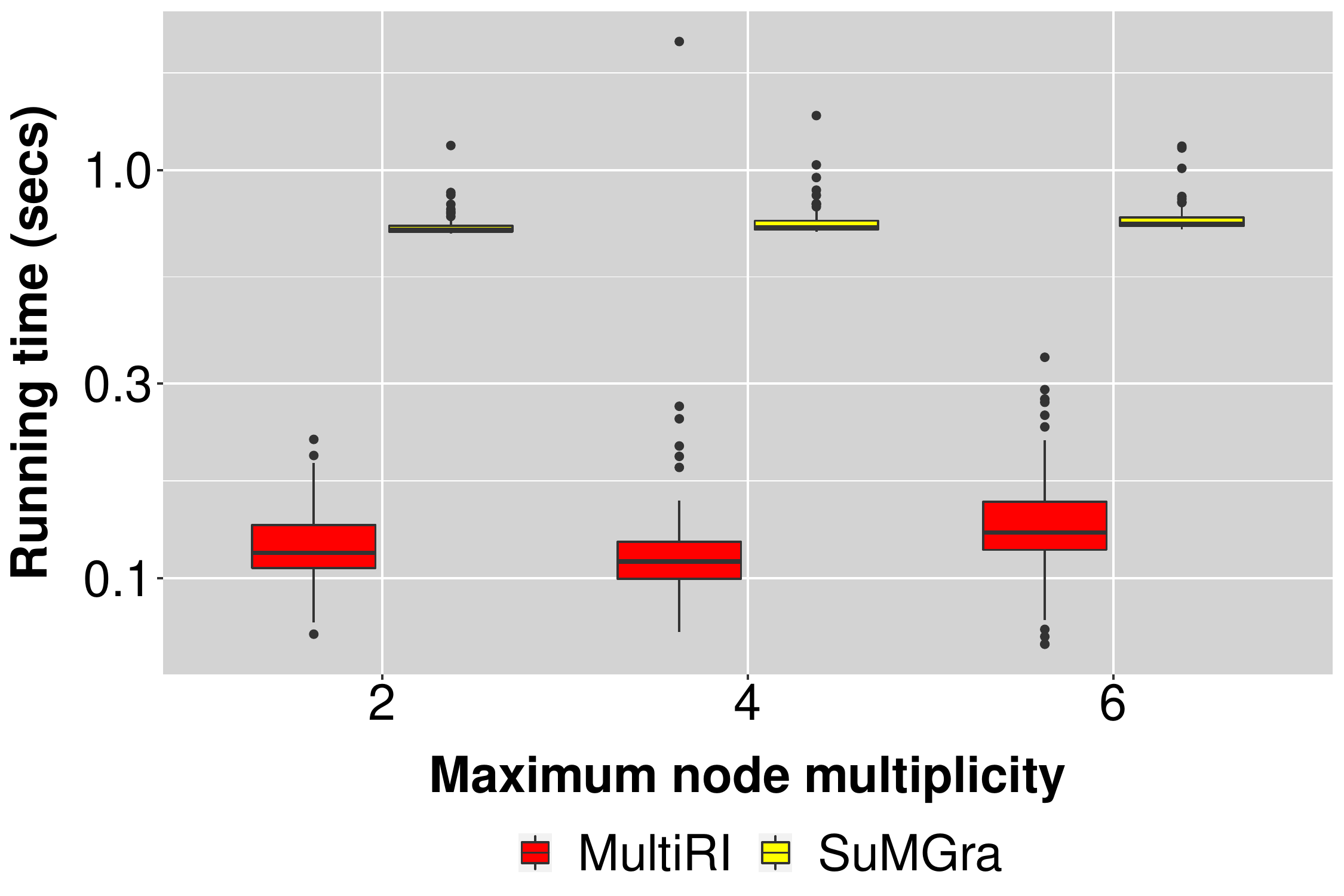}
\label{barabasi_multiplicity_node}}
\\
\subfloat[]
{\includegraphics[width=0.24\linewidth]{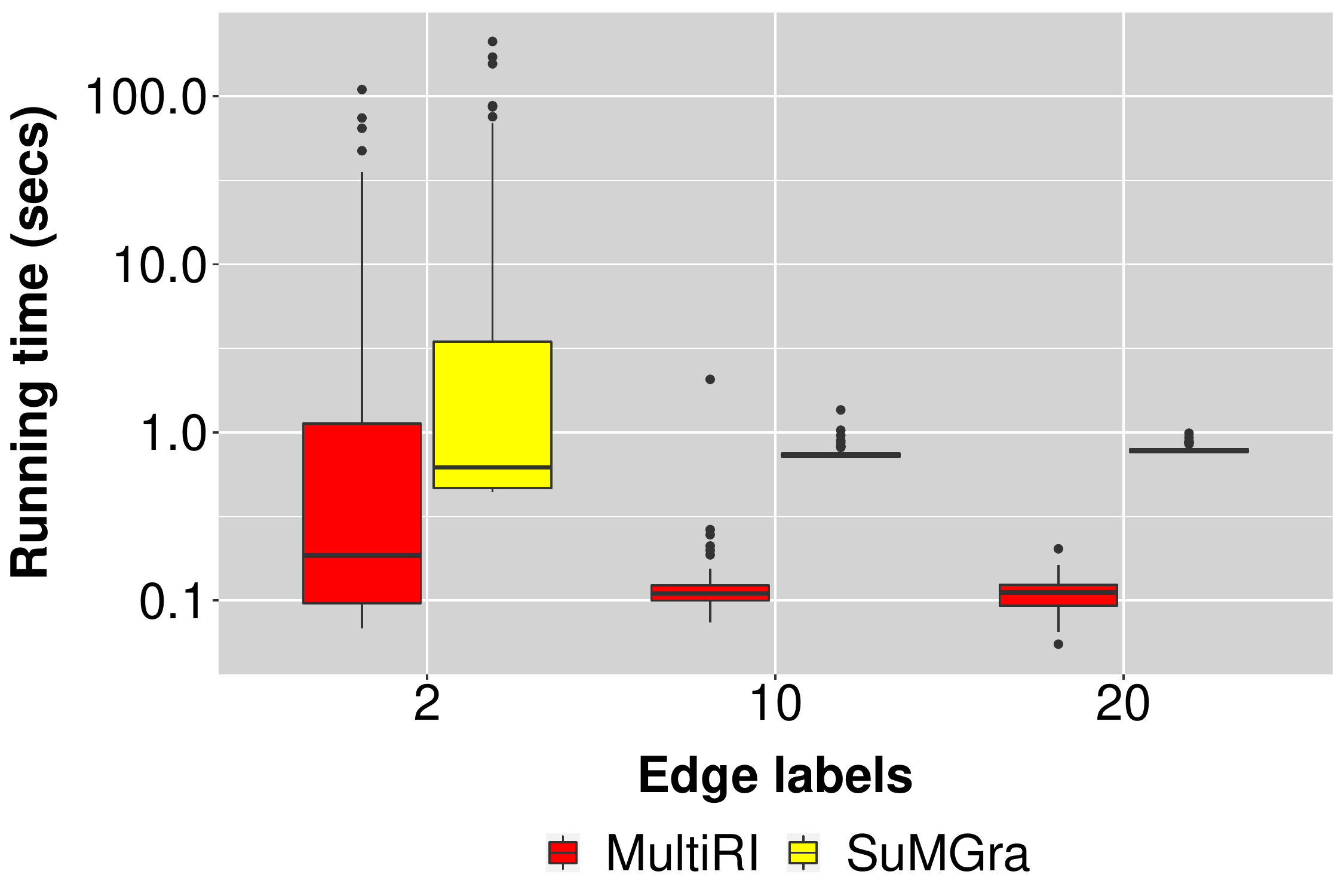}
\label{barabasi_num_edge_labs}}
\subfloat[]
{\includegraphics[width=0.24\linewidth]{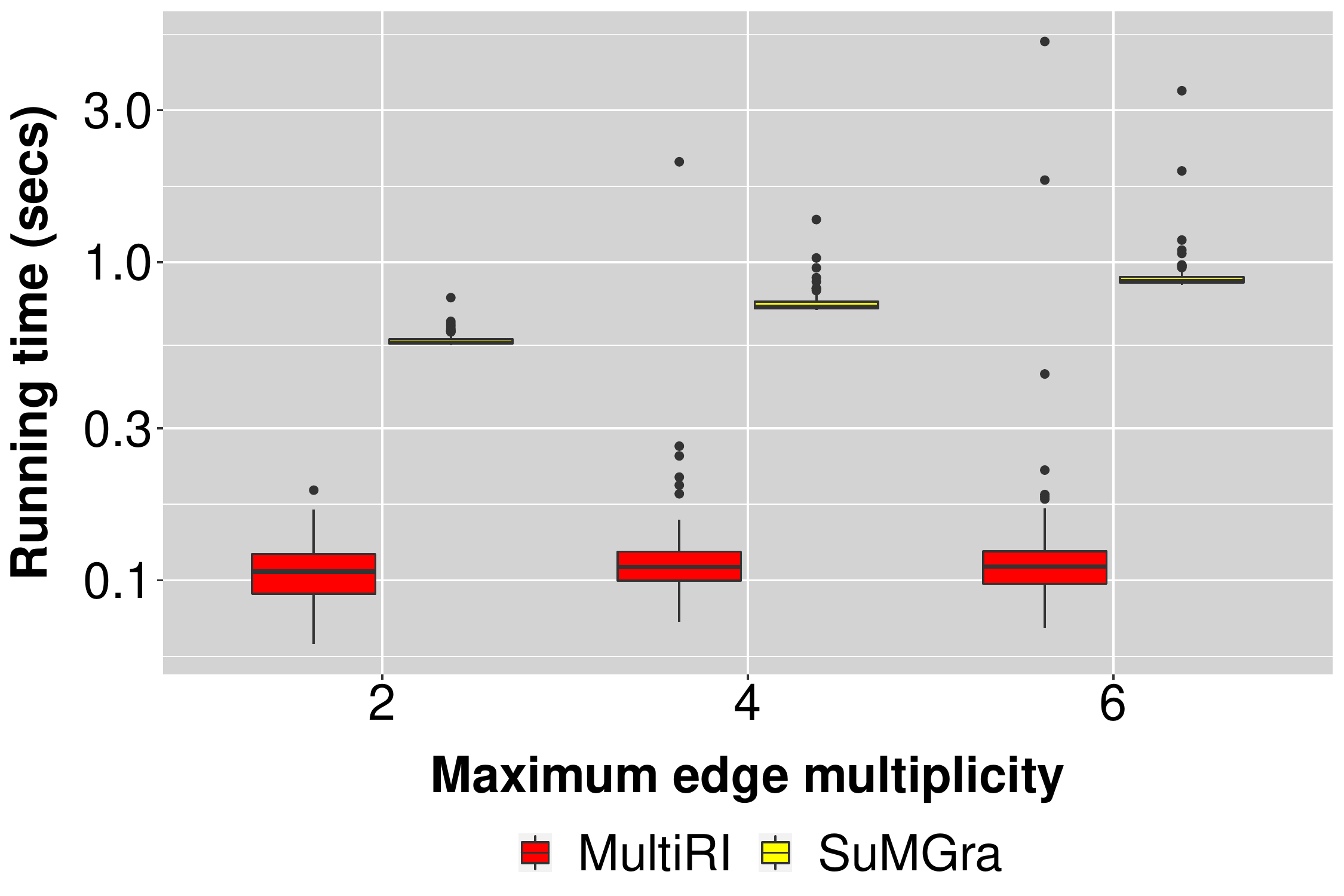}
\label{barabasi_multiplicity_edge}}
\subfloat[]
{\includegraphics[width=0.24\linewidth]{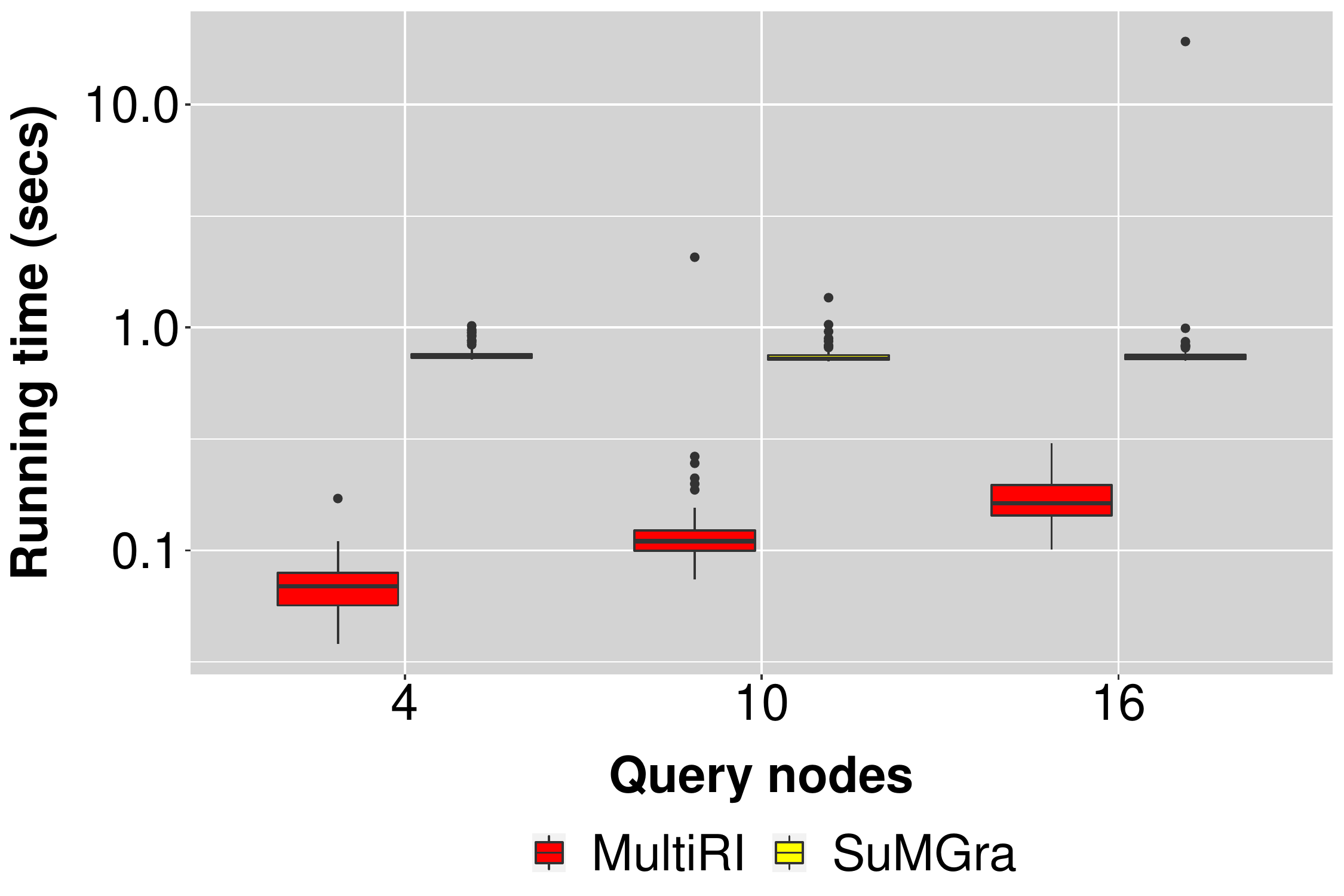}
\label{barabasi_query_size}}
\end{tabular}
\caption{Running times of MultiRI and SuMGra with varying values of: a) number of target nodes, b) target density, c) number of target node labels, d) maximum node multiplicity, e) number of target edge labels, f) maximum edge multiplicity and g) number of query nodes. Y-axes are shown in logarithmic scale to better highlight the differences between the compared algorithms. Results show that MultiRI is generally ten times faster than SuMGra.
}
\label{TimesSynth}
\end{figure*}

Fig. \ref{TimesSynth} depicts boxplots \high{of running times} for varying values of number of target nodes, target density, number of node labels, maximum node multiplicity, number of edge labels, maximum edge multiplicity and number of query nodes, respectively.

Experimental results show that MultiRI generally outperforms SuMGra by one order of magnitude. The largest difference between SuMGra and MultiRI can be observed when varying the number of target nodes (Fig. \ref{barabasi_num_nodes}), maximum node and edge multiplicity (Figs. \ref{barabasi_multiplicity_node} and \ref{barabasi_multiplicity_edge}) and number of query nodes (Fig. \ref{barabasi_query_size}). This shows the capability of MultiRI to perform well with large target graphs with a high multiplicity on both nodes and edges and large queries.

\begin{figure}[!ht]
\centering
\includegraphics[width=0.8\linewidth]{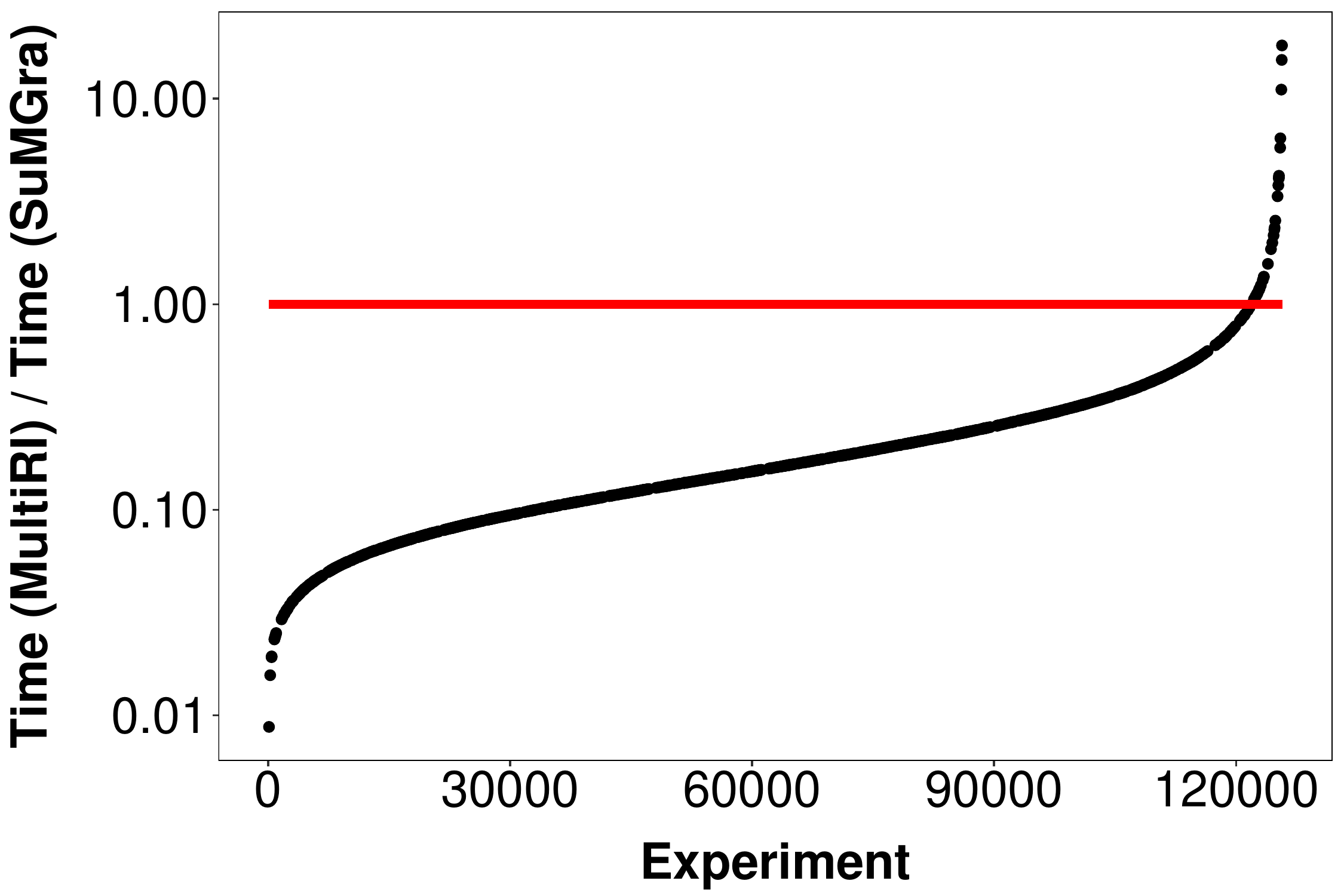}
\caption{Ratios between the running times of MultiRI and SuMGra for each experiment done with synthetic graphs. Experiments have been ordered based on the ratios. The red line indicate an equal running time for both algorithms. Ratios are reported on a logarithmic scale.}
\label{RatiosSynthetic}
\end{figure}

In Fig. \ref{RatiosSynthetic} we compare the performance of both algorithms on each experiment performed with synthetic graphs. We show the ratios between the running time of MultiRI and the running time of SuMGra for each experiment. Experiments are ordered based on the ratios. Compared to SuMGra, MultiRI is eight times faster on average and finishes before SuMGra in 97\% of the instances.

\begin{figure*}[!ht]
\centering
\begin{tabular}{c}
\subfloat[]
{\includegraphics[width=0.24\linewidth]{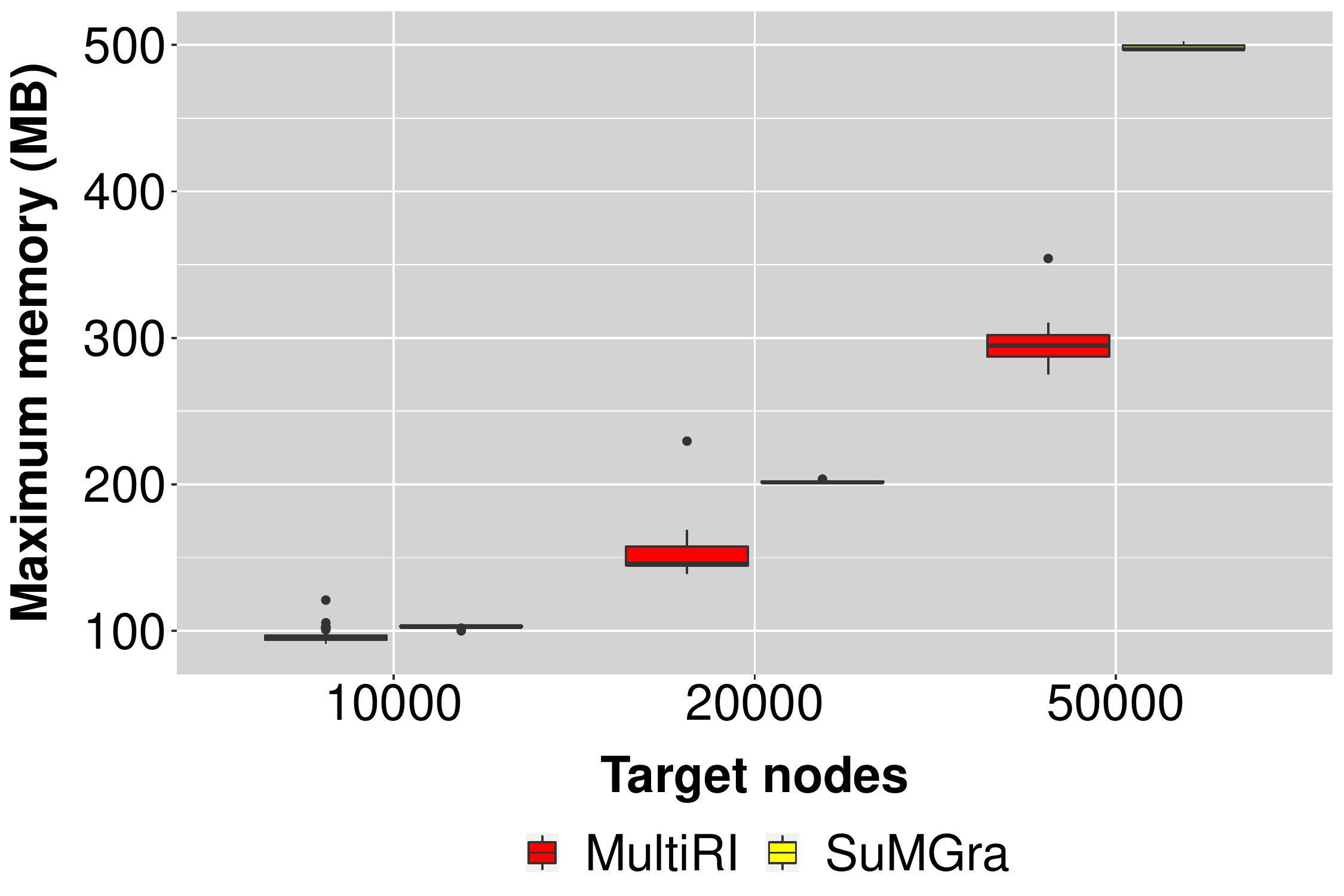}
\label{barabasi_num_nodes_memory}}
\subfloat[]
{\includegraphics[width=0.24\linewidth]{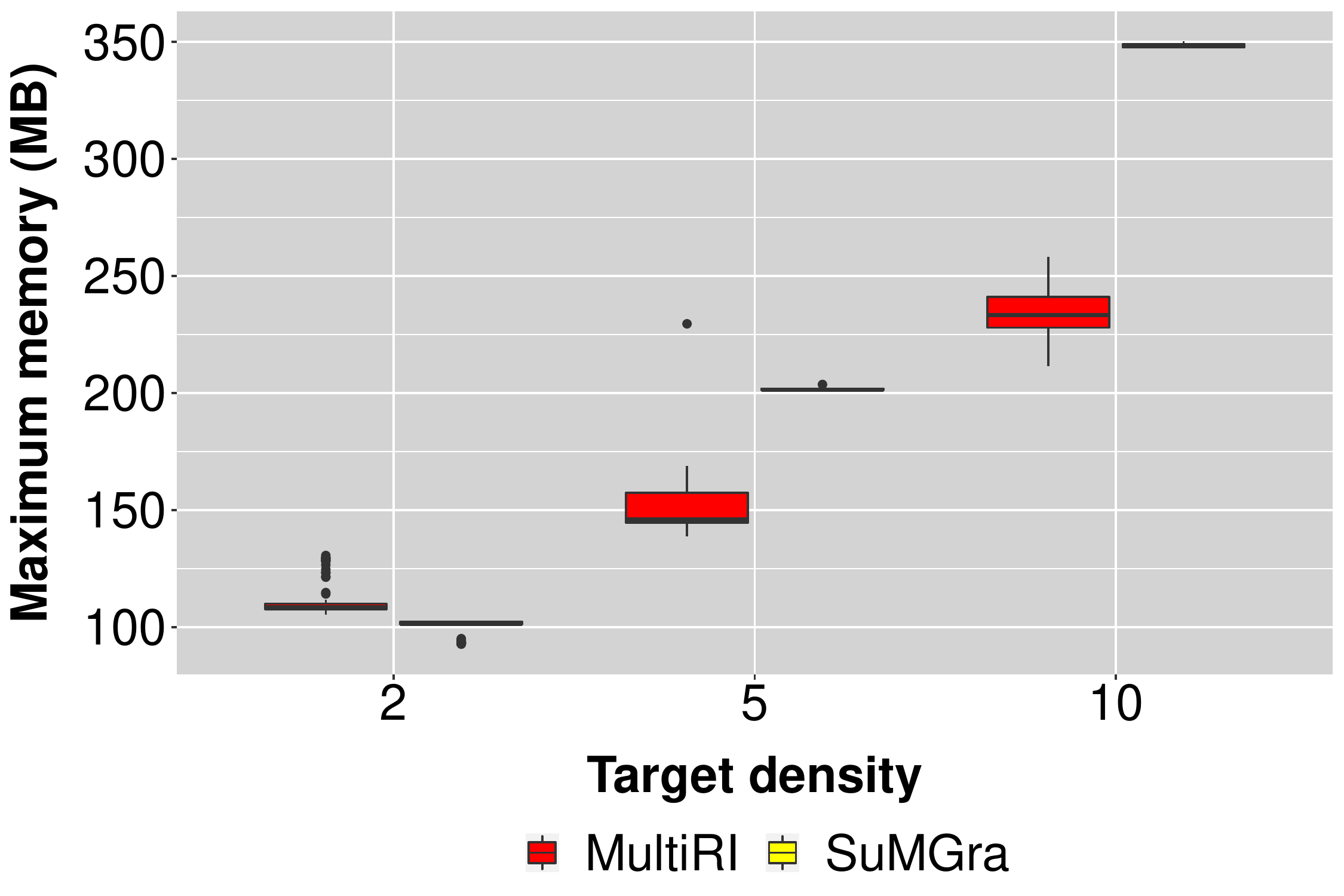}
\label{barabasi_density_memory}}
\subfloat[]
{\includegraphics[width=0.24\linewidth]{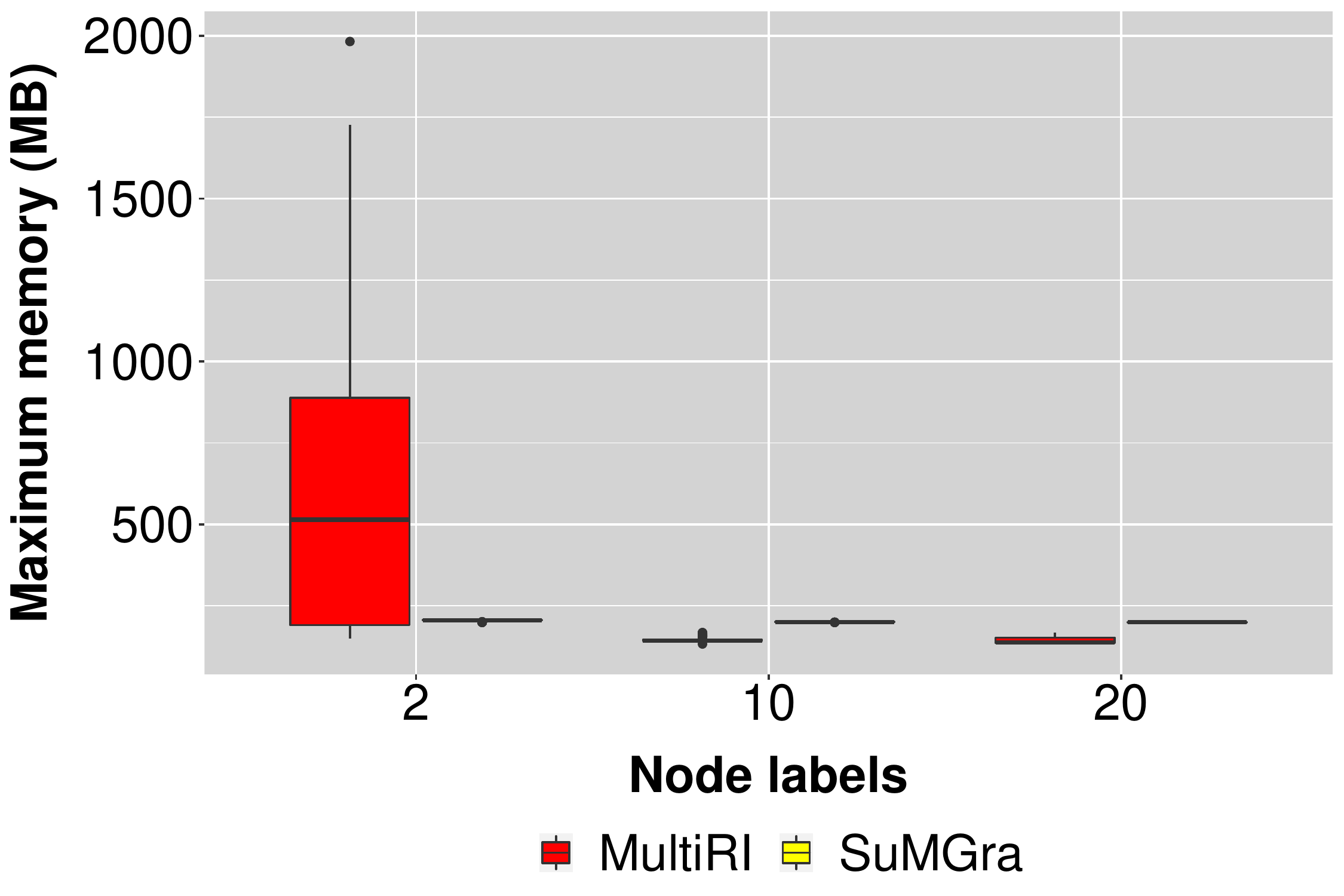}
\label{barabasi_num_node_labs_memory}}
\subfloat[]
{\includegraphics[width=0.24\linewidth]{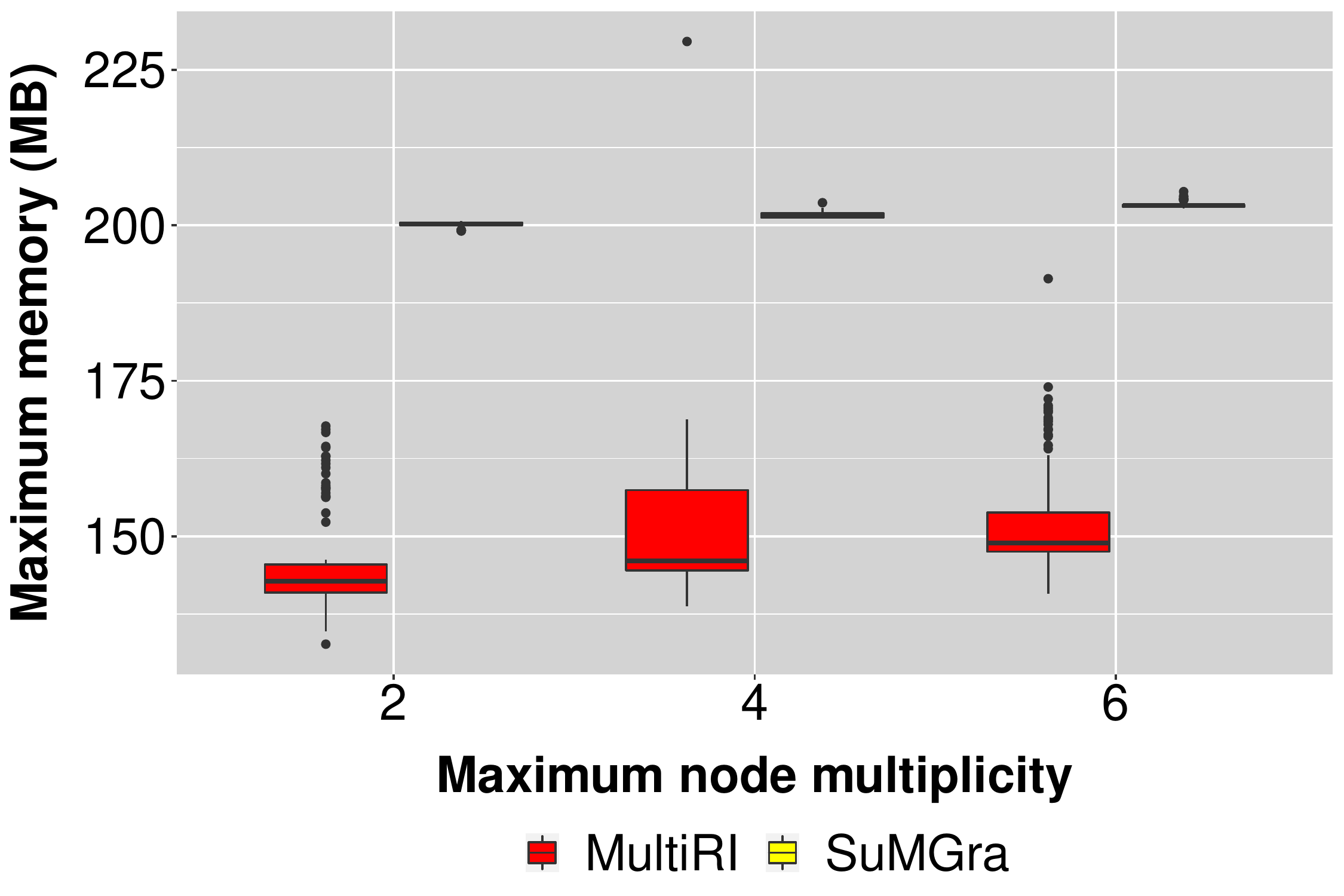}
\label{barabasi_multiplicity_node_memory}}
\\
\subfloat[]
{\includegraphics[width=0.24\linewidth]{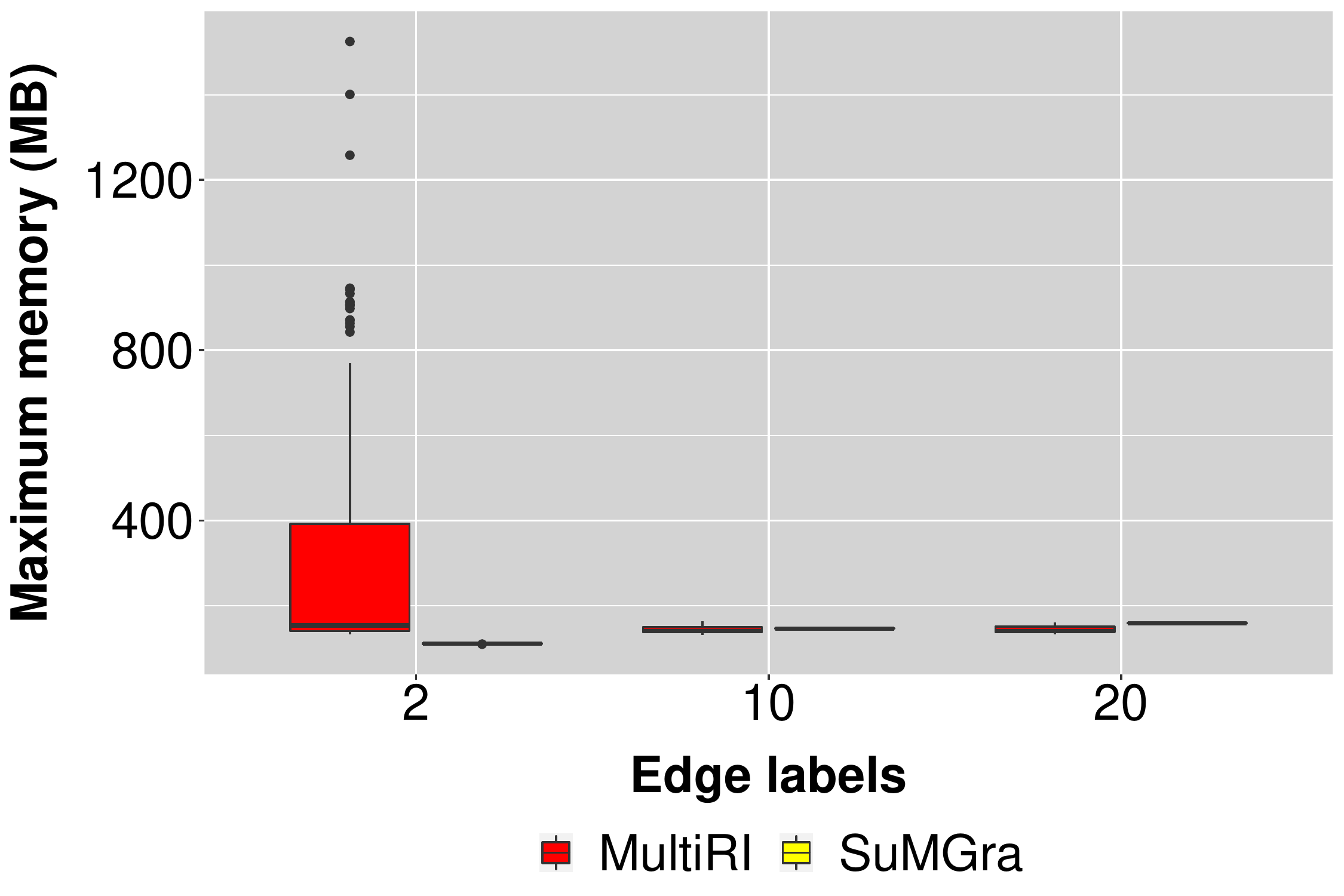}
\label{barabasi_num_edge_labs_memory}}
\subfloat[]
{\includegraphics[width=0.24\linewidth]{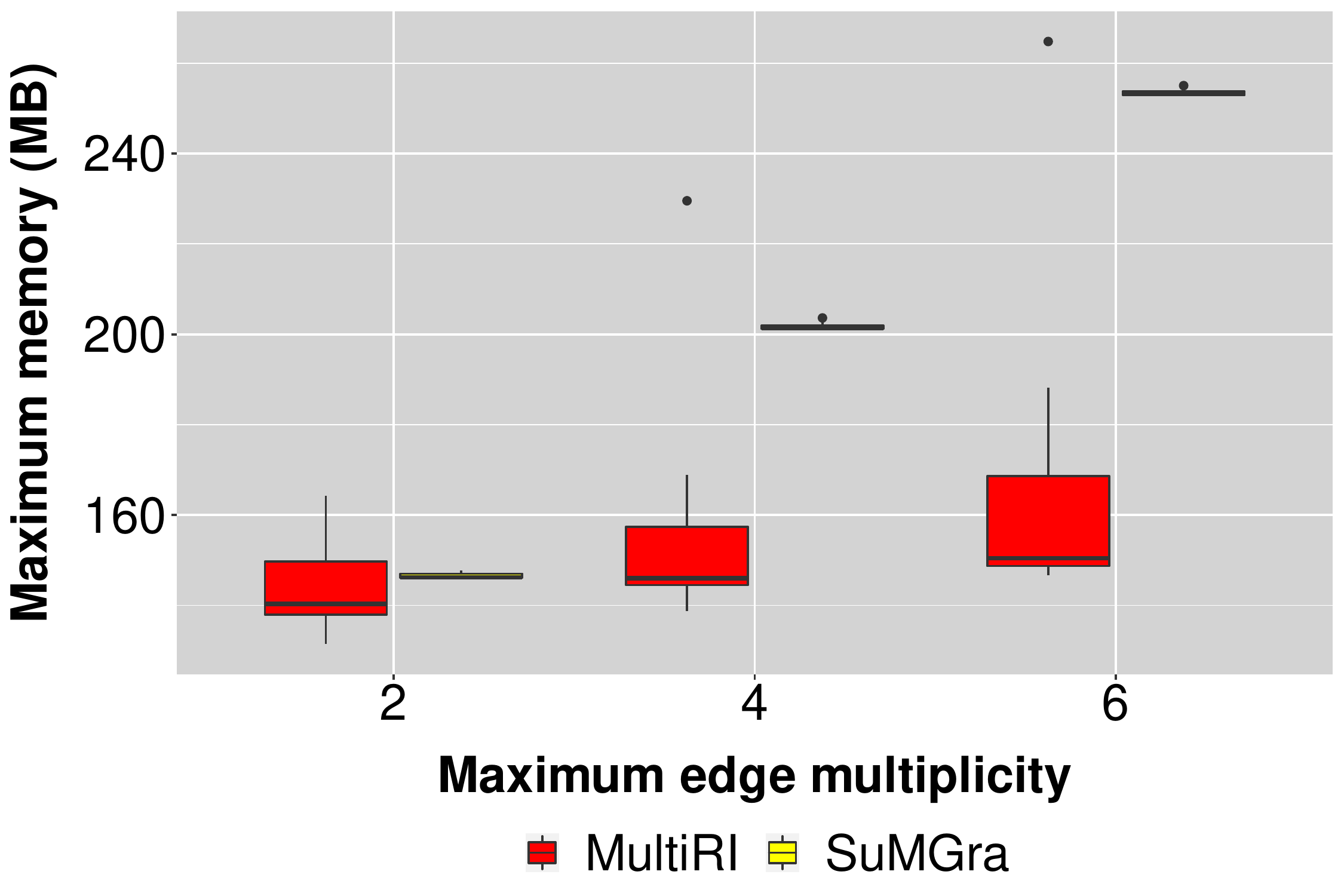}
\label{barabasi_multiplicity_edge_memory}}
\subfloat[]
{\includegraphics[width=0.24\linewidth]{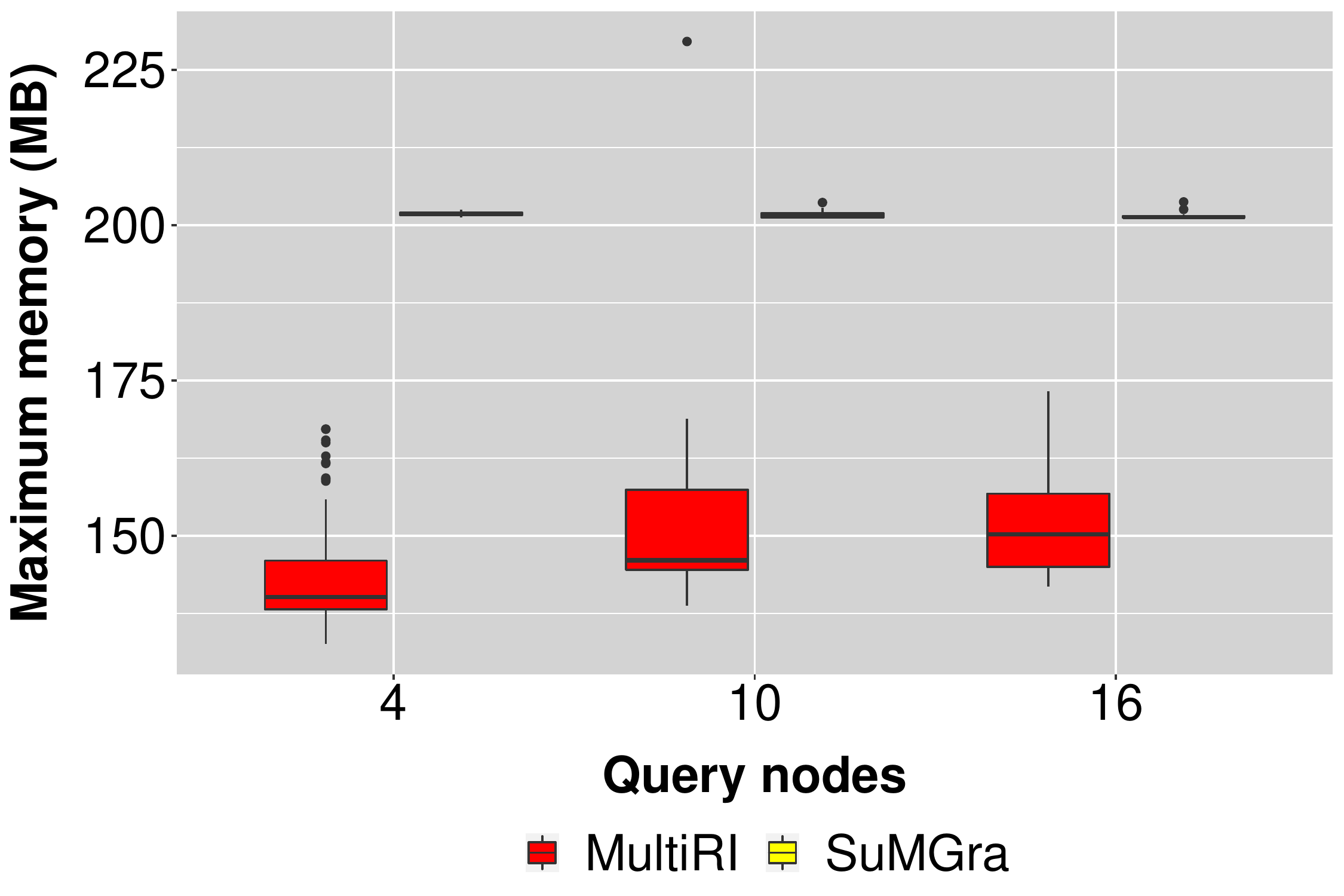}
\label{barabasi_query_size_memory}}
\end{tabular}
\caption{Memory usage of MultiRI and SuMGra with varying values of: a) number of target nodes, b) target density, c) number of target node labels, d) maximum node multiplicity, e) number of target edge labels, f) maximum edge multiplicity and g) number of query nodes.  Results show that MultiRI uses less memory than SuMGra on average.
}
\label{MemorySynth}
\end{figure*}

\high{Fig. \ref{MemorySynth} show boxplots of memory usage of both algorithms. MultiRI generally uses less memory than SuMGra, except for networks with few node or edge labels, where we can observe a lot of variability in memory usage. The difference between the two algorithms is more evident when we consider bigger and denser graphs (Figs. \ref{barabasi_num_nodes_memory} and \ref{barabasi_density_memory}) and graphs with high edge multiplicity (Fig. \ref{barabasi_multiplicity_edge_memory}). Interestingly, in these three cases both algorithms have an  increase of memory usage, while in all remaining cases there is no significant variation in memory consumption. This suggests that the spatial complexity of both MultiRI and SuMGra is mainly influenced on the size of target graph and its edge multiplicity. In particular the one of SuMGra looks to be exponentially dominated by such a factor.}

\subsection{\high{Comparison between MultiRI and SuMGra on real graphs}}

Next, we tested MultiRI and SuMGra on the real dataset. Since SuMGra works only on undirected graphs, the comparison between MultiRI and SuMGra was done by converting all directed edges to undirected edges. However, we also tested MultiRI alone on the two directed multigraphs of our dataset, namely \textsc{foldoc} and \textsc{swissleaks}, to evaluate its performance on directed multigraphs as well.

We randomly extracted 1,000 queries with 4 nodes, 1,000 queries with 10 nodes and 1,000 queries with 16 nodes from each graph, following the extraction procedure described in Section \ref{synthData}. Then we ran MultiRI and SuMGra on each query. Again, to plot running times, we considered only experiments in which both algorithms ended before a timeout of 5 minutes, \high{while for memory consumption we took into account all tests.}
Considering all experiments on real graphs, both algorithms ended before the timeout in 75.9\% of the cases. In 1.6\% of the experiments only MultiRI finished before the timeout, while in 1.3\% of the cases only SuMGra ended before the timeout. In 23.6\% of the experiments neither algorithm completed before the timeout.

\begin{figure}[!ht]
\centering
\begin{tabular}{c}
\subfloat[]
{\includegraphics[width=0.48\linewidth]{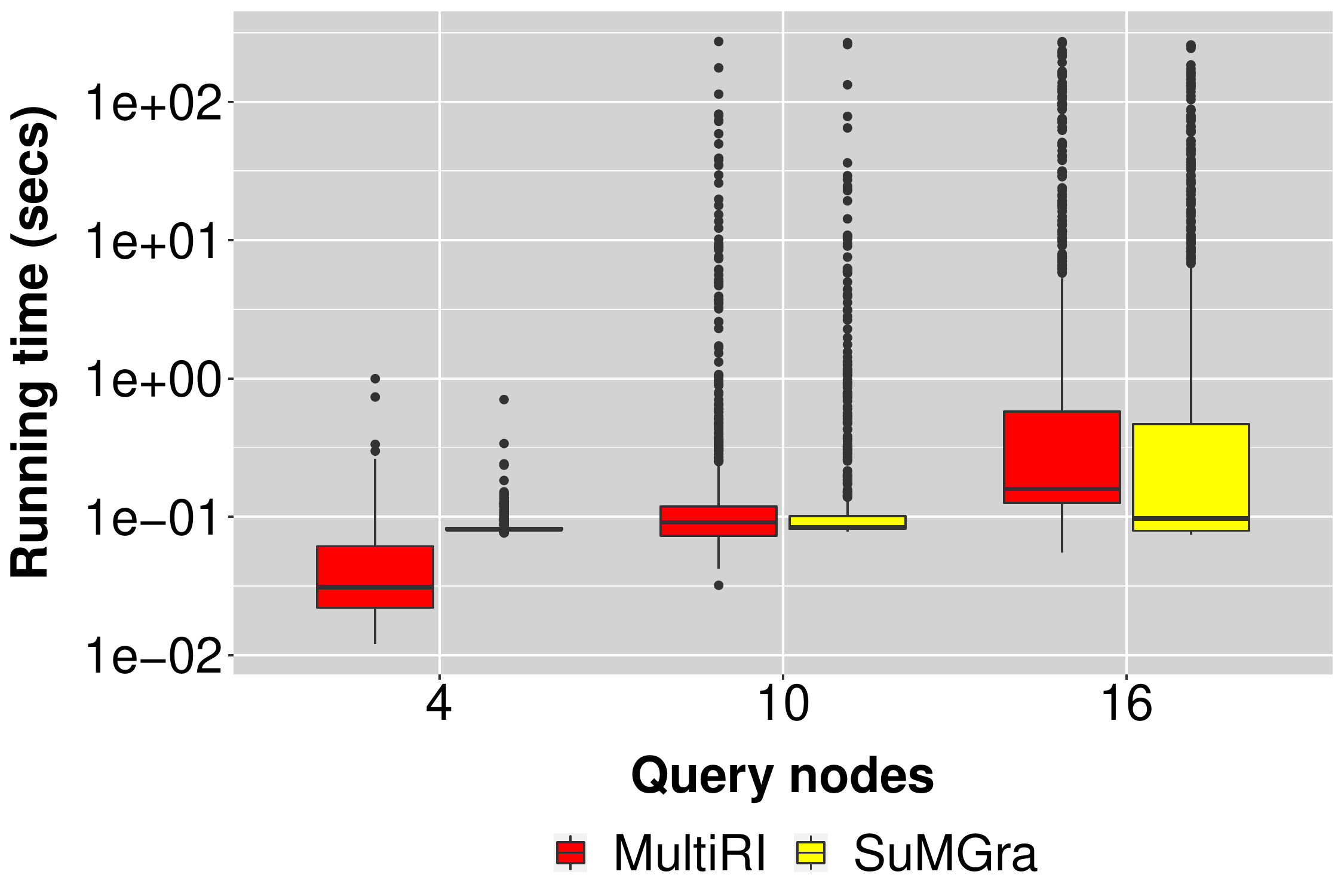}
\label{openflightsTimes}}
\subfloat[]
{\includegraphics[width=0.48\linewidth]{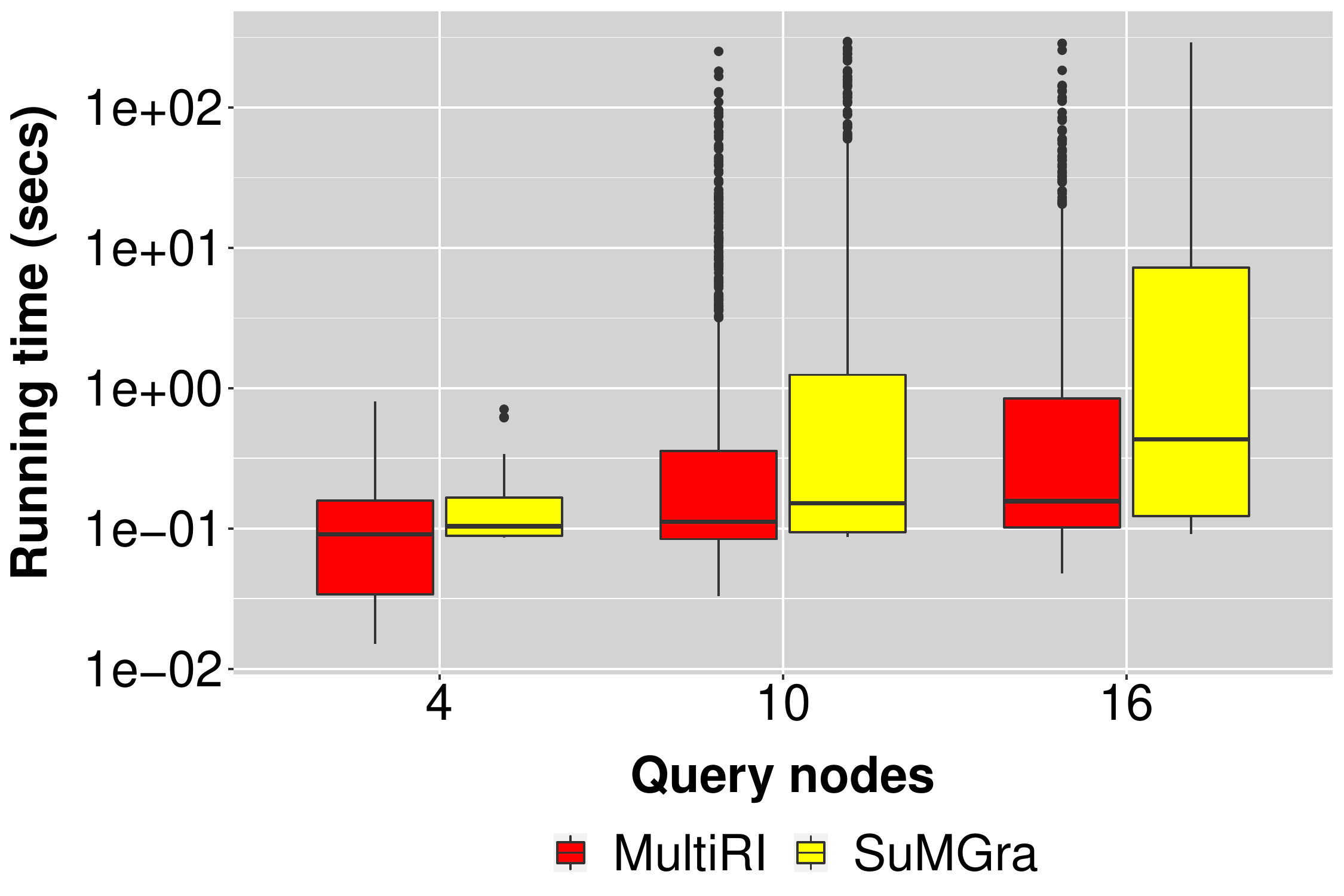}
\label{foldocTimes}}
\\
\subfloat[]
{\includegraphics[width=0.48\linewidth]{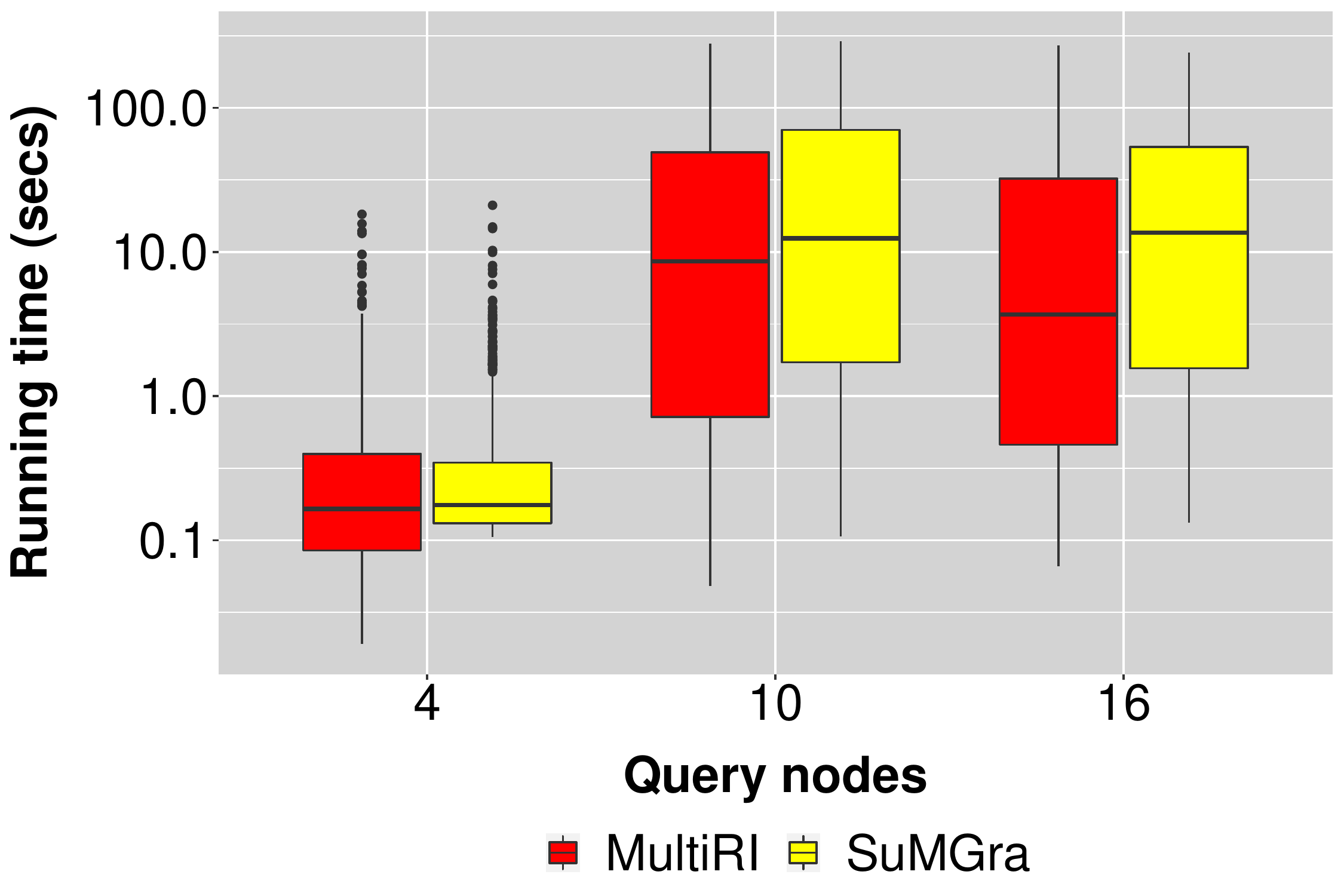}
\label{ppiyeastTimes}}
\subfloat[]
{\includegraphics[width=0.48\linewidth]{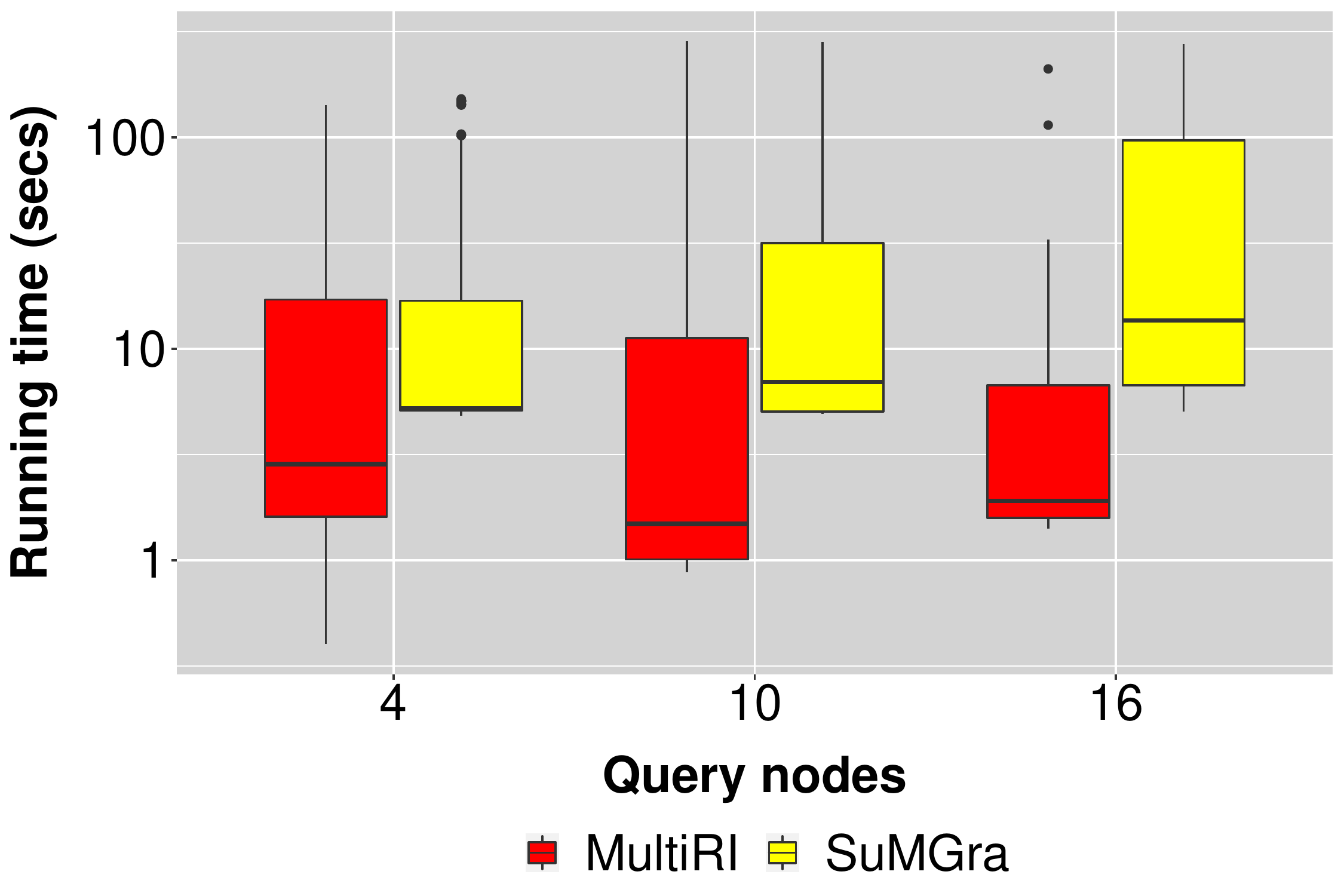}
\label{swissleaksTimes}}
\\
\subfloat[]
{\includegraphics[width=0.48\linewidth]{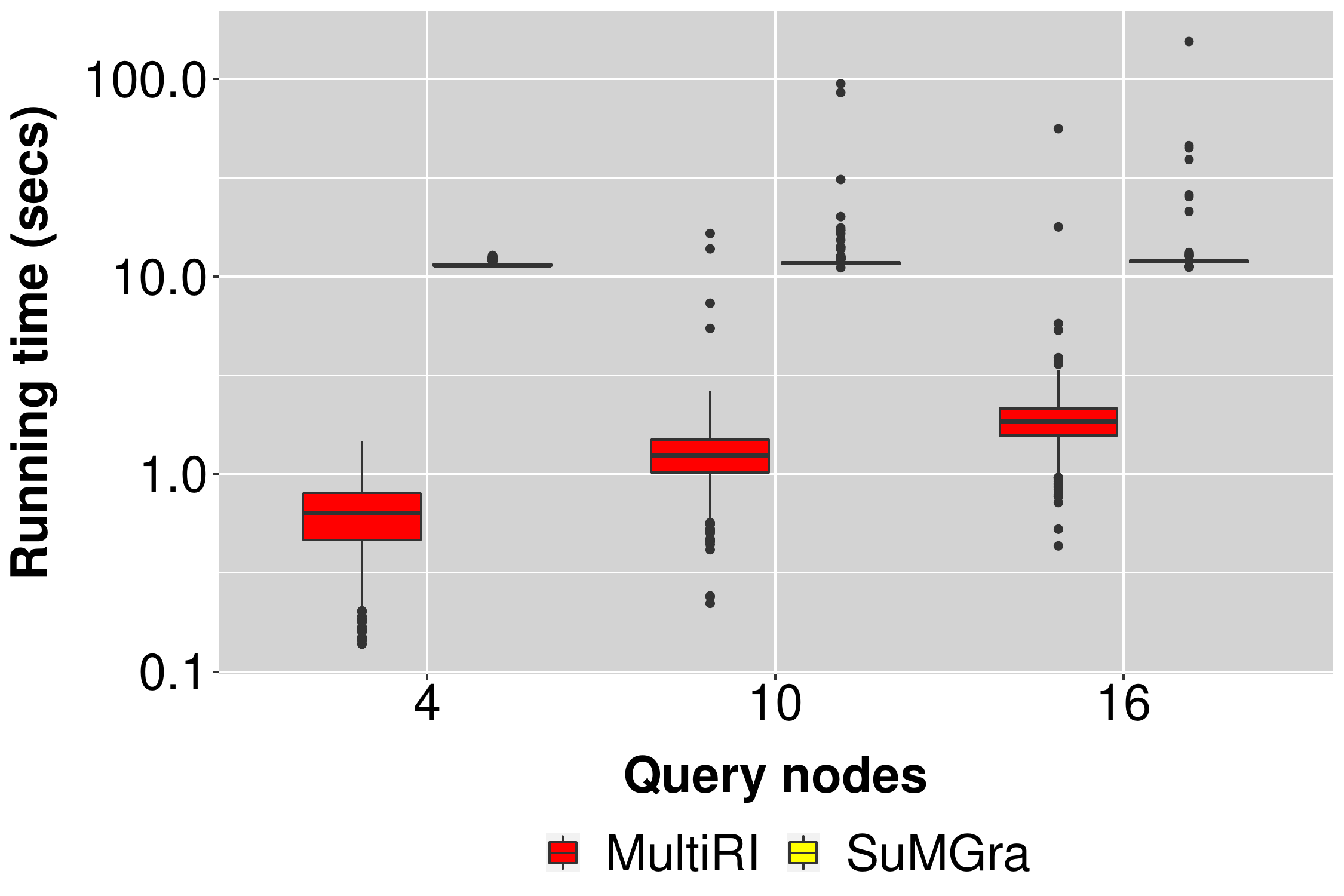}
\label{imdbTimes}}
\end{tabular}
\caption{Running times of MultiRI and SuMGra for the set of extracted queries with 4, 10 and 16 nodes in the following graphs: a) \textsc{openflights}, b) \textsc{foldoc}, c) \textsc{ppiyeast}, d) \textsc{swissleaks} and e) \textsc{imdb}. All graphs are treated as undirected. Y-axes are shown in logarithmic scale. Results show that MultiRI is approximately ten times faster than SuMGra on the two largest graphs.}
\label{TimesReal}
\end{figure}

Fig. \ref{TimesReal} shows box-plots of the running times for each real graph for the set of extracted queries with 4, 10 and 16 nodes, respectively. 

In the real dataset, SuMGra is almost as fast as MultiRI in the smallest graphs. However, MultiRI is approximately ten times faster than SuMGra in the largest graphs, i.e. \textsc{swissleaks} and \textsc{imdb}. The bottom line is that MultiRI performs well on large graphs.

\begin{figure}[!ht]
\centering
\begin{tabular}{c}
\subfloat[]
{\includegraphics[width=0.48\linewidth]{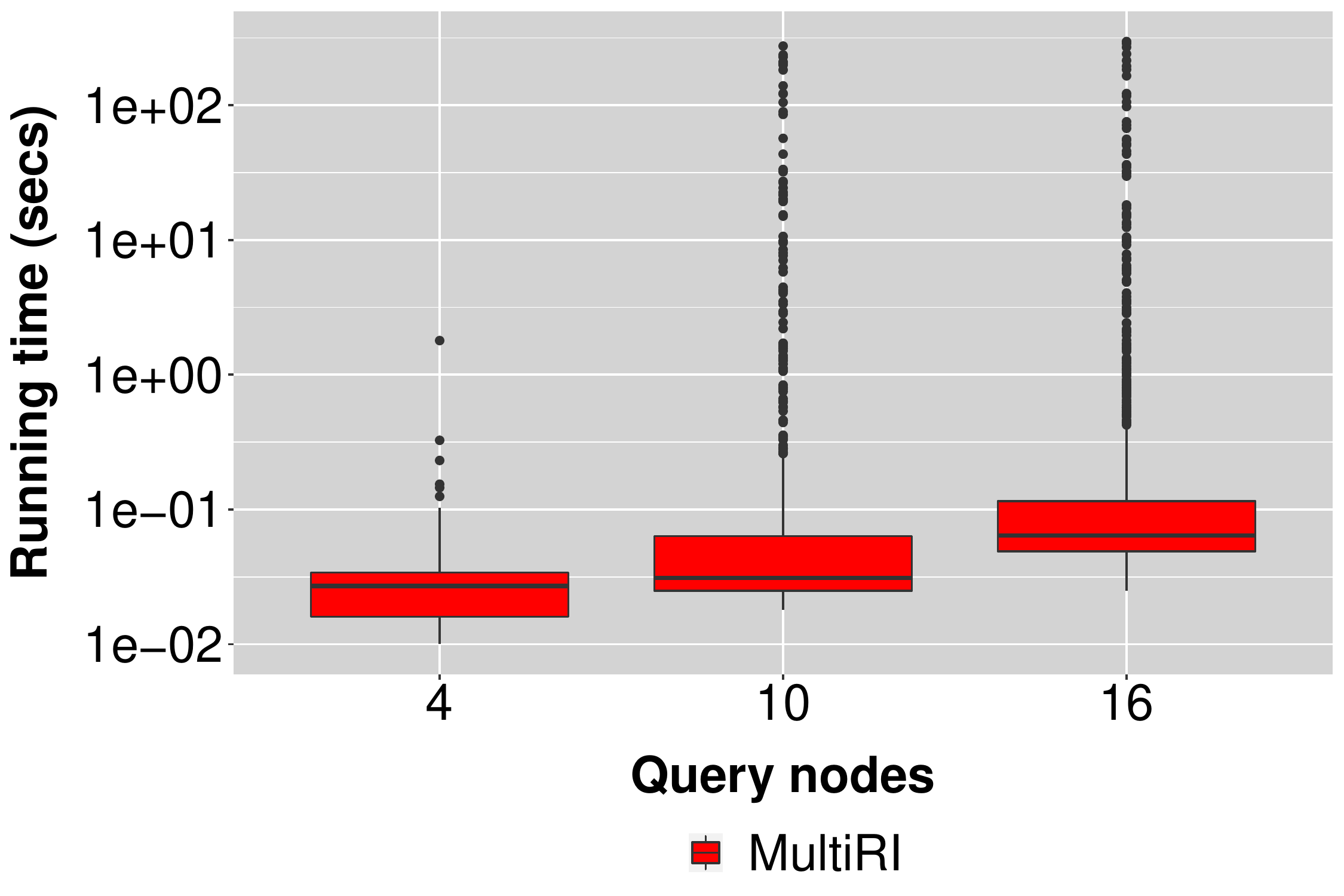}
\label{foldocDirected}}
\subfloat[]
{\includegraphics[width=0.48\linewidth]{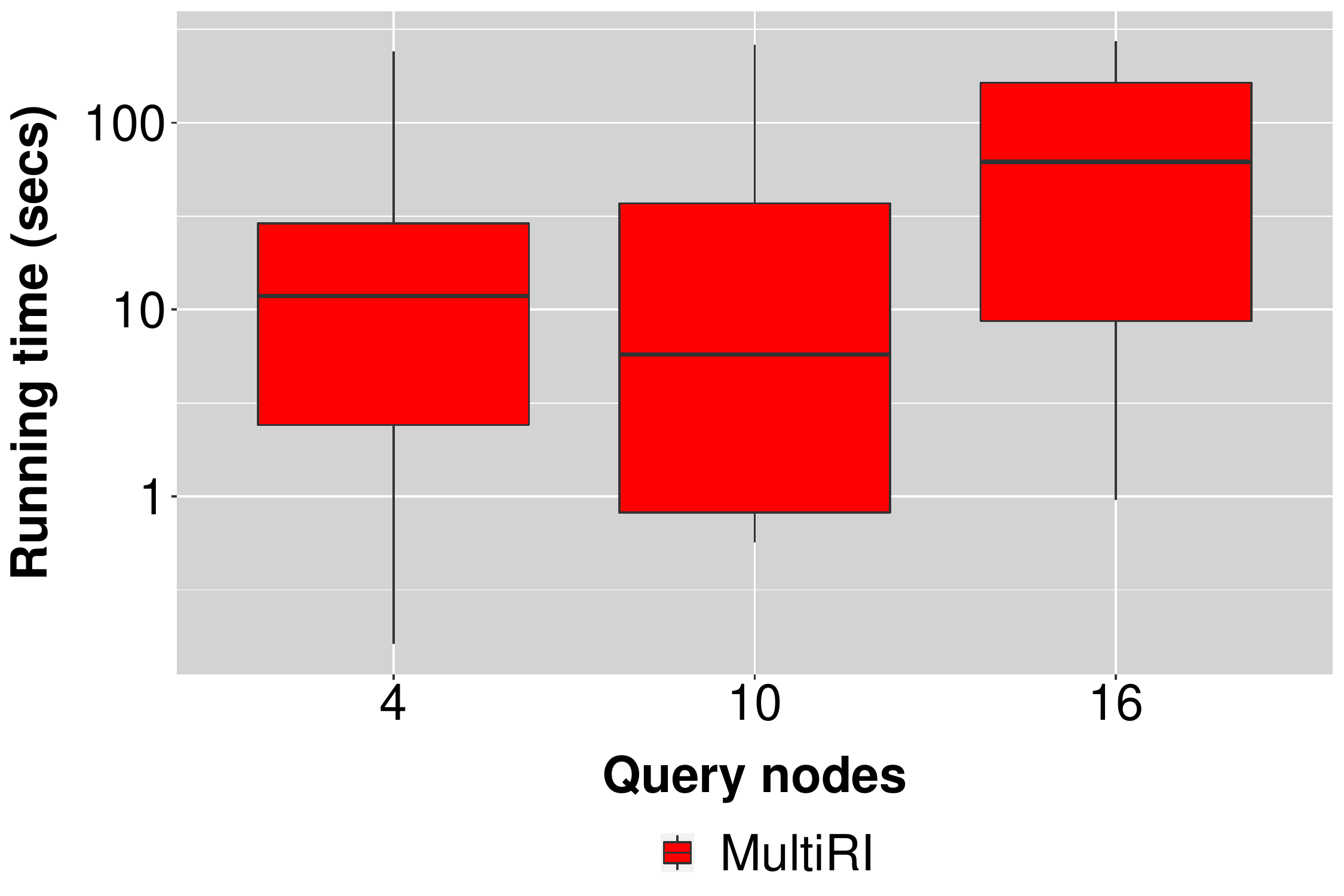}
\label{swissleaksDirected}}
\end{tabular}
\caption{Running times of MultiRI for the set of extracted queries with 4, 10 and 16 nodes in the directed multigraphs a) \textsc{foldoc} and b) \textsc{swissleaks}. the Y-axis is on a logarithmic scale.}
\label{TimesRealDirected}
\end{figure}

In Fig. \ref{TimesRealDirected} we show boxplots of running times of MultiRI for the two real directed multigraphs of our dataset, i.e. \textsc{foldoc} and \textsc{swissleaks}. Boxplots consider only experiments in which MultiRI ended before a timeout of 5 minutes. MultiRI completed 96\% and 27\% of the experiments before the timeout for \textsc{foldoc} and \textsc{swissleaks}, respectively.

\begin{figure}[!ht]
\centering
\includegraphics[width=0.8\linewidth]{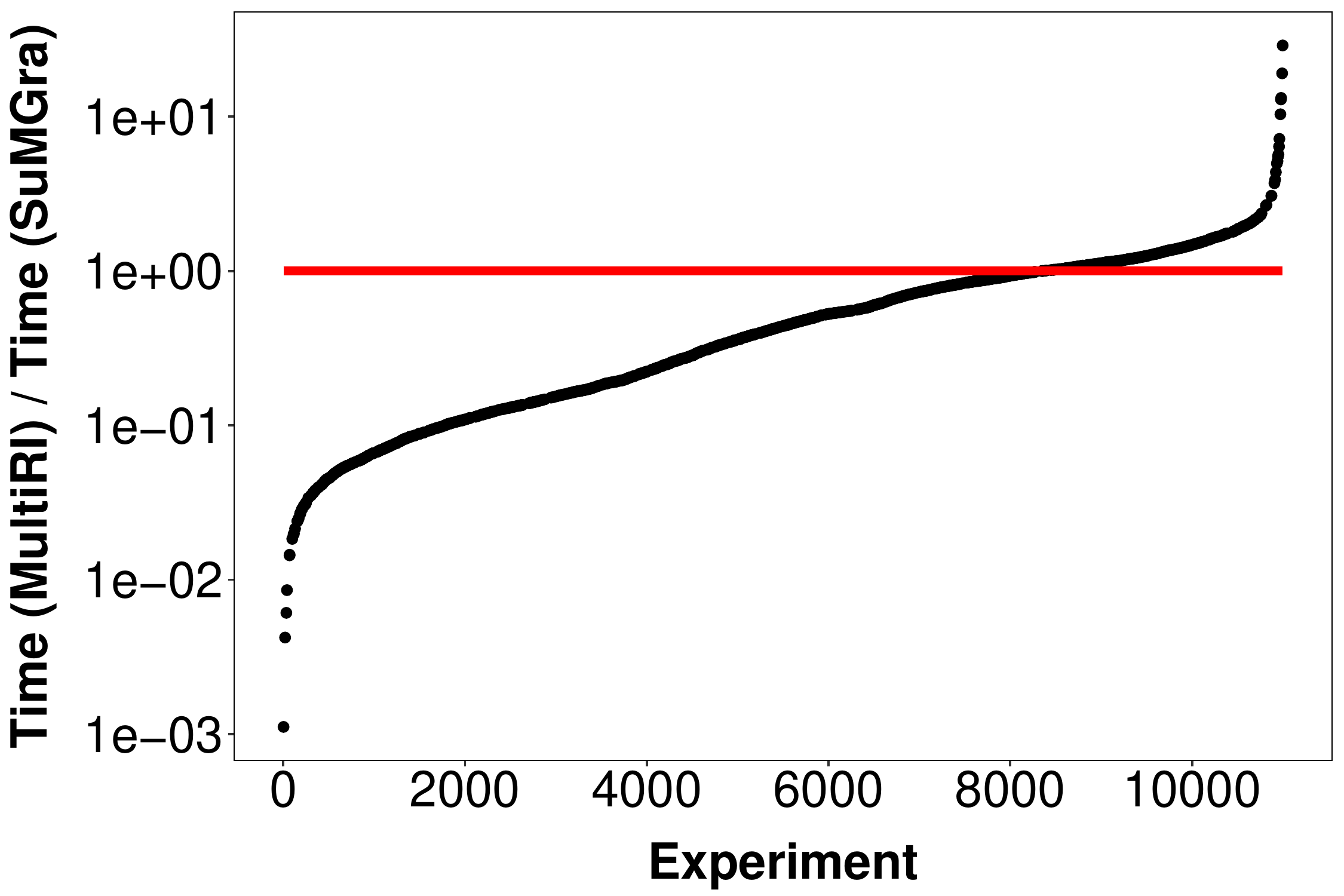}
\caption{Ratios between the running times of MultiRI and SuMGra for each experiment done with real graphs. All graphs are treated as undirected. Experiments have been ordered based on the ratios. The red line indicate an equal running time for both algorithms. Ratios are reported in logarithmic scale.}
\label{RatiosReal}
\end{figure}

\high{Again}, we \high{also} compared the performance of both algorithms on each experiment performed with real graphs. Fig. \ref{RatiosReal} shows the ratios between the running time of MultiRI and the running time of SuMGra for each experiment. The plot experiments are ordered based on the ratios. In the real networks, MultiRI is on average seven times faster than SuMGra and finishes before SuMGra in 76\% of the instances.

\begin{figure}[!ht]
\centering
\begin{tabular}{c}
\subfloat[]
{\includegraphics[width=0.48\linewidth]{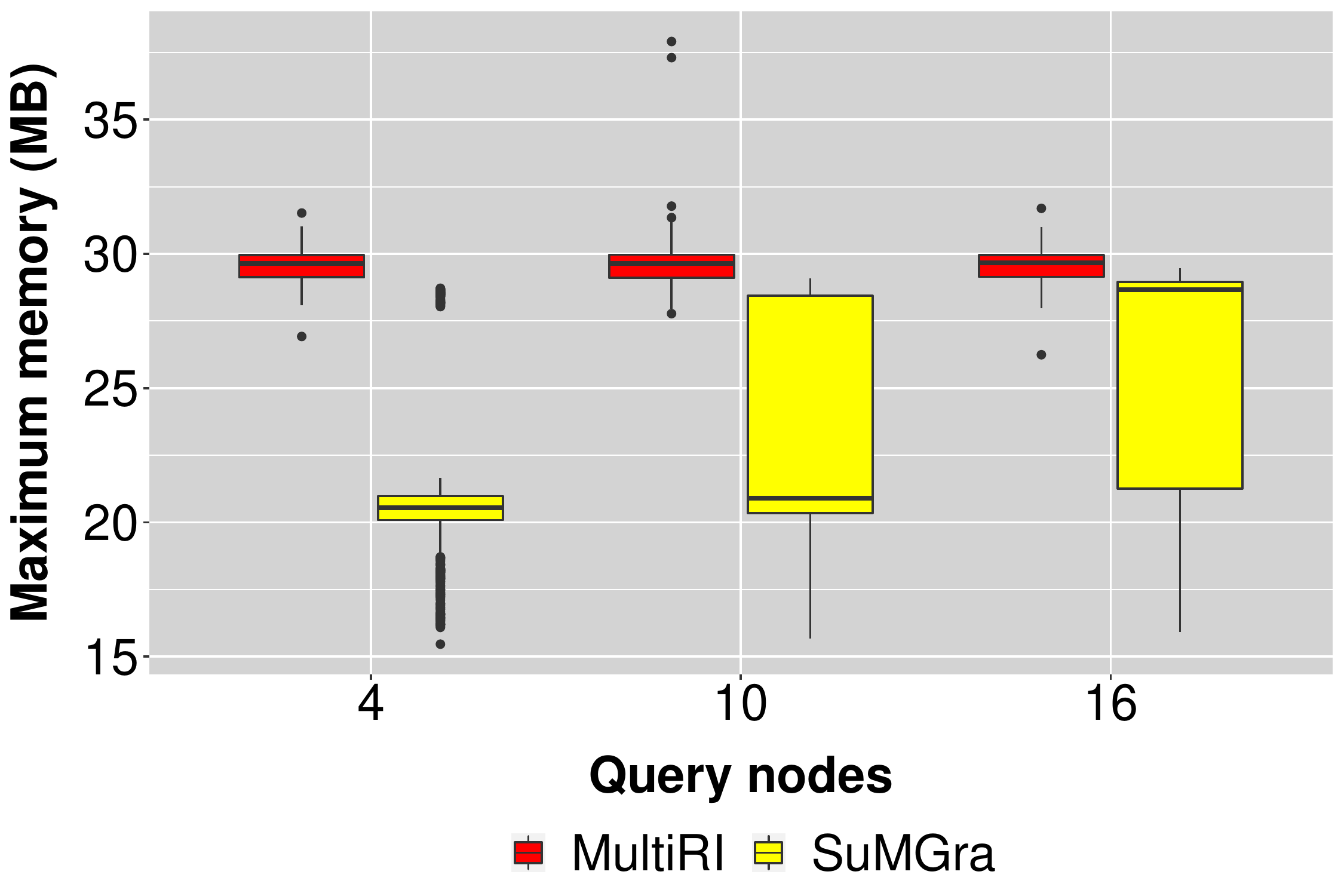}
\label{openflightsMemory}}
\subfloat[]
{\includegraphics[width=0.48\linewidth]{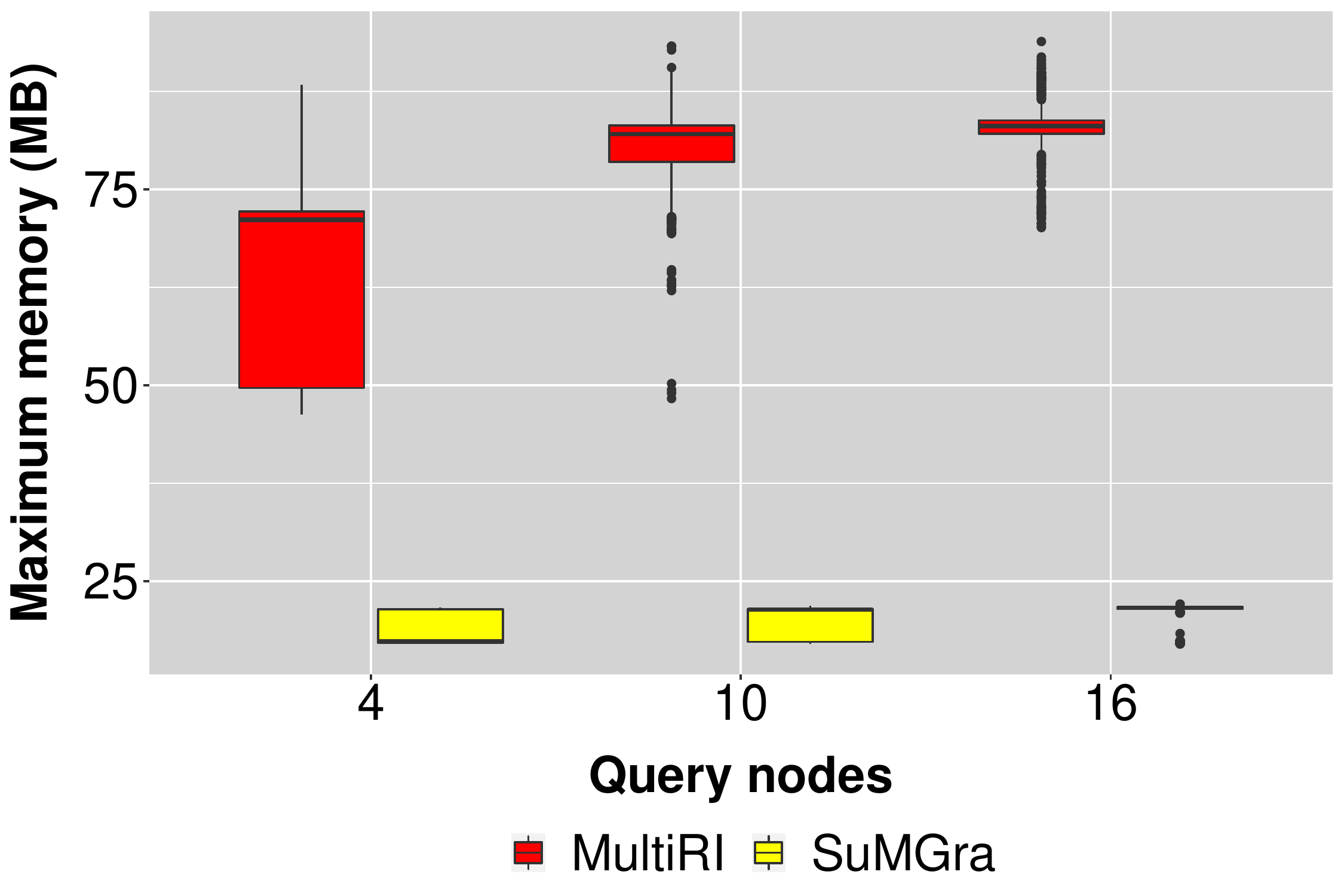}
\label{foldocMemory}}
\\
\subfloat[]
{\includegraphics[width=0.48\linewidth]{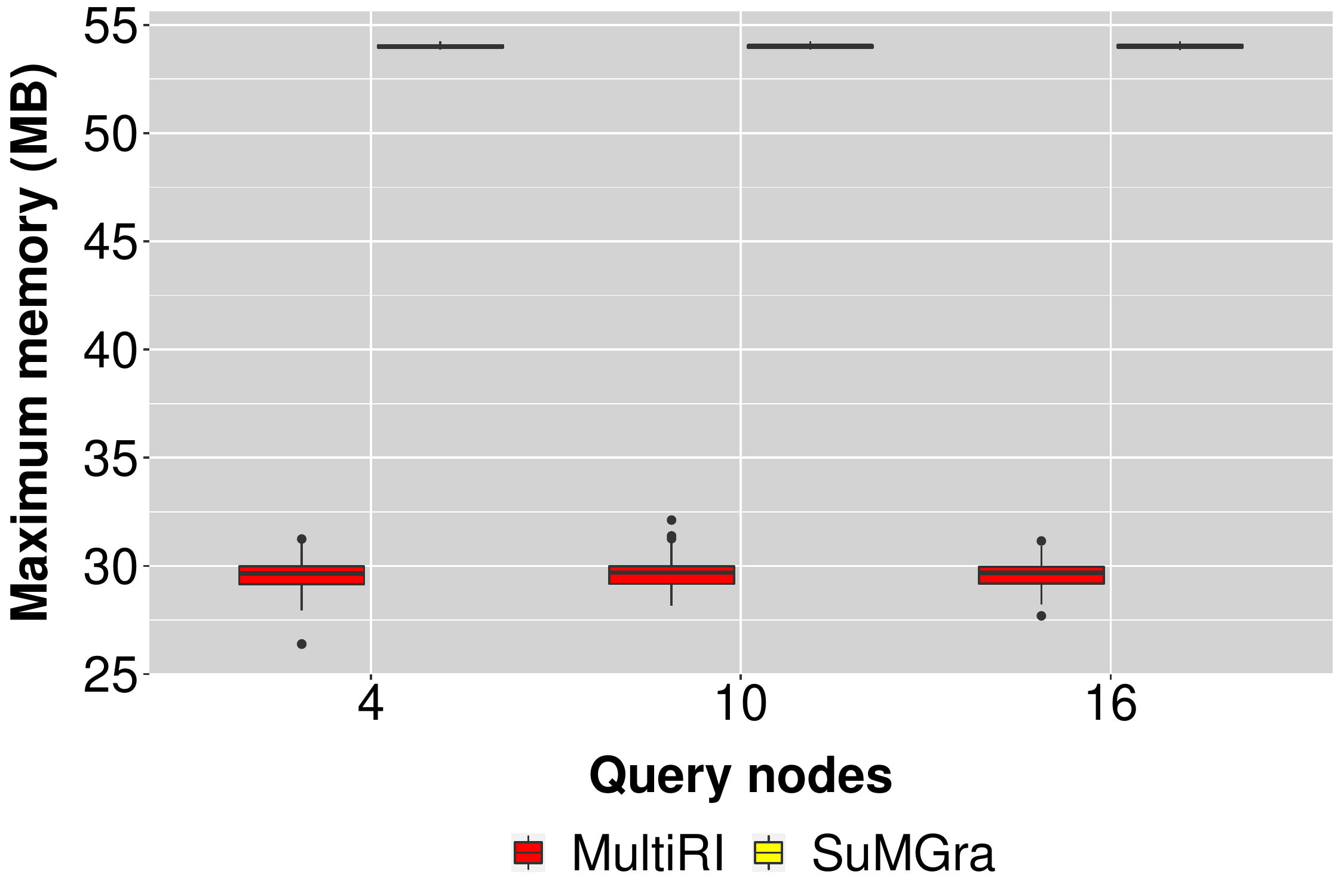}
\label{ppiyeastMemory}}
\subfloat[]
{\includegraphics[width=0.48\linewidth]{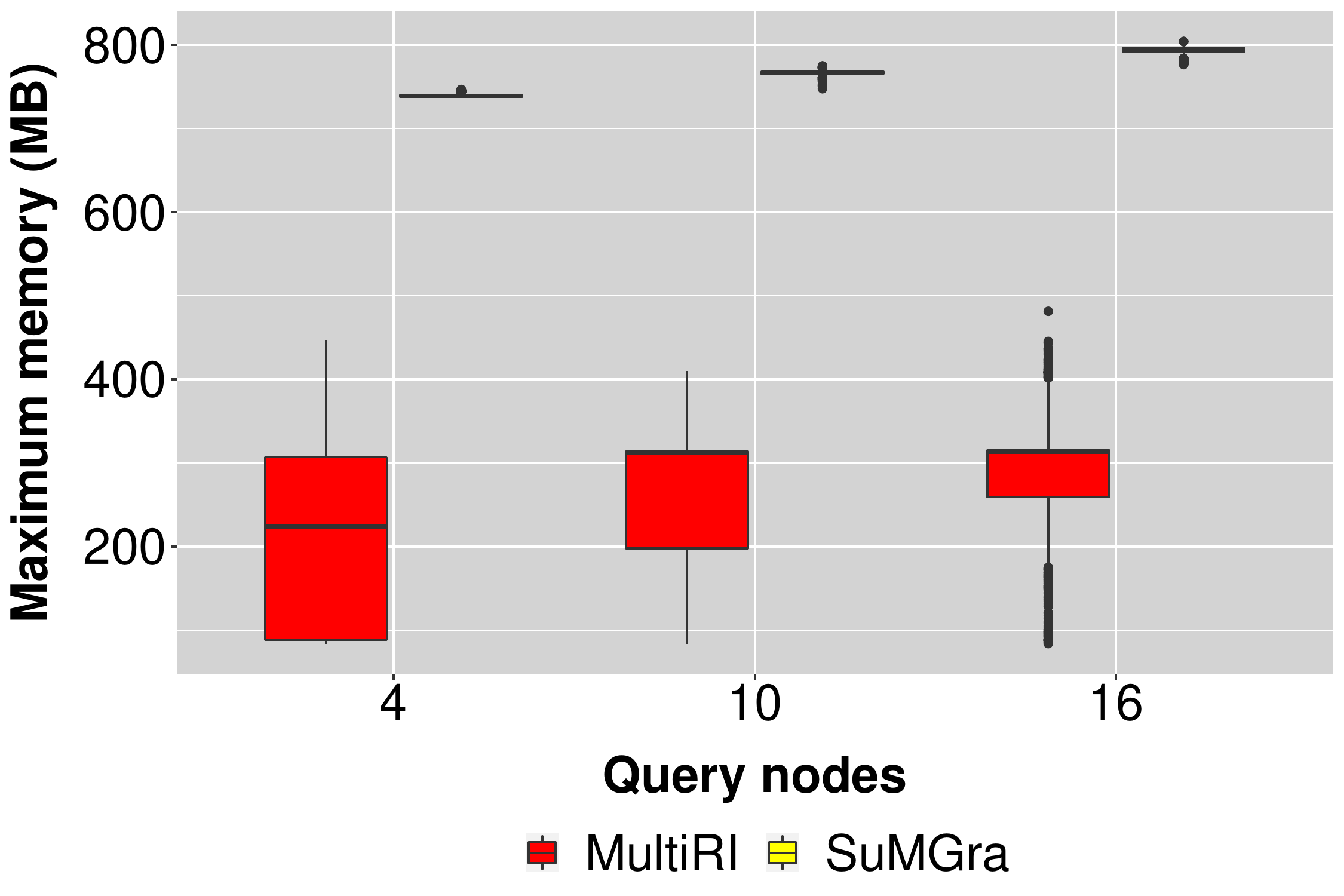}
\label{swissleaksMemory}}
\\
\subfloat[]
{\includegraphics[width=0.48\linewidth]{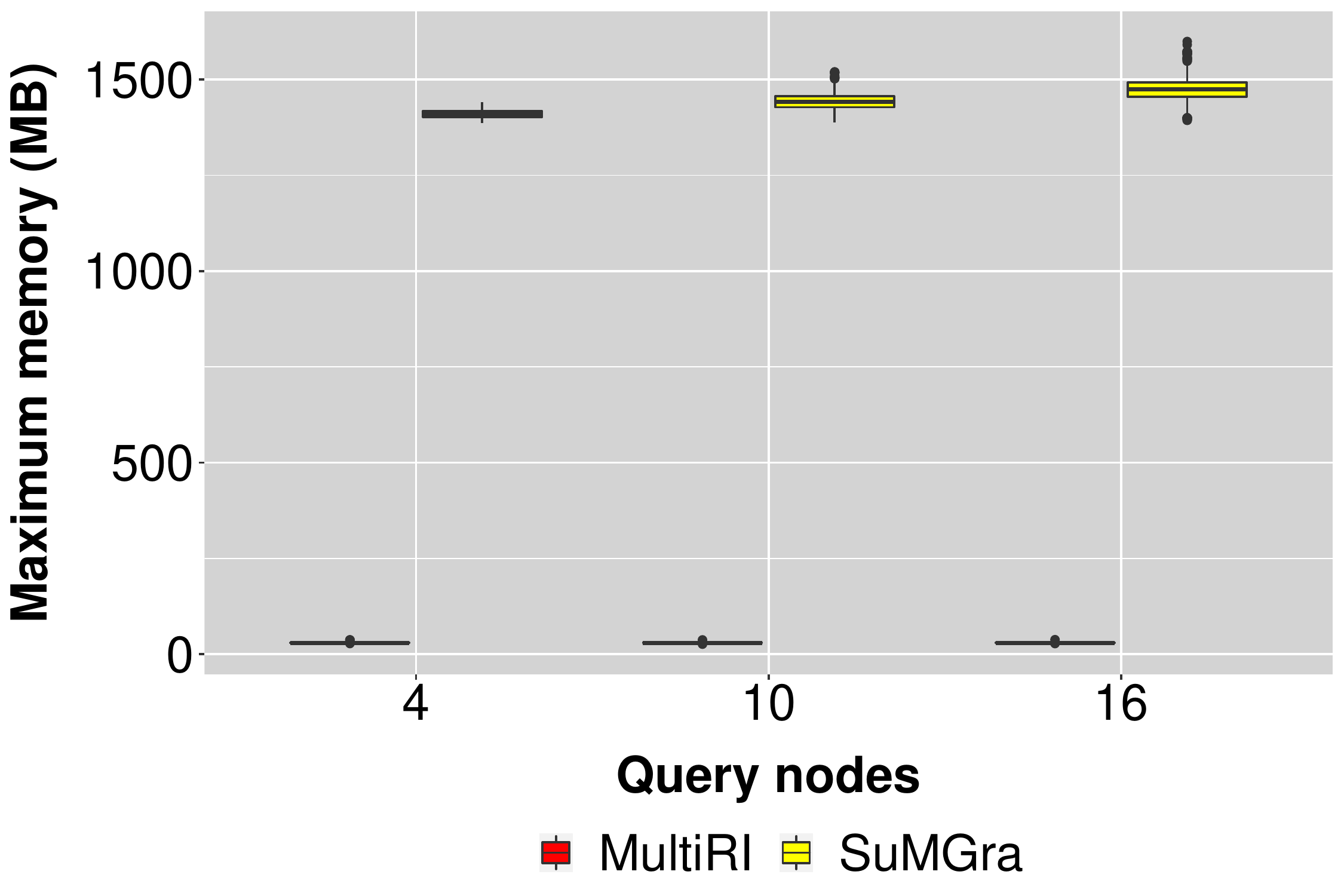}
\label{imdbMemory}}
\end{tabular}
\caption{Memory usage of MultiRI and SuMGra for the set of extracted queries with 4, 10 and 16 nodes in the following graphs: a) \textsc{openflights}, b) \textsc{foldoc}, c) \textsc{ppiyeast}, d) \textsc{swissleaks} and e) \textsc{imdb}. All graphs are treated as undirected. Results show that MultiRI uses  less memory than SuMGra on average.}
\label{MemoryReal}
\end{figure}

\high{In Fig. \ref{MemoryReal} we compare the memory usage of MultiRI and SuMGra for real graphs.} \high{Except for the two smallest networks, i.e. \textsc{openflights} and \textsc{foldoc}, MultiRI uses much less memory than SuMGra for bigger graphs, namely 4 times less for \textsc{ppiyeast} and \textsc{swissleaks} and 1000 times less for \textsc{imdb}. More interestingly, while SuMGra's memory usage mainly depends on graph size, edge multiplicity seems to be the main factor impacting on memory consumption for MultiRI. Indeed, we observe the highest memory usage of MultiRI (on average 300 MB) for \textsc{swissleaks}, for which $|\Gamma|=53$.}

\subsection{Scalability test}

We ran MultiRI and SuMGra on a large \high{collaboration multigraph} between people working in the show business (e.g. actors, writers, directors). The graph was extracted from the IMDB database. Nodes are people and edges connect two people if they worked together in at least one movie. People are labeled according to their main profession (i.e. 'actor', 'director', 'writer', 'producer', 'composer' and/or 'editor'), while edges are labeled according to the genres of the movies where they worked together (e.g. 'comedy', 'drama', 'thriller'). The final graph contains 2,508,369 nodes, 32,768,597 edges, 6 different node labels, 28 different edge labels, and maximum node and edge multiplicities equal to 3 and 22, respectively.

We designed a dataset of queries formed by small cliques with 3, 4 and 5 nodes, to understand if people tend to work together in movies of the same or different genres. In each clique, nodes have the same label, so they are all actors or directors for example, and edges can have all the same labels or mixed labels, considering all possible combinations of movie genres with repetitions. To avoid generating too many queries, we decided to focus on the five most popular genres, i.e. comedy, drama, action, thriller and horror. The final dataset includes 210 3-cliques, 1,260 4-cliques and 6,006 5-cliques.

We weren't able to get any results from SuMGra, since it required more than 32 GB to perform the matching. By contrast, MultiRI required 20 GB to complete all tasks. The average running time of MultiRI in computing the number of occurrences of 3-cliques, 4-cliques and 5-cliques was 23 secs, 64 secs and 169 secs, respectively. These results show that MultiRI is efficient and scalable with query size.

\subsection{\high{How Symmetry breaking conditions impact performances}}
\label{symmetryExp}

\high{In this subsection, we empirically show how symmetry breaking conditions affect the time complexity of MultiRI considering a set of small artificial graphs with varying density. Note that these results are general and hold for any subgraph matching algorithm, even though we applied here for our algorithm.}

\high{We built two datasets of artificial graphs generated using the Barabasi-Albert \cite{Barabasi1999} model.}

\high{The first dataset includes unlabeled simple graphs with $N=100$ nodes and varying density $d=10, 25, 40$. We built 10 networks for each value of $d$, for a total number of 30 networks. Since the efficacy of breaking conditions mainly depends on the topology of the query, we considered three different types of query topologies, namely paths, stars and cliques with $k=4, 5, 6$ nodes.}

\begin{figure}[!ht]
\centering
\begin{tabular}{c}
\subfloat[]
{\includegraphics[width=0.48\linewidth]{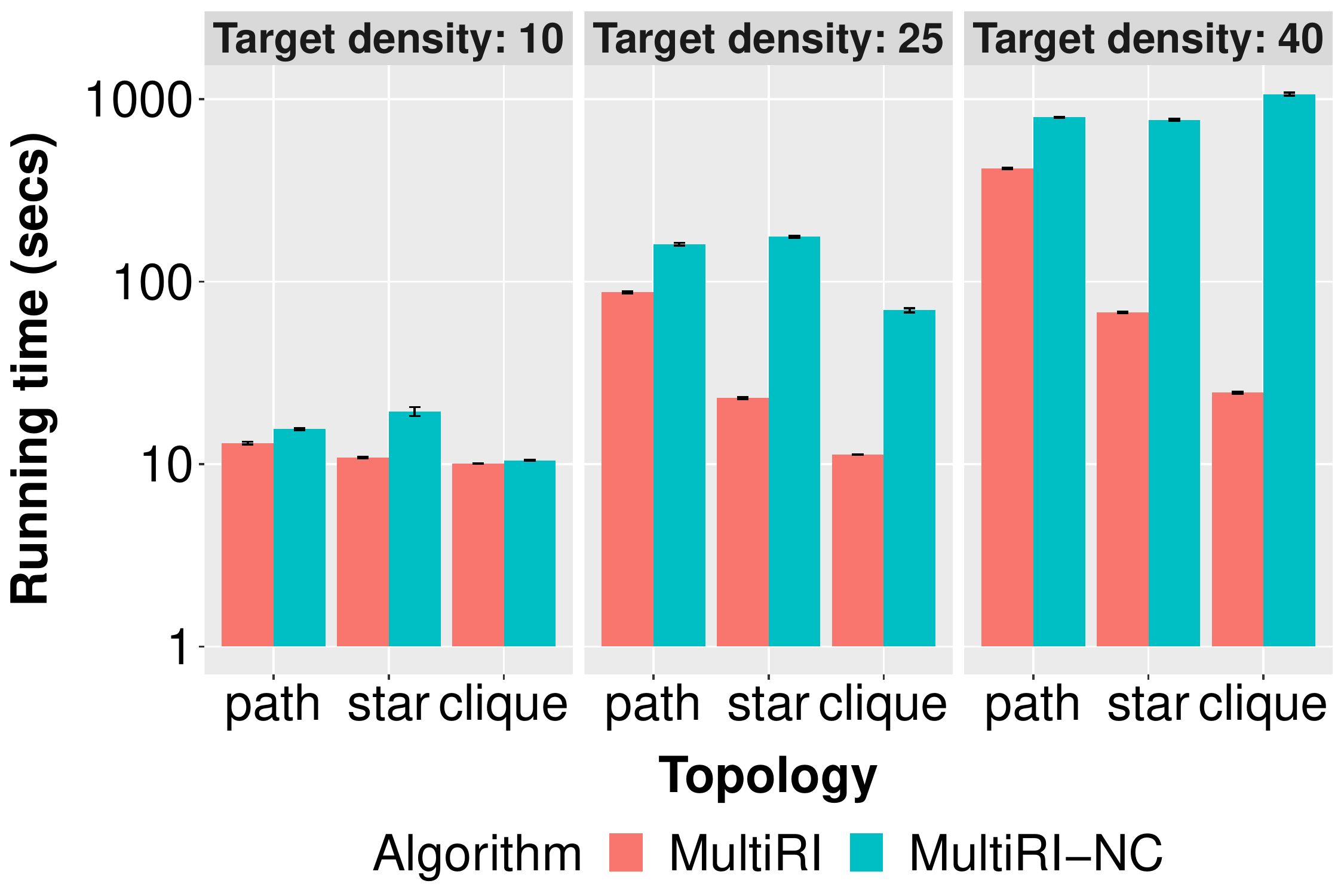}
\label{CompUnlabDensityTime}}
\subfloat[]
{\includegraphics[width=0.48\linewidth]{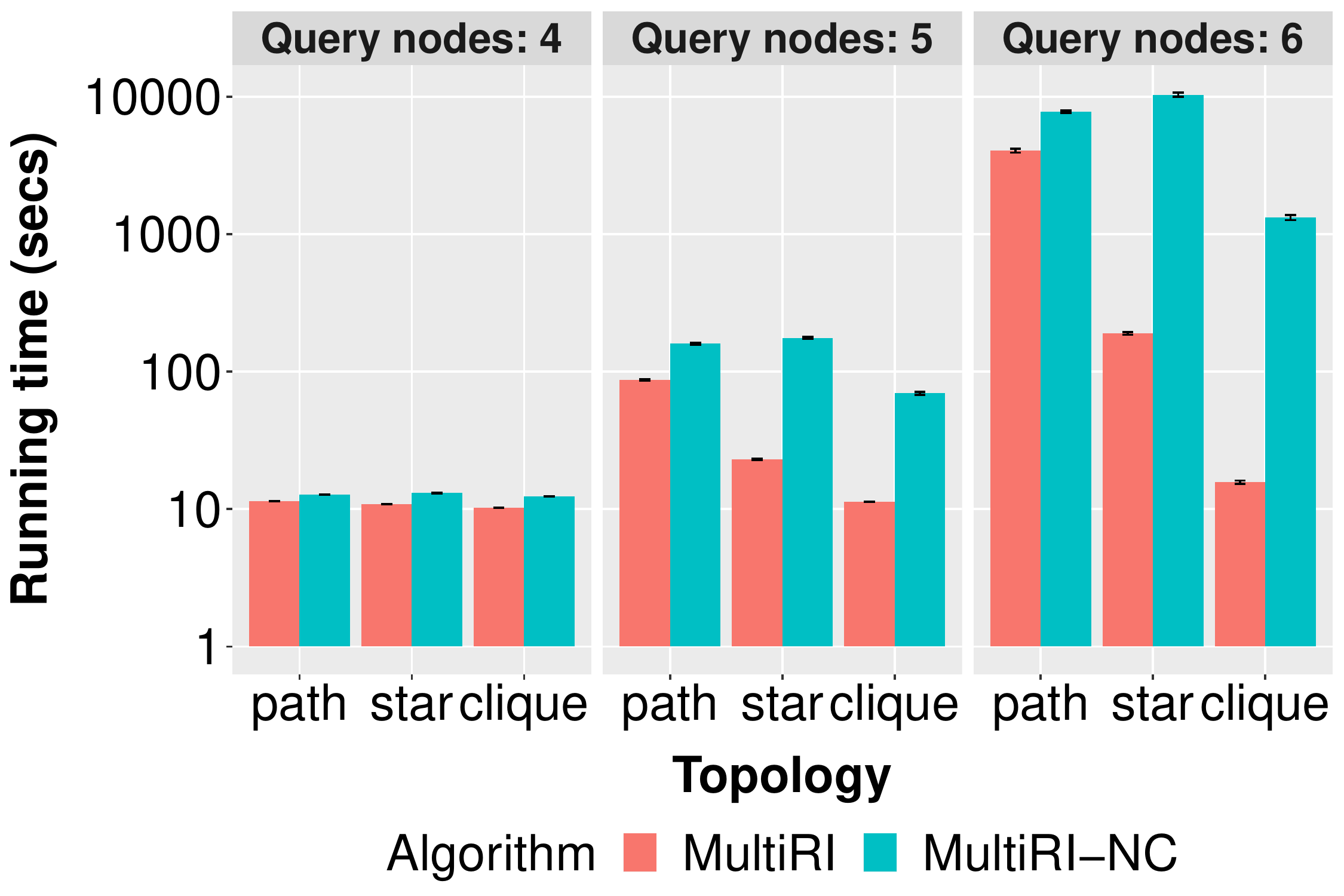}
\label{CompUnlabQueryTime}}
\end{tabular}
\caption{Running times of MultiRI with symmetry breaking conditions (MultiRI) vs MultiRI without breaking conditions (MultiRI-NC) in a dataset of artificial unlabeled Barabasi-Albert graphs with varying: a) target density and b) query size. Y-axes are on a logarithmic scale. The denser  the target and query, the greater  the difference between the two algorithms.}
\label{ComplexityUnlabeledTime}
\end{figure}

\begin{figure}[!ht]
\centering
\begin{tabular}{c}
\subfloat[]
{\includegraphics[width=0.48\linewidth]{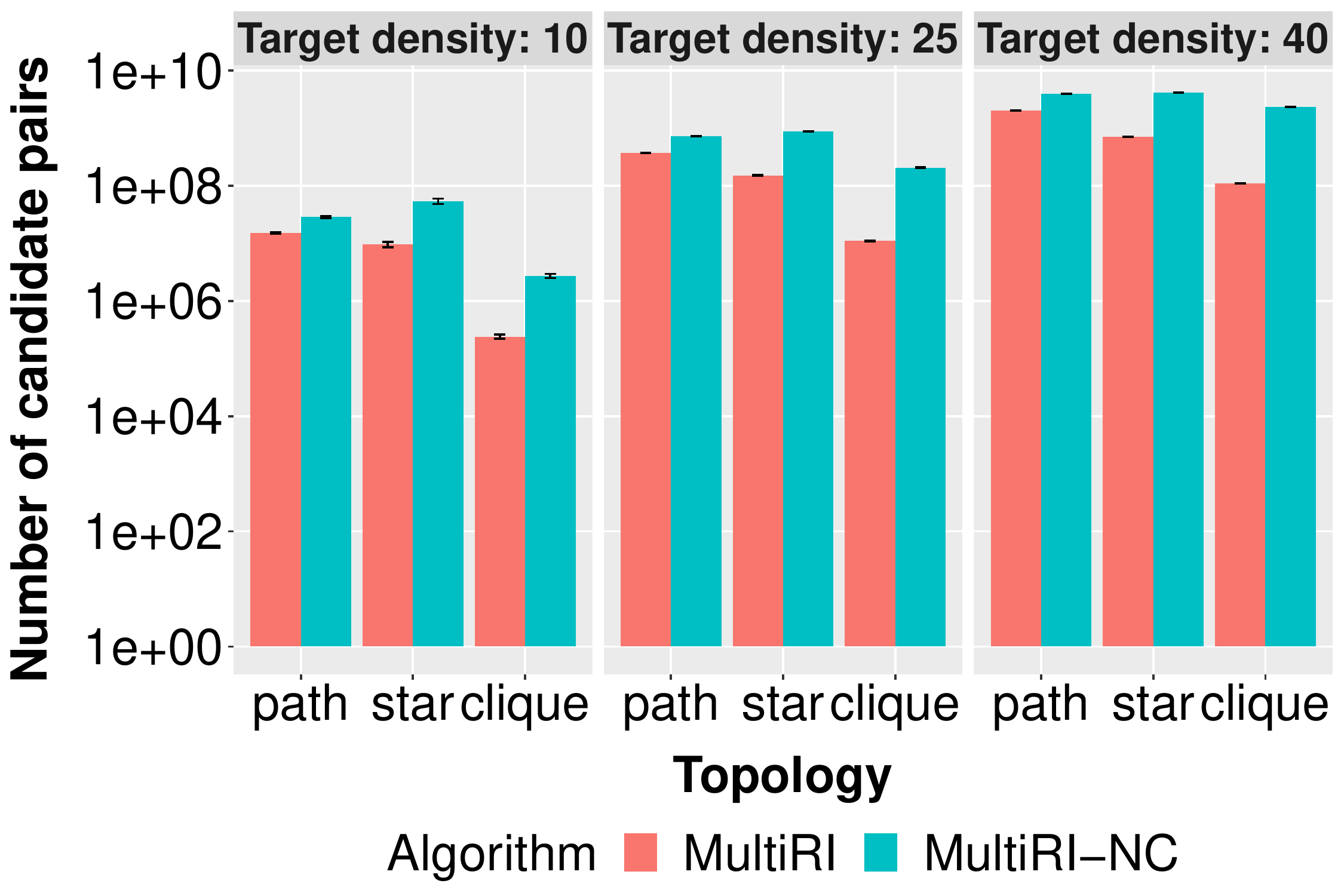}
\label{CompUnlabDensityStates}}
\subfloat[]
{\includegraphics[width=0.48\linewidth]{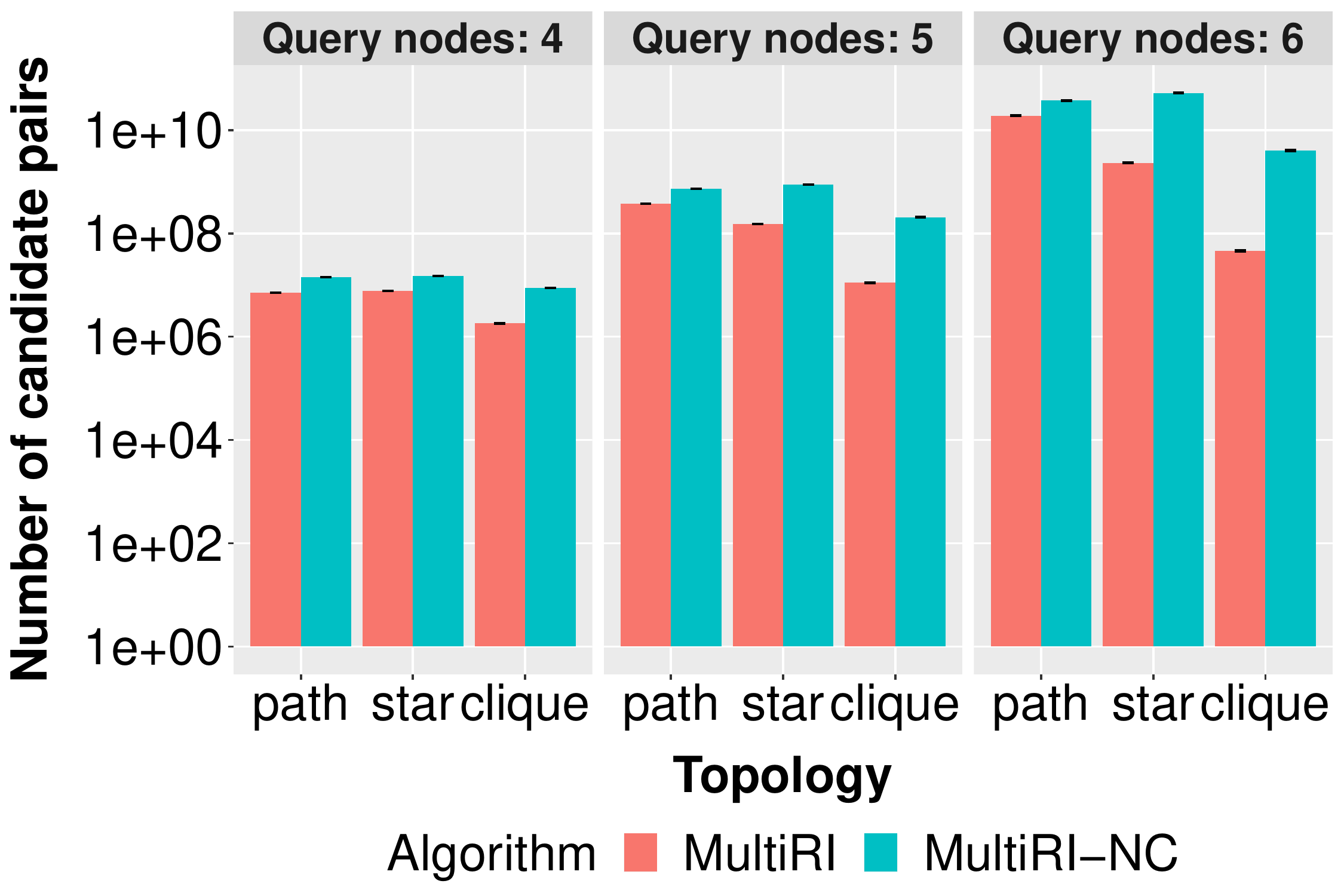}
\label{CompUnlabQueryStates}}
\end{tabular}
\caption{Number of candidate pairs examined by MultiRI with symmetry breaking conditions (MultiRI) vs MultiRI without breaking conditions (MultiRI-NC) in a dataset of artificial unlabeled Barabasi-Albert graphs with varying: a) target density and b) query size. Y-axes are shown in logarithmic scale. The denser are the target and query, the greater  the difference between the two algorithms.}
\label{ComplexityUnlabeledStates}
\end{figure}

\high{In Figs. \ref{ComplexityUnlabeledTime} and \ref{ComplexityUnlabeledStates} we compare the running times and the number of examined candidate pairs for matching of MultiRI with breaking conditions and MultiRI with no breaking conditions (called MultiRI-NC) with varying values of target density $d$ and query size $k$. For barplots with varying target density we took into account only queries with $k=5$ nodes, while for barplots with varying query size, we considered only targets with density $d=25$. All results show that breaking conditions reduce time more for dense query topologies than for sparse ones.
Specifically, independently of target density and query size, the reduction factor of running times and time complexity, derived theoretically in Section \ref{complexityAnalysis},  approximates the upper bounds 2 (for paths), $(k-2)!$ (for stars) and $(k-1)!$ (for cliques).}

\high{The second dataset comprises artificial Barabasi-Albert labeled graphs with $N=100$ nodes. Graphs were generated considering the following features in the target graph and three values for each parameter:}

\begin{itemize}
\item \high{Density $d$: 10, 25 and 40, that is the ratio between the number of edges and the number of nodes (multiple edges between two nodes are counted only once)};
\item Number of \high{distinct} node labels $\sigma$: 2, 5 and 8;
\item Maximum node multiplicity (i.e. maximum number of labels that any single node has)  $NM$: 1, 2 and 4;
\item Number of edge labels $\gamma$: 2, 5 and 8;
\item Maximum edge multiplicity (i.e. maximum number of edges there can be between two nodes) $EM$: 1, 2 and 4.
\end{itemize}

\high{We generated 10 graphs for each combination of values of these features. We excluded from the analysis all the combinations with $\sigma < NM$ and $\gamma < EM$. Labels for a node $u$ are numeric values between 1 and $\sigma$ and are chosen as follows: first, the node multiplicity is set randomly and uniformly to a  value $m$ between 1 and $NM$, then $m$ distinct labels are assigned to $u$ randomly and uniformly without replacement. Edge labels are numeric values between 1 and $\gamma$ and are chosen similarly to node labels.}

\high{We randomly extracted from each network queries with three different types of topologies (paths, stars and cliques) and varying number of nodes ($k=4,5,6$), by using a random walk approach similar to the one described in Subsection \ref{synthData}. For each kind of topology and value of $k$, we extracted 10 queries from each graph. So, the final dataset includes 1,050 artificial labeled graphs and 94,500 queries.}

Again, we compared both the running times and the number of examined pairs of MultiRI with and without breaking conditions on varying values of each feature.

\begin{figure}[!ht]
\centering
\begin{tabular}{c}
\subfloat[]
{\includegraphics[width=0.48\linewidth]{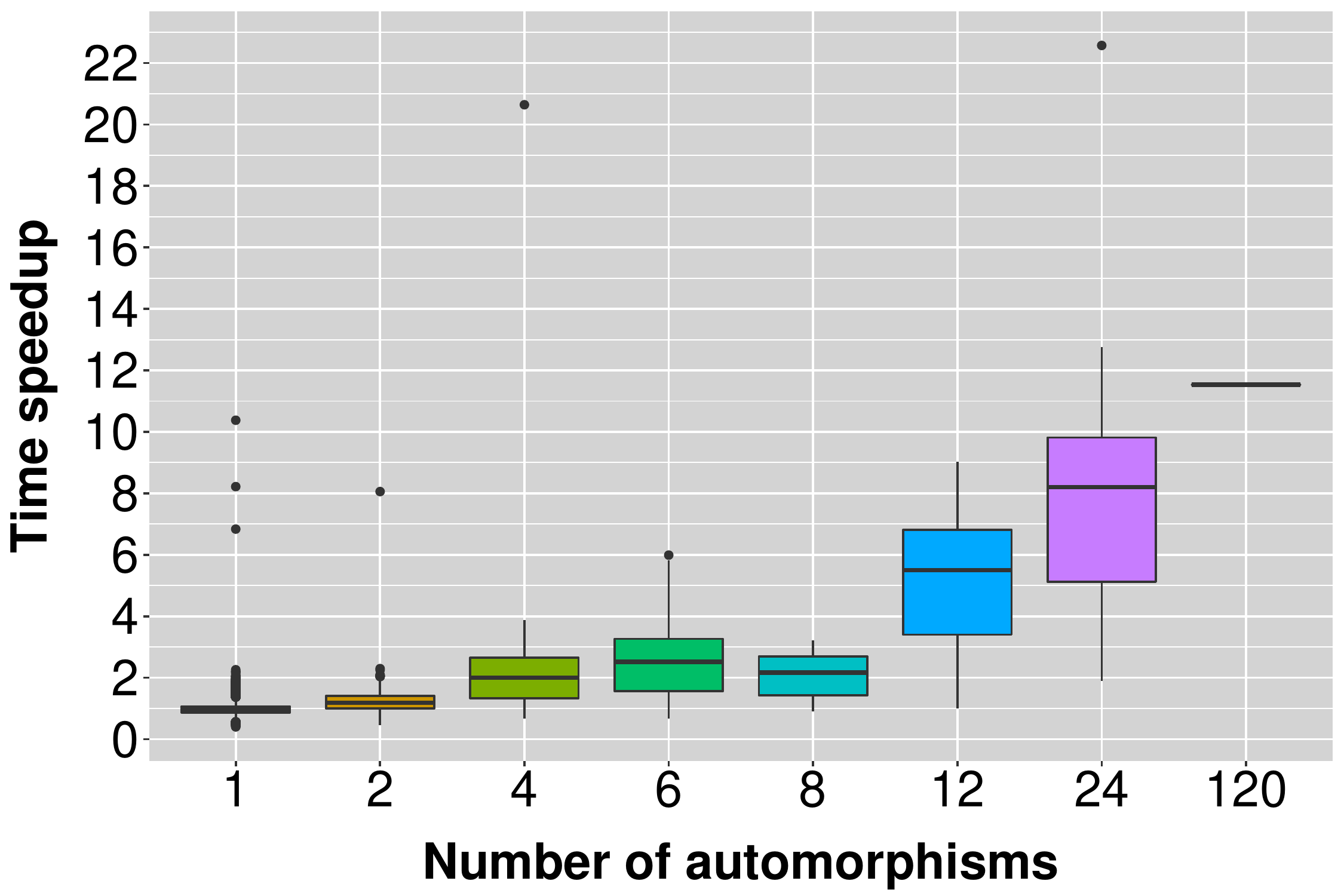}
\label{CompLabAutoTimes}}
\subfloat[]
{\includegraphics[width=0.48\linewidth]{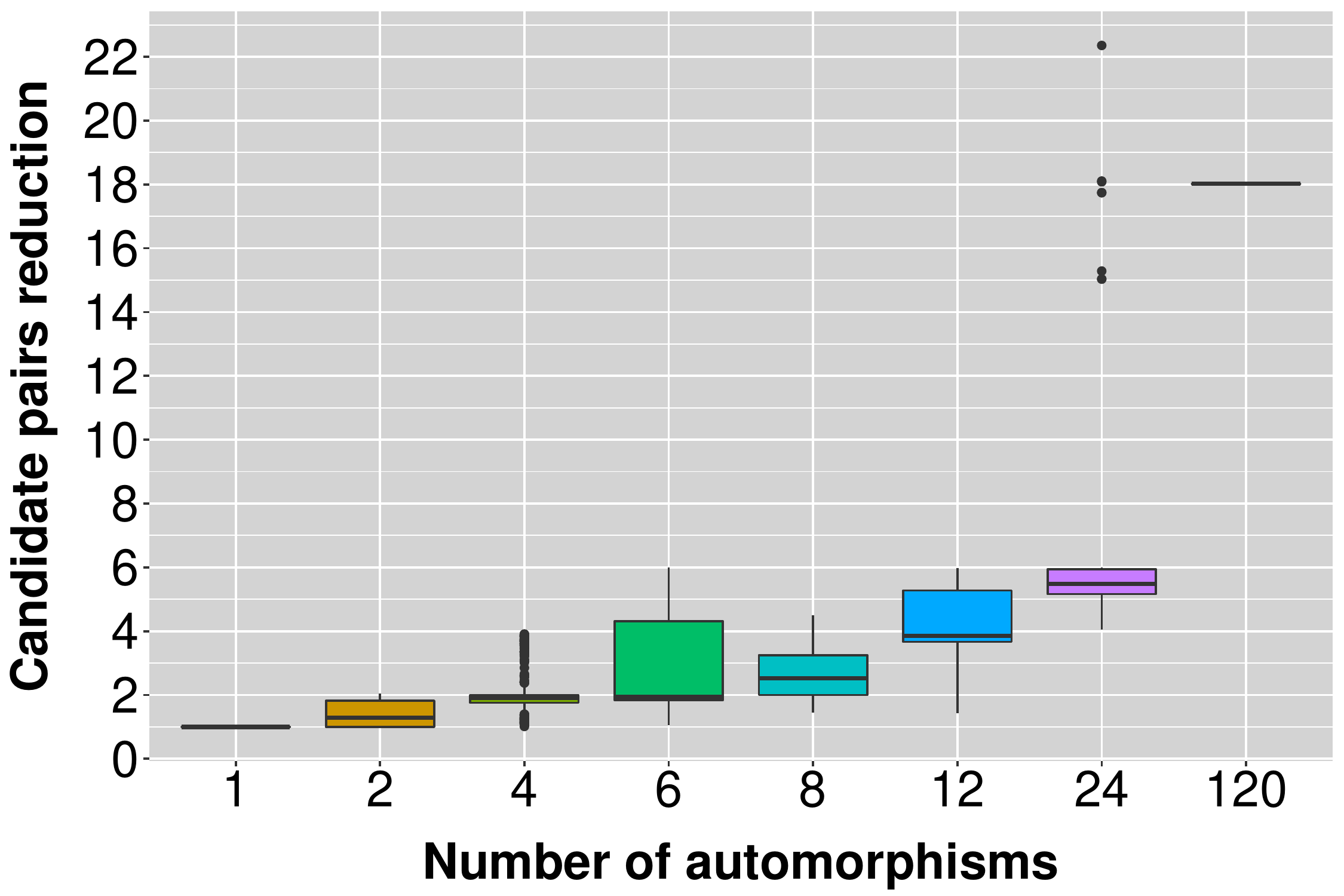}
\label{CompLabAutoStates}}
\end{tabular}
\caption{a) Average speedup and b) average reduction of candidate pairs for matching of MultiRI with breaking condition (MultiRI) with respect to MultiRI without breaking conditions (MultiRI-NC) with varying number of automorphisms of artificial labeled queries. The speedup exponentially increases with greater numbers of automorphisms.}
\label{ComplexityLabeledAuto}
\end{figure}

\high{Results at first glad say that on average there is no significant difference between the two versions of MultiRI (on both running time and number candidate pairs), even with few labels on nodes and edges and low multiplicities. On the other hand, they present very high standard deviation, implying that there is a lot of variability in results. 
Since in this case queries with the same topology may have different number of automorphisms depending on labels and multiplicities of nodes and edges, we grouped queries by their numbers of automorphisms. Then we built boxplots on the time speedup and explored candidate pairs reduction (Fig. \ref{ComplexityLabeledAuto}). These results confirm that breaking conditions strongly impact performances when we consider queries with a high number of automorphisms even on labeled multi-graphs.}

\section{Conclusion}
\label{conclusion}

\high{Labeled multigraphs, i.e. graphs allowing multiple labels on nodes and multiple labeled edges between two nodes) are a natural way to represent information. In fact, they fit applications like movie databases in which actors can act in many genres and in which a given actor and other actor may be related in many ways.}

MultiRI is a new algorithm and system for subgraph matching in \high{multigraphs} and allows queries on such graphs. MultiRI incorporates a few innovative algorithmic ideas including the use of filters based on lightweight  compatibility domains and symmetry breaking conditions.
The algorithm has been widely tested using a benchmark of both artificial and real multigraphs. 
Our analysis shows improvements of about a factor of ten with respect to the state-of-the art, across a variety of graphs and queries, \high{with a limited usage of memory.} MultiRI also enables scaling to multigraphs with millions of nodes and edges.

Many useful applications are already within reach of MultiRI as the real graphs of our experiments illustrate. Very large \high{multigraphs} will require access to secondary storage and parallelism. Those are subjects of future work.

\ifCLASSOPTIONcaptionsoff
  \newpage
\fi



%

%
\begin{IEEEbiography}[{\includegraphics[width=1in,height=1.25in,clip,keepaspectratio]{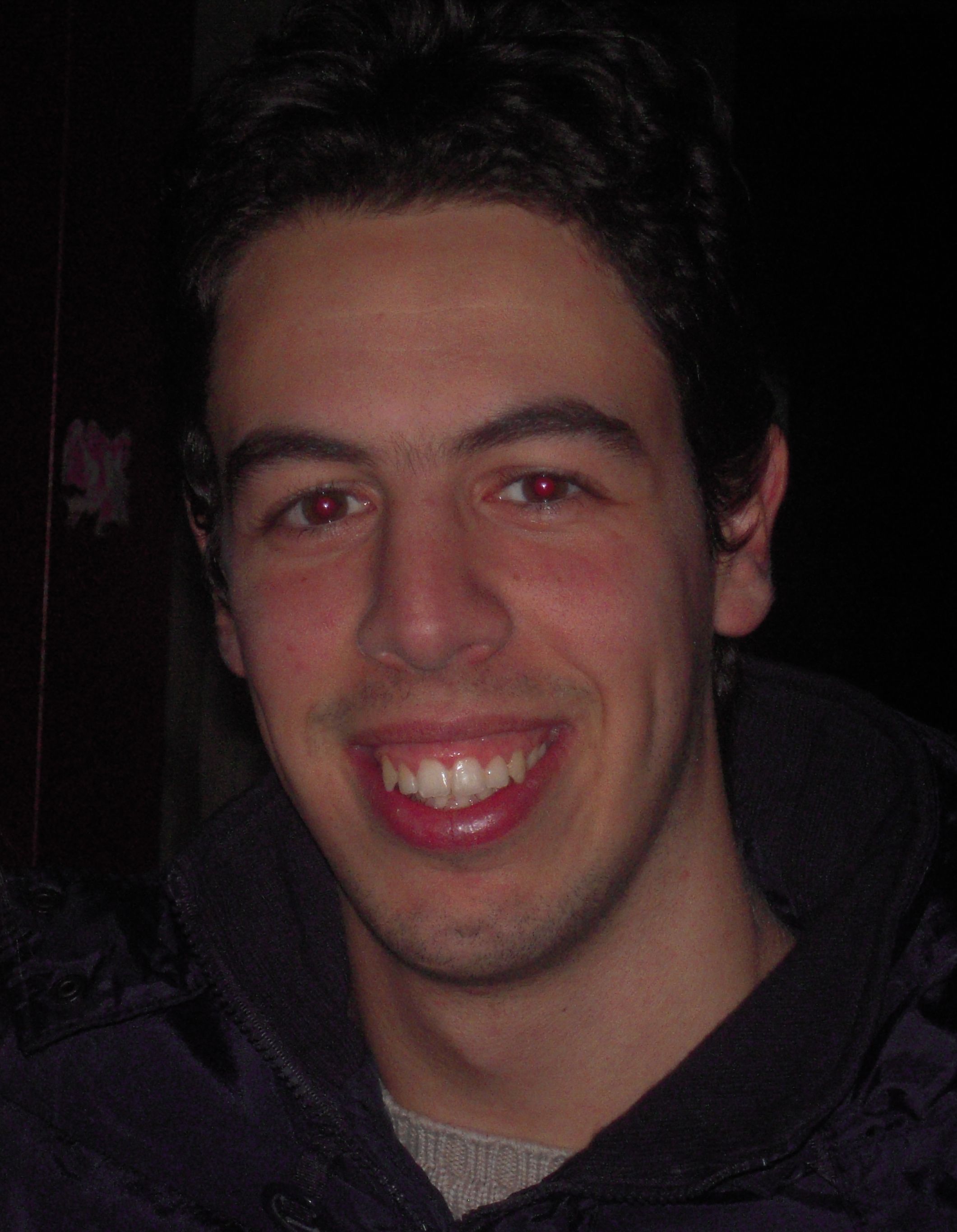}}]{Giovanni Micale}
is an Assistant Professor at the Department of Clinical and Experimental Medicine of University of Catania. He took his master degree in Computer Science at University of Catania and his PhD in Computer Science at University of Pisa in 2015. His research interests mainly focus on network analysis, with applications to biological and social networks. In particular, he works on network alignment, graph matching and network motifs finding.
\end{IEEEbiography}
\begin{IEEEbiography}[{\includegraphics[width=1in,height=1.25in,clip,keepaspectratio]{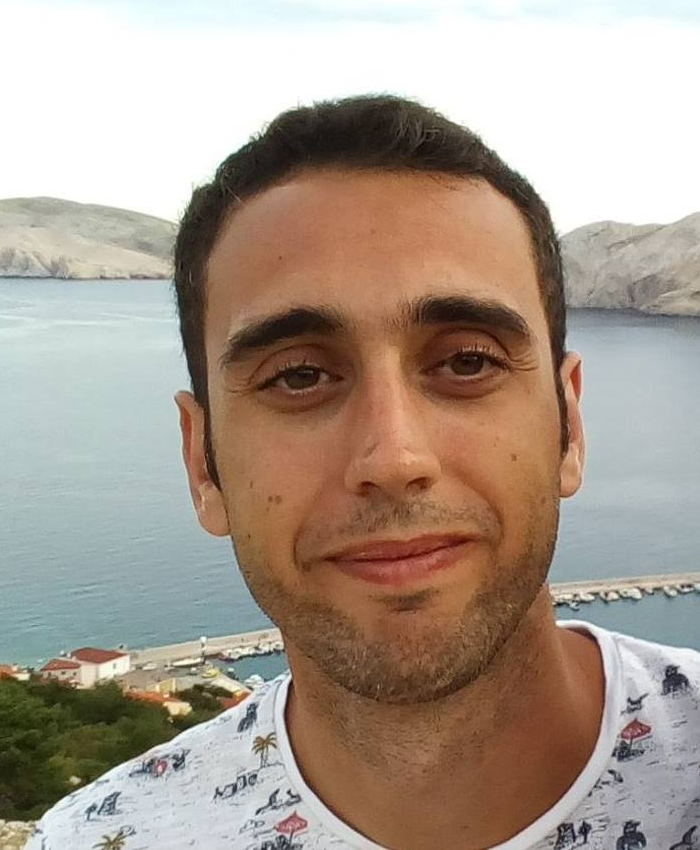}}]{Vincenzo Bonnici} received the master degree in Computer Science from the University of Catania (Italy), and the PhD degree in Computer Science from the University of Verona (Italy). His master thesis was focused on solving the subgraph isomorphism problem for biological networks. For the PhD thesis, he extended his interest to other bioinformatic fields such as genome sequence analysis and omic data integration. He has been abroad for his scientific studies and training at the Institute for Genomics and Bioinformatics, University of California, Irvine (USA), and at the Fondazione per la Ricerca e la Cura dei Linfomi nel Ticino, Istituto Oncologico della Svizzera Italiana, Bellinzona (CH). He is currently a research fellow at the Department of Computer Science of the University of Verona. He is a temporary professor for the Bachelor Degree in Computer Science and student mentor for the Master degree in Medical Bioinformatics. He his reviewer for several scientific journal published by IEEE, Elsevier, Frontiers, Oxford University Press and Springer. He is  editor for the MDPI journals Future Internet and Energies, and review editor for the Applied Mathematics and Statistics of Forntiers.
\end{IEEEbiography}
\begin{IEEEbiography}[{\includegraphics[width=1in,height=1.25in,clip,keepaspectratio]{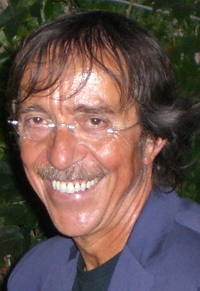}}]{Alfredo Ferro}
is Full Professor of Computer Science/Bioinformatics at the Department of Clinical and Experimental Medicine, University of Catania, Italy. He has been Director of Graduate Studies in Computer Science at the University of Catania from 1996 to 2002. He is the Director (since 1989) of the J.T.Schwartz International School for Scientific Research (Lipari School). His present research area includes Database, Data Mining and Algorithms with applications to Bioinformatics and Networking. In particular his research group has been mainly working on Computational RNAi and Network Biology. He has co-authored over 80 journal papers.
\end{IEEEbiography}
\begin{IEEEbiography}[{\includegraphics[width=1in,height=1.25in,clip,keepaspectratio]{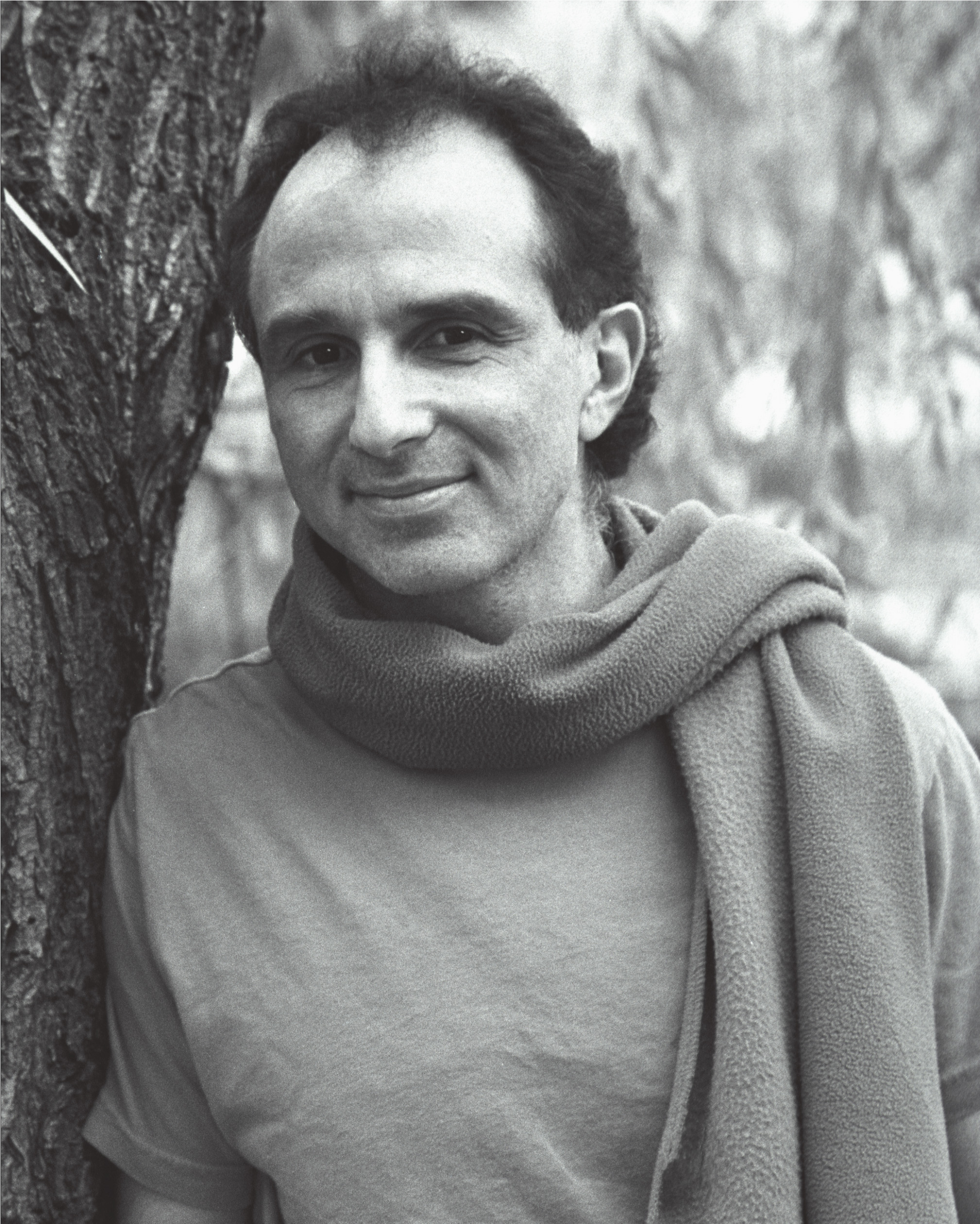}}]{Dennis Shasha}
is the Julius Silver Professor of Computer Science at New York University, an Associate Director of NYU WIRELESS, and an INRIA international chair. He has worked on pattern recognition on graphs and trees since 1988 and on general data-intensive problems since 1984. Because he likes to type, he has written
six books of puzzles about a mathematical detective named
Dr. Ecco, a biography about
great computer scientists,
and a book about the future of computing.
He has also written five technical books about database tuning, biological
pattern recognition, time series, DNA computing, resampling statistics,
and causal inference in molecular networks.He has co-authored
over 85 journal papers,
80 conference papers, and 25 patents.
He has written the puzzle column for various
publications including Scientific American, Dr. Dobb's Journal,
and currently the Communications of the ACM.
\end{IEEEbiography}
\begin{IEEEbiography}[{\includegraphics[width=1in,height=1.25in,clip,keepaspectratio]{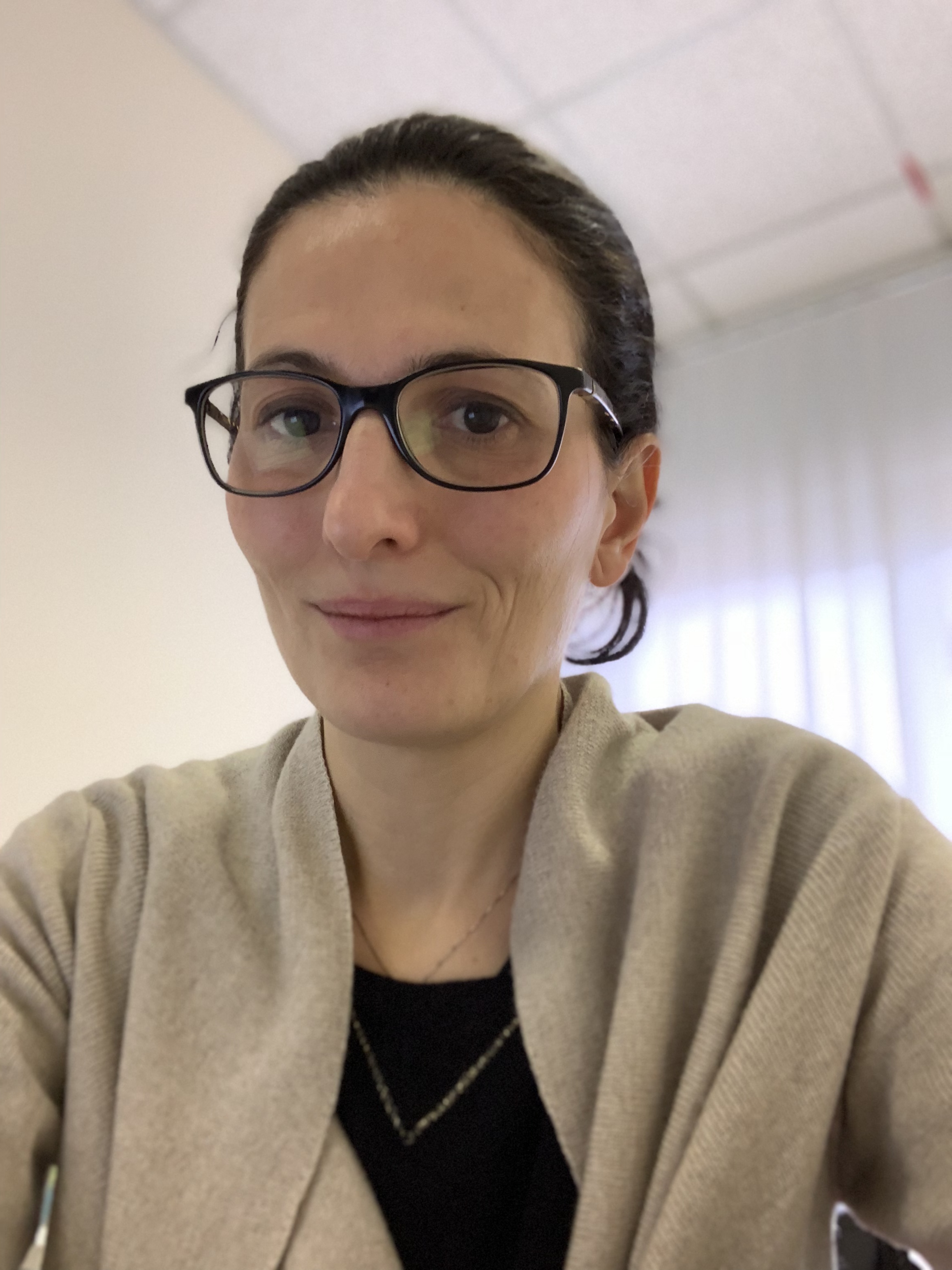}}]{Rosalba Giugno} is Associate Professor at the Department of Computer Science, University of Verona since May 2016.  She is the referent of the Master Degree in Medical Bioinformatics at University of Verona. She is the PI of the InfOmics laboratory at University of Verona. Her research is focused on algorithms for graphs and biological networks, integration and analysis of biomolecular data, modeling of biological systems, and RNA interference. She is author of 130 scientific publications, 70 in international journals. She is member of the Steering  Committee of Internal Conferences and Journal such as the  Jacob T. Schwartz International School for Scientific Research in BioInformatics and Computational Biology and in Social Complex Systems. Since 2017 she is the Director of Infolife laboratory made by 35 Italian Universities Research Units. The strategic rationale of the lab is to bring together all of investigative strategies developed over the last twenty years in Bioinformatics and related computer science research community, going by the algorithms, models, up to the formal systems, to form the bulk required to address critical issues in medical applications.
\end{IEEEbiography}
\begin{IEEEbiography}[{\includegraphics[width=1in,height=1.25in,clip,keepaspectratio]{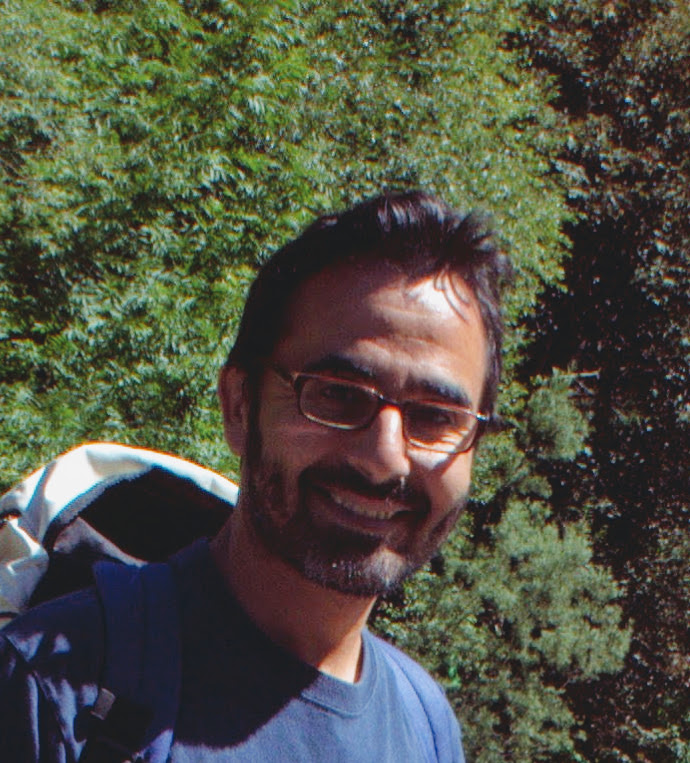}}]{Alfredo Pulvirenti}
is associate professor of Computer Science at the Department of Clinical and Experimental Medicine of University of Catania, Italy.  He is member of the PhD in “Complex Systems for Physical, Socio-economic and Life Sciences” at the University of Catania. He his member of the Steering Committee of ” Jacob T. Schwartz International School for Scientific Research”. Since the beginning of his carrier he has worked in the field of Data Mining and Bioinformatics. In the last few years his research has been focused on graph analysis problems, recommendation systems, data integration and high-throughput biological data mining in general. He also works with Istituto Nazionale di Geofisica e Vulcanologia (INGV) on the development of artificial intelligence models for light-weight seismic stations. He has co-authored over than 110 scientific publications, 75 in international journals.
\end{IEEEbiography}
\vfill
\clearpage
\pagestyle{empty}
\begin{appendix}
\label{appendix}
\renewcommand\thefigure{\thesection\arabic{figure}}
\setcounter{figure}{0}
\begin{figure}[b!]
\begin{algorithmic}[1]
\renewcommand{\algorithmicrequire}{\textbf{Procedure:}}
\REQUIRE \textsc{ComputeDomains}
\renewcommand{\algorithmicrequire}{\textbf{Input:}}
\renewcommand{\algorithmicensure}{\textbf{Output:}}
\renewcommand{\algorithmicprint}{\textbf{break}}
\REQUIRE $Q$: query, $T$: target
\ENSURE $Dom$: set of compatibility domains of nodes in $Q$
\FORALL{$q \in V_Q$} 
    \STATE{$Dom(q) := \emptyset$} 
\ENDFOR
\FORALL{$t \in V_T$}
    \FORALL{$q \in V_Q$}
        \IF{$\sigma_Q(q) \subseteq \sigma_T(t) \land \deg(q) \leq \deg(t)$}
            \STATE{$Dom(q) := Dom(q) \cup \{t\}$}
        \ENDIF
    \ENDFOR
\ENDFOR
\FORALL{$q' \in V_Q$}
    \FORALL{$t' \in Dom(q')$}
        \FORALL{$q'' \in V_Q : (q',q'') \in E_Q$}
            \IF{$ \not\exists\,\,t''\,\,\in Dom(q'') : (t',t'') \in E_T$}
                \STATE{$Dom(q') := Dom(q') \setminus \{t'\}$}
            \ENDIF
        \ENDFOR
    \ENDFOR
\ENDFOR
\RETURN $Dom$
\end{algorithmic}
\caption{Computation of compatibility domains.}
\label{domainsalgo}
\end{figure}
\begin{figure}[b!]
\begin{algorithmic}[1]
\renewcommand{\algorithmicrequire}{\textbf{Procedure:}}
\REQUIRE \textsc{OrderQueryNodes}
\renewcommand{\algorithmicrequire}{\textbf{Input:}}
\renewcommand{\algorithmicensure}{\textbf{Output:}}
\renewcommand{\algorithmicprint}{\textbf{break}}
\REQUIRE $Q$: query
\ENSURE $\mu$: ordered list of $Q$ nodes
\STATE $\mu :=$ empty
\STATE $\mathcal{U} := V_Q$
\WHILE{$|\mu|<|V_Q|$}
    \FORALL{$q \in \mathcal{U}$}
        \STATE $V_{q,vis} := V_{q,neig} := V_{q,unv} := \emptyset$
        \FORALL{$q' \in V_Q$}
            \IF{$q' \in \mu$}
                \IF{$q' \in N(q)$}
                    \STATE $V_{q,vis} := V_{q,vis} \cup \{q'\}$
                \ELSIF{$q' \in N(\mathcal{U} \cap N(q))$}
                    \STATE $V_{q,neig} := V_{q,neig} \cup \{q'\}$
                \ENDIF
            \ELSIF{$q' \in N(q) \land q' \notin N(\mu)$}
                \STATE $V_{q,unv} := V_{q,unv} \cup \{q'\}$
            \ENDIF
        \ENDFOR
    \ENDFOR
    \STATE $MAX_{vis} := \mathop{\mathrm{arg\,max}}_{q \in \mathcal{U}} |V_{q,vis}|$
    \STATE $MAX_{neig} := \mathop{\mathrm{arg\,max}}_{q \in MAX_{vis}} |V_{q,neig}|$
    \STATE $q_{max} := \mathrm{random}(\mathop{\mathrm{arg\,max}}_{q \in MAX_{neig}} |V_{q,unv}|)$
   \STATE append$(\mu,q_{max})$
    \STATE $\mathcal{U} := \mathcal{U} \setminus \{q_{max}\}$
\ENDWHILE
\RETURN $\mu$
\end{algorithmic}
\caption{Computation of the ordering of query nodes.}
\label{orderalgo}
\end{figure}
\begin{figure}[h!]
\begin{algorithmic}[1]
\renewcommand{\algorithmicrequire}{\textbf{Procedure:}}
\REQUIRE \textsc{SubgraphMatching}
\renewcommand{\algorithmicrequire}{\textbf{Input:}}
\renewcommand{\algorithmicensure}{\textbf{Output:}}
\renewcommand{\algorithmicprint}{\textbf{define}}
\REQUIRE $V_Q$: set of query nodes, $Dom$: compatibility domains, $\mathcal{C}$: set of breaking conditions, $\mu$: ordering of query nodes.
\ENSURE $Matches$: list of matches
\PRINT $f:$ partial mapping
\PRINT $\mathcal{M}:$ partial match
\FORALL{$q \in V_Q$}
\STATE $f(q) :=$ undefined
\ENDFOR
\STATE $\mathcal{M} := \emptyset$
\STATE $q :=$ first$(\mu)$
\STATE $Cand(q) := Dom(q)$
\STATE $Matches := \emptyset$
\STATE $Matches :=$ \textsc{Match}$(V_Q,Dom,\mu,\mathcal{C},f,\mathcal{M},Matches,$\\ \hspace{1.3in}
$q,Cand)$
\RETURN $Matches$
\end{algorithmic}
\caption{Sub-Multigraph matching.}
\label{submatchingalgo}
\end{figure}
\begin{figure}[h!]
\begin{algorithmic}[1]
\renewcommand{\algorithmicrequire}{\textbf{Procedure:}}
\REQUIRE \textsc{Match}
\renewcommand{\algorithmicrequire}{\textbf{Input:}}
\renewcommand{\algorithmicensure}{\textbf{Output:}}
\renewcommand{\algorithmicprint}{\textbf{break}}
\REQUIRE $V_Q$: set of query nodes, $Dom$: compatibility domains, $\mu$: ordering of query nodes, $\mathcal{C}$: set of breaking conditions, $f$: matching function, $\mathcal{M}$: partial match, $Matches$: set of matches, $q$: query node, $Cand$: sets of candidate nodes for each query node
\FORALL{$t \in Cand(q)$}
\STATE $feasible := \textsc{CheckFeasibility}(q,t,\mathcal{C},f,\mathcal{M})$
\IF{$feasible ==$ true}
\STATE $\mathcal{M} := \mathcal{M} \cup \{(q,t)\}$
\STATE $f(q) := t$
\IF{$|\mathcal{M}|=|V_Q|$}
\STATE $Matches := Matches \cup \{\mathcal{M}\}$
\STATE $f(q) :=$ undefined
\STATE $\mathcal{M} := \mathcal{M} \setminus \{(q,t)\}$
\ELSE
\STATE $q' := $next$(\mu,q)$
\STATE $Cand(q') := N(t) \cap Dom(q')$
\STATE $Matches := $ $\textsc{Match}(V_Q,Dom,\mu,\mathcal{C},f,\mathcal{M},$\\ \hspace{1.2in} $Matches,q',Cand)$
\ENDIF
\ENDIF
\ENDFOR
\renewcommand{\algorithmicprint}{\textbf{stop}}
\IF{$|\mathcal{M}|>0$}
\STATE $f(q) :=$ undefined
\STATE $\mathcal{M} := \mathcal{M} \setminus \{(q,t)\}$
\ELSE
\RETURN $Matches$
\ENDIF
\end{algorithmic}
\caption{Recursive \textsc{Match} procedure within MultiRI.}
\label{matchalgo}
\end{figure}

\begin{lemma}
\label{lemmaBreaking}
Given a query $Q$ having $k$ nodes, Algorithm of Fig. \ref{breakingalgo} ends when only one automorphism of $Q$ is left and returns the list of all symmetry breaking conditions of $Q$.
\end{lemma}

\begin{proof}
The algorithm starts by computing $A$, the automorphism matrix of $Q$ having $l$ rows and $k$ columns.
Let's order the column headers in ascending order according to node ids.
Each row contains the query nodes listed according to the automorphism it represents. For each pair of nodes $q_1$ and $q_2$ in an orbit, the columns corresponding to $q_1$ and $q_2$ will have the same set of nodes.
If $l \geq 2$, then one or more orbits will have $h \geq 2$ nodes. Among those orbits we will select $Orb = \{q_1,q_2,\cdots,q_i,\cdots,q_h\}$ the one with the minimum node id. Let $q_i$ be that node. 
There will be $h-1$ symmetry breaking conditions with respect to $q_i$ of the form  $\{q_i \prec q_j\}$ with $j \neq i$. The algorithm accumulates such breaking conditions in a set $C$. 
For each of them, there exist at least two rows $x$ and $y$ of the matrix in which $q_i$ and $q_j$ are in the same column and in at least one of these rows $q_j$ is to the left of $q_i$. The rows in which $q_j$ is to the left of $q_i$ will be then discarded.
After discarding these rows, the algorithm iterates by re-computing the orbits based on the remaining rows of $A$.  
The algorithm ends when the matrix $A$ has only one row ($l=1$). The resulting set $C$ will be a set of breaking conditions that are sufficient to reduce this matrix to one row, the identity automorphism (in which every query node maps to itself). 
\end{proof}

\begin{lemma}
\label{lemmaMatching}
Let $S$ be an occurrence of the query $Q$ in $T$. At the end of the matching process, Algorithm of Fig. \ref{matchalgo} returns subgraph $S$ no more than once. 
\end{lemma}

\begin{proof}
Let $l$ be the number of automorphisms of query $Q$ and $s_1,s_2,\cdots,s_k$ be the ids of nodes of $S$. The same occurrence $S$ can be matched to every automorphism of $Q$.

Let $q_1, q_2,\cdots, q_k$ be the processing order of the query nodes.
Condition 1 of Feasibility Rules ensures that, for each, $q_i$ and $q_j$ in $V_Q$ such that $q_i \prec q_j$ and $(q_j,M(q_j)) \in M$ we have $id(M(q_i)) < id(M(q_j))$. 

Now, assume that there are two  nodes in $S$, namely $s_i$ and $s_j$, that can be matched with the same query node $q$. Since the mapping is injective, there must be at least another query node $q'$ which can be mapped to $s_i$ or $s_j$. This means that in $Q$ there exist two automorphisms in which $q$ and $q'$ are matched. 

Two distinct cases may arise: 
\begin{inparaenum}[(i)]
    \item $q \prec q'$ (or vice-versa $q' \prec q$) in $C$. 
In this case the algorithm will apply the SBC rule on node identifiers $s_i$ and $s_j$ resulting in one of the following two inequalities: $s_i < s_j$ or $s_j< s_i$. One of these will violate the SBC rule and the corresponding partial matching will be discarded.
\item  There is no breaking condition involving $q$ and $q'$. In this case both partial matches $\cdots,(q,s_i),\cdots,(q',s_j),\cdots$ and $\cdots,(q,s_j),\cdots,(q',s_i),\cdots$  would be possible and  both matches would be returned by the algorithm. This implies that there is an automorphism of the query $Q$ in which $q$ maps to $q'$ without a corresponding breaking condition.
However, if at the and of the algorithm we get both matches, then this contradicts Lemma 1 which ensures that the breaking conditions in $C$ are enough to yield only the identity automorphism. Therefore this case cannot arise.
\end{inparaenum}

The same reasoning applies to every pair of nodes in $S$ that can be matched to the same query node, so at the end of the matching process there is only one possible mapping between query nodes and occurrence nodes. Therefore the set of target nodes $S$ will be returned only once.
\end{proof}

\begin{theorem}
\label{theorem}
For each occurrence $S$ of the query $Q$ in $T$, the Algorithm of Fig. \ref{matchalgo} will returns $S$ exactly once.
\end{theorem}
\begin{proof}
The algorithm of Fig. \ref{matchalgo}, constructs the occurrence $S$ by checking all the possible suitable matches.  When it eliminates one match violating a breaking conditions, thanks to Lemma \ref{lemmaMatching}, there exists at least another (which will be constructed in the next iterations) one with same nodes and edges which will do not violate them. Therefore $S$ will be returned only once. 

\end{proof}

\end{appendix}

\end{document}